\documentclass[fleqn,usenatbib]{mnras}

\usepackage{longtable,lscape}
\newcommand{\be}{\begin{equation}}
\newcommand{\ee}{\end{equation}}
\newcommand{\bea}{\begin{eqnarray}}
\newcommand{\eea}{\end{eqnarray}}
\newcommand{\Om}{\Omega_{m0}}
\newcommand{\Ok}{\Omega_{k0}}
\newcommand{\wX}{w_{\rm X}}
\newcommand{\om}{$\Omega_{m0}$}
\newcommand{\ok}{$\Omega_{k0}$}
\newcommand{\wx}{$w_{\rm X}$}
\newcommand{\hii}{H\,\textsc{ii}}
\newcommand{\mq}{Mg\,\textsc{ii} QSO}
\newcommand{\mii}{Mg\,\textsc{ii}}
\newcommand{\cq}{C\,\textsc{iv} QSO}
\newcommand{\civ}{C\,\textsc{iv}}
\newcommand{\rfe}{${\cal R}_{\rm{Fe\,\textsc{ii}}}$}
\newcommand{\Feii}{Fe\,\textsc{ii}}
\newcommand{\obh}{\Omega_{b}h^2}
\newcommand{\och}{\Omega_{c}h^2}
\newcommand{\onh}{\Omega_{\nu}h^2}
\newcommand{\obhs}{$\Omega_{b}h^2$}
\newcommand{\ochs}{$\Omega_{c}h^2$}
\newcommand{\hunit}{$\rm{km \ s^{-1} \ Mpc^{-1}}$}
\newcommand{\lcdm}{$\Lambda$CDM}
\newcommand{\pcdm}{$\phi$CDM}
\usepackage[margin=5pt]{subfig}
\usepackage{booktabs}
\usepackage[figuresright]{rotating}
\usepackage{textcase}
\usepackage{array}
\usepackage{caption}
\usepackage{threeparttable}
\usepackage{scalerel}
\usepackage{tikz}
\usetikzlibrary{svg.path}

\definecolor{orcidlogocol}{HTML}{A6CE39}
\tikzset{
  orcidlogo/.pic={
    \fill[orcidlogocol] svg{M256,128c0,70.7-57.3,128-128,128C57.3,256,0,198.7,0,128C0,57.3,57.3,0,128,0C198.7,0,256,57.3,256,128z};
    \fill[white] svg{M86.3,186.2H70.9V79.1h15.4v48.4V186.2z}
                 svg{M108.9,79.1h41.6c39.6,0,57,28.3,57,53.6c0,27.5-21.5,53.6-56.8,53.6h-41.8V79.1z M124.3,172.4h24.5c34.9,0,42.9-26.5,42.9-39.7c0-21.5-13.7-39.7-43.7-39.7h-23.7V172.4z}
                 svg{M88.7,56.8c0,5.5-4.5,10.1-10.1,10.1c-5.6,0-10.1-4.6-10.1-10.1c0-5.6,4.5-10.1,10.1-10.1C84.2,46.7,88.7,51.3,88.7,56.8z};
  }
}

\newcommand\orcidicon[1]{\href{https://orcid.org/#1}{\mbox{\scalerel*{
\begin{tikzpicture}[yscale=-1,transform shape]
\pic{orcidlogo};
\end{tikzpicture}
}{|}}}}

\usepackage[T1]{fontenc}

\DeclareRobustCommand{\VAN}[3]{#2}
\let\VANthebibliography\thebibliography
\def\thebibliography{\DeclareRobustCommand{\VAN}[3]{##3}\VANthebibliography}

\usepackage{graphicx}	
\usepackage{amsmath}	
\usepackage{amssymb}	
\usepackage{newtxtext,newtxmath}

\title[\civ\ radius-luminosity and cosmology]{Standardizing reverberation-measured \civ\ time-lag quasars, and using them with standardized \mii\ quasars to constrain cosmological parameters}

\author[Cao et al.]{
Shulei Cao$^{\orcidicon{0000-0003-2421-7071}{1}}$\thanks{E-mail: shulei@phys.ksu.edu},
Michal Zaja\v{c}ek$^{\orcidicon{0000-0001-6450-1187}{2}}$\thanks{E-mail: zajacek@mail.muni.cz},
Swayamtrupta Panda$^{\orcidicon{0000-0002-5854-7426}{3,4}}$\thanks{E-mail: panda@cft.edu.pl}\thanks{CNPq Fellow},
Mary Loli Mart\'inez-Aldama$^{\orcidicon{0000-0002-7843-7689}{5}}$\thanks{E-mail: mmartinez@das.uchile.cl},
\newauthor \hspace{0.1mm}
Bo\.zena Czerny$^{\orcidicon{0000-0001-5848-4333}{3}}$\thanks{E-mail: bcz@cft.edu.pl},
Bharat Ratra$^{\orcidicon{0000-0002-7307-0726}1}$\thanks{E-mail: ratra@phys.ksu.edu}
\\
$^{1}$Department of Physics, Kansas State University, 116 Cardwell Hall, Manhattan, KS 66506, USA\\
$^{2}$Department of Theoretical Physics and Astrophysics, Faculty of Science, Masaryk University, Kotl\'a\v{r}sk\'a 2, 611 37 Brno, Czech Republic\\
$^{3}$Center for Theoretical Physics, Polish Academy of Sciences, Al.\ Lotnik\'{o}w 32/46, 02-668 Warsaw, Poland\\
$^{4}$Laborat\'orio Nacional de Astrof\'isica - MCTIC, R. dos Estados Unidos, 154 - Na\c{c}\~oes, Itajub\'a - MG, 37504-364, Brazil\\
$^{5}$Departamento de Astronomía, Universidad de Chile, Camino del Observatorio 1515, Santiago, Chile\\
}

\date{Accepted XXX. Received YYY; in original form ZZZ}

\pubyear{2022}

\begin{document}
\label{firstpage}
\pagerange{\pageref{firstpage}--\pageref{lastpage}}
\maketitle

\begin{abstract}
We use 38 \civ\ quasar (QSO) reverberation-mapped (RM) observations, which span eight orders of magnitude in  luminosity and the redshift range $0.001064 \leq z \leq 3.368$, to simultaneously constrain cosmological-model and QSO radius-luminosity ($R-L$) relation parameters in six cosmological models, using an improved technique that more correctly accounts for the asymmetric errors bars of the time-lag measurements. We find that $R-L$ relation parameters are independent of the cosmological models used in the analysis and so the $R-L$ relation can be used to standardize the \civ\ QSOs. The \civ\ QSO cosmological constraints are consistent with those from \mii\ QSOs, allowing us to derive joint \civ\ + \mii\ QSO cosmological constraints which are consistent with currently accelerated cosmological expansion, as well as consistent with cosmological constraints derived using better-established baryon acoustic oscillation (BAO) and Hubble parameter [$H(z)$] measurements. When jointly analyzed with $H(z)$ + BAO data, current \civ\ + \mii\ QSO data mildly tighten current $H(z)$ + BAO data cosmological constraints.
\end{abstract}




\section{Introduction}

The well-established observed currently accelerated expansion of the Universe has motivated many theoretical cosmological models. In the standard general-relativistic spatially-flat \lcdm\ cosmological model \citep{peeb84} dark energy in the form of a time-independent cosmological constant $\Lambda$ powers the currently accelerated cosmological expansion and contributes $\sim70\%$ of the current cosmological energy budget, with non-relativistic cold dark matter (CDM) and baryonic matter contributing $\sim25\%$ and $\sim5\%$, respectively. Although the flat \lcdm\ model makes predictions consistent with most observations \citep[see, e.g.][]{scolnic_et_al_2018, Yuetal2018, planck2018b, eBOSS_2020}, it has some potential observational discrepancies \citep{DiValentinoetal2021a, PerivolaropoulosSkara2021, Abdallaetal2022} that motivate us to also study dynamical dark energy models as well as models with non-zero spatial curvature.

There are better-established cosmological probes, such as baryon acoustic oscillation (BAO), type Ia supernova apparent magnitude, and Hubble parameter [$H(z)$] measurements that, when jointly analyzed, provide fairly restrictive cosmological parameter constraints \citep[see, e.g.][]{CaoRatra2022}. Cosmological probes that are now under active development can tighten these constraints. Amongst these developing probes are \hii\ starburst galaxy apparent magnitude data that reach to redshift $z \sim 2.4$ \citep{Mania_2012, Chavez_2014, GM2021, CaoRyanRatra2020, CaoRyanRatra2021, Johnsonetal2022, Mehrabietal2022}, quasar (QSO) angular size measurements that reach to $z \sim 2.7$ \citep{Cao_et_al2017a, Ryanetal2019, Zhengetal2021, Lian_etal_2021, CaoRyanRatra2022}, QSO flux observations that reach to $z \sim 7.5$ \citep{RisalitiLusso2015, RisalitiLusso2019, KhadkaRatra2020a, KhadkaRatra2020b, KhadkaRatra2021, KhadkaRatra2022, Lussoetal2020, ZhaoXia2021, Rezaeietal2022, Luongoetal2021, Leizerovichetal2021, Colgainetal2022, DainottiBardiacchi2022},\footnote{Note however that the \cite{Lussoetal2020} QSO flux compilation assumes a model for the QSO UV--X-ray correlation that is invalid above redshifts $z \sim 1.5-1.7$ so this is the limit to which the \cite{Lussoetal2020} QSOs can be used to determine cosmological constraints \citep{KhadkaRatra2021, KhadkaRatra2022}.} gamma-ray burst data that reach to $z \sim 8.2$ \citep{Wang_2016, Wangetal_2021, Dirirsa2019, Demianskietal_2021, KhadkaRatra2020c, Huetal_2021, Khadkaetal_2021b, LuongoMuccino2021, Caoetal_2021, CaoDainottiRatra2022b, CaoKhadkaRatra2022, CaoDainottiRatra2022, DainottiNielson2022, Liuetal2022}, and --- the main subject of our paper --- reverberation-mapped (RM) QSO data that reach to $z \sim 3.4$ \citep[][and this paper]{Czernyetal2021, Zajaceketal2021, Khadkaetal_2021a, Khadkaetal2022a, Khadkaetal2021c}. 

\citet{Khadkaetal_2021a, Khadkaetal2022a, Khadkaetal2021c} derived cosmological constraints from RM H$\beta$ and \mq\ data. In this paper, we show that RM \cq\ data, that extend to higher $z$, are standardizable and derive the first cosmological constraints from \civ\ data. 

With its ionization potential energy of $\sim 64.5$ eV, \civ\ ($\lambda$1549) belongs to the high-ionization line (HIL) component of the QSO broad-line region \citep[BLR; ][]{1988MNRAS.232..539C,2005MNRAS.356.1029B,Karasetal2021}, which can partially form an outflow that is manifested by the blueshifted centroid of the line \citep[see e.g.][]{2021MNRAS.503.3145B} as well as by the frequent blueward line-emission asymmetry \citep{2005MNRAS.356.1029B}. It is therefore not yet established whether all of the \civ\ material is virialized and if the standard reverberation mapping (hereafter RM\footnote{Depending on the context, we use the abbreviation RM for both reverberation mapping as a method and for reverberation-mapped quasars interchangeably.}) of HILs can lead to the reliable measurements of the SMBH masses, as was previously done using the low-ionization lines (LILs) for several hundreds of objects (mainly using the Balmer line H$\beta$ and the resonance \mii\ line).

However, the BLR radius-luminosity ($R-L$) relationship for the \civ\ line, in the flat $\Lambda$CDM model, appears to now be well established with a significant positive correlation and a relatively small dispersion \citep{2007ApJ...659..997K,2018ApJ...865...56L,2019ApJ...887...38G,2021ApJ...915..129K}, which allows for the possibility of using this relation for constraining cosmological parameters as we previously did for H$\beta$ and \mii\ lines \citep{2019ApJ...883..170M, 2019FrASS...6...75P, Zajaceketal2021,Czernyetal2021,Khadkaetal_2021a,Khadkaetal2022a,Khadkaetal2021c}.

In this paper, we use 38 high-quality \cq\ measurements, which span eight orders of magnitude in luminosity $\sim 10^{40-48}$ erg s$^{-1}$ and the redshift range $0.001064 \leq z \leq 3.368$, to constrain, for the first time, cosmological-model and $R-L$ relation parameters, in six general relativistic dark energy cosmological models, using a more correct technique to account for the asymmetric errors of the time-lag measurements. We find that the \civ\ $R-L$ relation parameters are independent of cosmological model, so \cq\ data are standardizable through the \civ\ $R-L$ relation. \mq\ data were also found to be standardizable through the \mii\ $R-L$ relation \citep{Khadkaetal_2021a, Khadkaetal2022a}. We find that cosmological constraints from \civ\ and \mii\ QSO data are mutually consistent and are also consistent with those from $H(z)$ + BAO data. Although the cosmological constraints from the joint analysis of \civ\ + \mii\ QSO data are weak, jointly analyzing \civ\ + \mii\ data with $H(z)$ + BAO data results in a mild ($<0.1\sigma$) tightening of the $H(z)$ + BAO cosmological constraints.

This paper is organized as follows. We briefly introduce the cosmological models/parametrizations we study in Section \ref{sec:model}. In Sections \ref{sec:data} and \ref{sec:analysis}, we outline the data sets and the analysis methods we use, respectively. Our constrained cosmological parameter and $R-L$ relation parameter results are presented and discussed in Sections \ref{sec:results} and \ref{sec:Discussion}. We summarize our conclusions in Section \ref{sec:conclusion}.

\section{Cosmological models}
\label{sec:model}

We use various combinations of data to simultaneously constrain cosmological model parameters and \civ\ and \mq\ $R-L$ relation parameters in six spatially-flat and non-flat dark energy cosmological models.\footnote{For discussions of spatial curvature observational constraints see \citet{Chenetal2016}, \citet{Ranaetal2017}, \citet{Oobaetal2018a, Oobaetal2018b}, \citet{ParkRatra2019a, ParkRatra2019b}, \citet{DESCollaboration2019}, \citet{Lietal2020}, \citet{Handley2019}, \citet{EfstathiouGratton2020}, \citet{DiValentinoetal2021b}, \citet{Vagnozzietal2020, Vagnozzietal2021}, \citet{KiDSCollaboration2021}, \citet{ArjonaNesseris2021}, \citet{Dhawanetal2021}, \citet{Renzietal2021}, \citet{Gengetal2022}, \citet{WeiMelia2022}, \citet{MukherjeeBanerjee2022}, \citet{Glanvilleetal2022}, and references therein.}
The Hubble parameter $H(z)$, discussed below, is used to make theoretical predictions in these cosmological models. 

In the cosmological models here, we assume one massive and two massless neutrino species with the non-relativistic neutrino physical energy density parameter $\onh=\sum m_{\nu}/(93.14\ \rm eV)=0.06\ \rm eV/(93.14\ \rm eV)$, where $h$ is the Hubble constant $H_0$ in units of 100 \hunit. The non-relativistic matter density parameter $\Om = (\onh + \obh + \och)/{h^2}$, where \obhs\ and \ochs\ are the current values of the observationally-constrained baryonic and cold dark matter physical energy density parameters, respectively.\footnote{In the \cq\ and \mq\ alone cases, the current value of the baryonic matter energy density parameter and the Hubble constant are set to $\Omega_b=0.05$ and $H_0=70$ \hunit, respectively, because these data alone are unable to constrain $\Omega_b$ and $H_0$.} Including neutrino species is more accurate even though it only has a mild effect on the constraints from \mii\ and \civ\ data.

In the \lcdm\ models the Hubble parameter
\be
\label{eq:EzL}
    H(z, \textbf{\emph{p}}) = H_0\sqrt{\Om\left(1 + z\right)^3 + \Ok\left(1 + z\right)^2 + \Omega_{\Lambda}},
\ee
where the cosmological parameters $\textbf{\emph{p}}=\{\Ok,\Om,H_0\}$ with $\Ok$ being the spatial curvature energy density parameter and the cosmological constant dark energy density parameter $\Omega_{\Lambda} = 1 - \Om - \Ok$. In the flat \lcdm\ model the constrained cosmological parameters are $H_0$, \obhs, and \ochs\ (only $\Omega_c$ is constrained in the \cq\ and \mq\ alone cases), whereas in the non-flat \lcdm\ model one additional cosmological parameter, \ok, is constrained.

In the XCDM parametrizations 
\be
\label{eq:EzX}
    H(z, \textbf{\emph{p}}) = H_0\sqrt{\Om\left(1 + z\right)^3 + \Ok\left(1 + z\right)^2 + \Omega_{\rm X0}\left(1 + z\right)^{3\left(1 + \wX\right)}},
\ee
where the cosmological parameters $\textbf{\emph{p}}=\{\Ok,\Om,H_0,\wX\}$ with \wx\ being the X-fluid equation of state parameter ($\wX=-1$ correspond to \lcdm\ models). The current value of the dynamical dark energy density parameter of the X-fluid $\Omega_{\rm X0} = 1 - \Om - \Ok$. In the flat XCDM parameterization the constrained cosmological parameters are $H_0$, \obhs, \ochs, and \wx\ (only $\Omega_c$ and \wx\ are constrained in the \cq\ and \mq\ alone cases), whereas in the non-flat XCDM parametrization \ok\ is constrained as well. 

In the \pcdm\ models \citep{peebrat88,ratpeeb88,pavlov13}\footnote{For discussions of \pcdm\ observational constraints see  \cite{Zhaietal2017}, \cite{ooba_etal_2018b, ooba_etal_2019}, \cite{park_ratra_2018, park_ratra_2019b, park_ratra_2020}, \cite{SolaPercaulaetal2019}, \cite{Singhetal2019}, \cite{UrenaLopezRoy2020}, \cite{SinhaBanerjee2021}, \cite{Xuetal2021}, \cite{deCruzetal2021}, \cite{Jesusetal2021}, and references therein.}
\be
\label{eq:Ezp}
    H(z, \textbf{\emph{p}}) = H_0\sqrt{\Om\left(1 + z\right)^3 + \Ok\left(1 + z\right)^2 + \Omega_{\phi}(z,\alpha)},
\ee
where the cosmological parameters $\textbf{\emph{p}}=\{\Ok,\Om,H_0,\alpha\}$ and the scalar field, $\phi$, dynamical dark energy density parameter,
\be
\label{Op}
\Omega_{\phi}(z,\alpha)=\frac{1}{6H_0^2}\bigg[\frac{1}{2}\dot{\phi}^2+V(\phi)\bigg],
\ee
is determined by numerically solving the Friedmann equation \eqref{eq:Ezp} and the equation of motion of the scalar field
\be
\label{em}
\ddot{\phi}+3H\dot{\phi}+V'(\phi)=0.
\ee 
Here we assume an inverse power-law scalar field potential energy density
\be
\label{PE}
V(\phi)=\frac{1}{2}\kappa m_p^2\phi^{-\alpha}.
\ee
In these equations, an overdot and a prime denote a derivative with respect to time and $\phi$, respectively, $m_p$ is the Planck mass, $\alpha$ is a positive constant ($\alpha = 0$ correspond to \lcdm\ models), and $\kappa$ is a constant that is determined by the shooting method implemented in the Cosmic Linear Anisotropy Solving System (\textsc{class}) code \citep{class}. In the flat \pcdm\ model the constrained cosmological parameters are $H_0$, \obhs, \ochs, and $\alpha$ (only $\Omega_c$ and $\alpha$ are constrained in the \cq\ and \mq\ alone cases), whereas in the non-flat \pcdm\ model \ok\ is also constrained.

\section{Data}
\label{sec:data}

In this work, we use \cq, \mq, and $H(z)$ + BAO data, as well as combinations of these data sets, to constrain cosmological-model and QSO $R-L$ relation parameters. These data sets are summarized next, with the emphasis on the \cq\ sample.

\begin{itemize}
\item[]{\it \cq\ data.} Here we describe the sample of high-quality 38 \civ\ RM QSOs. The first time-lag measurements of the broad \civ\ line were inferred by \citet{2005ApJ...632..799P, 2006ApJ...641..638P} for 4 sources. The source NGC4151 was monitored by \citet{2006ApJ...647..901M}. The \civ\ time-lag for the intensively monitored NGC5548 was determined by \citet{2015ApJ...806..128D}. \citet{2018ApJ...865...56L} performed RM for 17 high-luminosity QSOs for over 10 years, out of which 8 QSOs were reported to have statistically significant ($>1\sigma$) \civ\ time-lag measurements. \citet{2019MNRAS.487.3650H} report 2 \civ\ detections for quasars at $z=1.905$ and $z=2.593$ from the photometric Dark Energy Survey Supernova Program (DES-SN) and the spectroscopic Australian Dark Energy Survey (Oz-DES). Within the Sloan Digital Sky Survey Reverberation Mapping project (SDSS-RM), \citet{2019ApJ...887...38G} determined \civ\ time-lag measurements for 48 QSOs with an average false-positive rate of $10\%$. Of these, 16 QSOs pass the highest-quality criteria, in the redshift range of $1.4<z<2.8$ and the monochromatic luminosity range $44.5<\log{[L_{1350}\,({\rm erg\,s^{-1}})]}<45.6$. \citet{2019ApJ...883L..14S} showed that adding more photometric data points and spectroscopic data ($9+5$ years of photometric and spectroscopic measurements, respectively, in comparison to $4+4$ years of spectroscopic and photometric monitoring performed by \citeauthor{2019ApJ...887...38G}, \citeyear{2019ApJ...887...38G}) results in the significant detection of 3 more \civ\ time-lag measurements. Using 20 years of photometric and spectrophotometric data, \citet{2021ApJ...915..129K} report significant \civ\ time-delays for 3 QSOs at redshifts $z=2.172$, $2.646$, and $3.192$.\\

\begin{figure}
    \centering
    \includegraphics[width=\columnwidth]{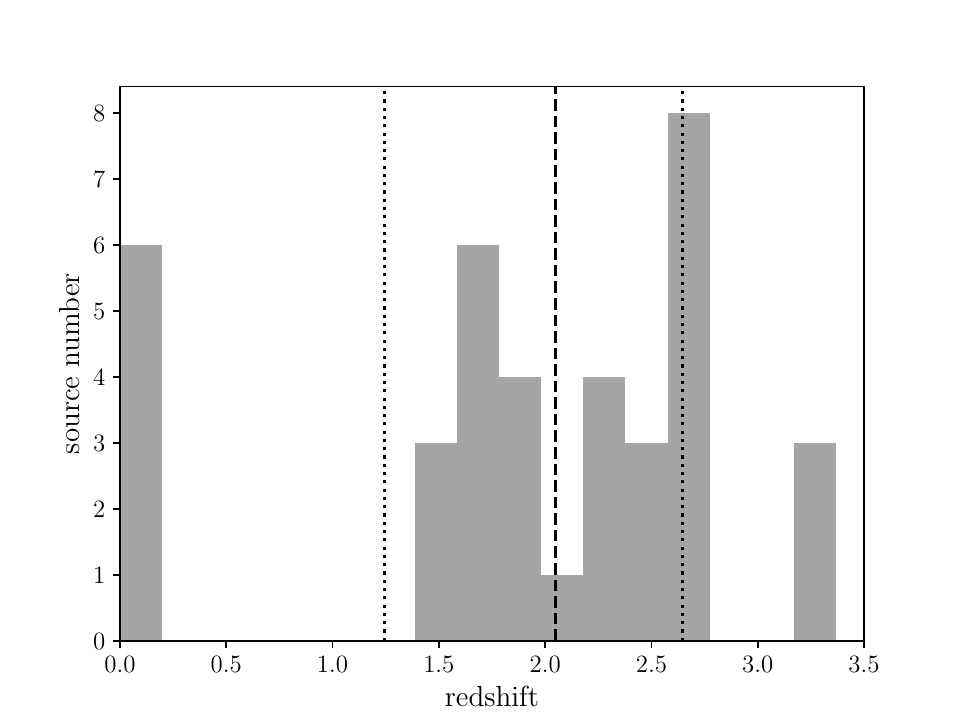}
    \caption{The redshift distribution for the ``golden'' sample of 38 \civ\ QSOs from \citet{2021ApJ...915..129K}. The dashed line stands for the redshift median at $2.048$, while the dotted lines represent 16\% and 84\% percentiles at $1.244$ and $2.647$, respectively. The histogram bin size is $\Delta z\simeq 0.2$.}
    \label{fig_redshift}
\end{figure}

\citet{2021ApJ...915..129K} compiled (with a few corrections) available \civ\ RM sources, finding 38 QSOs with significantly measured \civ\ time-delays, including their 3 QSOs. The time-delays were determined by using either the interpolated cross-correlation function (ICCF) or the $z$-transformed dicrete correlation function (zDCF) or a combination of both, which ensures a relative homogeneity of the sample in terms of the time-delay methodology in comparison with, e.g., the H$\beta$ sample where several different methods are applied and combined, see e.g. \citet{2019AN....340..577Z} or \citet{2020A&A...642A..59R} for overviews. We refer to this sample of 38 \civ\ QSOs as ``golden'' and it covers the redshift range $0.001064 \leq z \leq 3.368$, with the median redshift of $2.048$ and 16-\% and 84-\% percentile redshifts of $1.244$ and $2.647$, respectively. The redshift distribution is shown in Fig.~\ref{fig_redshift}. The sources belonging to the golden sample are listed in Table~\ref{tab:civdata}, including their redshift, flux density at 1350\,\AA\,, monochromatic luminosity at 1350\,\AA\, (computed assuming a flat $\Lambda$CDM model with $H_0=70\,{\rm km\,s^{-1}\,Mpc^{-1}}$, $\Om=0.3$, and $\Omega_{\Lambda}=0.7$), and the rest-frame \civ\ time-lag $\tau$ (typically with asymmetrical error bars). 

The correlation between the rest-frame \civ\ time-delay and the UV monochromatic luminosity at $1350\,$\AA\, is significant, with the Pearson correlation coefficient $r=0.898$ ($p=2.082 \times 10^{-14}$) and the Spearman rank-order correlation coefficient $s=0.799$ ($p=1.751 \times 10^{-9}$). Given the large correlation coefficient, we fit the golden dataset with the power-law relation $\log{\tau}=\beta_{\rm C}+\gamma_{\rm C} \log{(L_{1350}/10^{44}\,{\rm erg\,s^{-1}})} $, where $\log\equiv\log_{10}$, and find the best-fit intercept $\beta_{\rm C}=1.04 \pm 0.07$ and the best-fit slope $\gamma_{\rm C}=0.42 \pm 0.03$, for which the individual time-delay errors were neglected in the Levenberg-Marquardt algorithm. This results in an intrinsic scatter of $\sigma=0.32$ dex and $\chi^2=31.8$ (36 degrees of freedom). When we consider individual symmetrized time-delay errors (see the discussion in the paragraph below eq.\ (\ref{eq:sigma_mq})), we obtain $\gamma_{\rm C}=0.56\pm 0.04$ and $\beta_{\rm C}=0.98\pm 0.07$ with $\sigma=0.41$ dex and $\chi^2=16.6$. See Fig.~\ref{fig_civ_radius_luminosity} for the best-fitting relations, which were inferred using the \texttt{curve\_fit} function from the \texttt{scipy} library. To add information about the accretion state of each QSO from our sample in the $R-L$ relation, we estimate the Eddington ratio $\lambda_{\rm Edd}=L_{\rm bol}/L_{\rm Edd}$, where $L_{\rm bol}=\kappa_{\rm bol}L_{1350}$ is the bolometric luminosity calculated using the luminosity-dependent bolometric factor $\kappa_{\rm bol}$ according to \citet{2019MNRAS.488.5185N}, and $L_{\rm Edd}$ is the Eddington luminosity \citep{2014A&A...565A..17Z,2017FoPh...47..553E,2020ApJ...903..140Z}. To obtain $L_{\rm Edd}$, the supermassive black hole (SMBH) mass was calculated using the virial relation $M_{\bullet}=f_{\rm vir} c\tau \text{FWHM}^2/G$, where the virial factor $f_{\rm vir}$ was estimated using the fitted formula that inversely scales with the full width at half maximum, FWHM, see \citet{2018NatAs...2...63M}. The FWHM and $\tau$ values were adopted from the compilation of \citet{2021ApJ...915..129K}. In Fig.~\ref{fig_civ_radius_luminosity} each source is coloured by the corresponding value of $\log{\lambda_{\rm Edd}}$.

\begin{figure}
    \centering
    \includegraphics[width=\columnwidth]{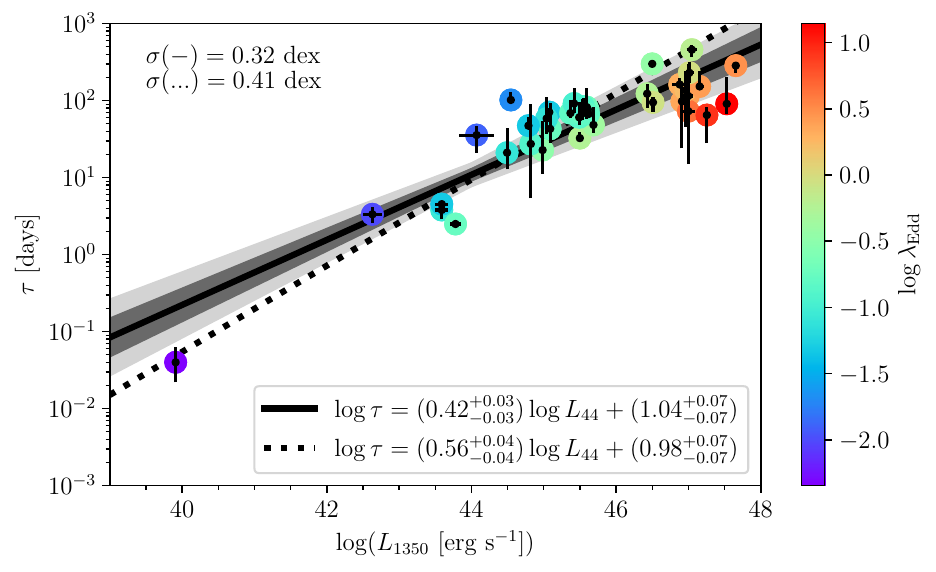}
    \caption{Radius-luminosity relationship for the ``golden'' sample of 38 \civ\ QSOs \citep[see also][]{2021ApJ...915..129K}. The luminosity at 1350\,\AA\, is based on the flat $\Lambda$CDM model (with $H_0=70\,{\rm km\,s^{-1}\,Mpc^{-1}}$, $\Om=0.3$, and $\Omega_{\Lambda}=0.7$). The points are colour-coded with respect to the calculated Eddington ratio $\lambda_{\rm Edd}=L_{\rm bol}/L_{\rm Edd}$ of the source. When we neglect individual time-delay errors, the best-fit relation (black solid line) determined by the Levenberg-Marquardt algorithm has a slope of $\gamma_{\rm C}=0.42\pm 0.03$ and a scatter of $\sigma=0.32$ dex. The dark and light gray areas around the best-fit solid line are one and two sigma confidence regions. Considering individual time-delay errors yields the best-fit relation (dotted line) with a larger slope of $\gamma_{\rm C}=0.56\pm 0.04$ and the scatter of data around this relation is $\sigma=0.41$ dex.}
    \label{fig_civ_radius_luminosity}
\end{figure}

 Apart from the ``golden'' \civ\ AGN sample reported by \citet{2021ApJ...915..129K}, there are additional lower-quality \civ\ time-lag detections reported in literature. \citet{2019ApJ...887...38G} report additional 32 \civ\ detections with a lower quality rating of 1, 2, and 3.\footnote{\citet{2019ApJ...887...38G}  assign the quality flag of 1 to the lowest-quality time-lag detections, while 5 is the highest quality rating. There are different factors considered in the rating scheme, such as the clear correlation between the continuum and \civ\ emission-line light curves, overall consistency between JAVELIN, CREAM, and ICCF time-lag detections, and the presence of multiple significant time-lag peaks.} As a follow-up of the Oz-DES RM project, \citet{2022MNRAS.509.4008P} designed a quality-cut methodology and out of 6 test sources, 2 at redshifts $1.93$ and $2.74$ pass all the quality criteria. These additional 34 sources will be considered in a future study.

\item[]{\it \mq\ data.} As listed in table A1 of \cite{Khadkaetal_2021a}, the \mq\ sample consists of 78 QSOs and spans the redshift range $0.0033 \leq z \leq 1.89$. Detailed descriptions of \mq\ data can be found in \cite{Khadkaetal_2021a} and \cite{2020ApJ...903...86M}, where it was shown that they obey the $R-L$ relation, with measured quantities being time-delay $\tau^{\prime}$ and QSO flux $F_{3000}$ measured at 3000 \(\text{\r{A}}\); see also \cite{2022arXiv220111062P} for an updated \mq\ $R-L$ relation in the fixed flat $\Lambda$CDM cosmology. We note that the \mq\ sample is relatively homogeneous since 57 significant time-delay detections (hence 73\% of \mq\ sources) were determined by \citet{2020ApJ...901...55H}, who applied a consistent time-delay method to all the sources based on JAVELIN \citep{2011ApJ...735...80Z}, which was compared with the CREAM results that were generally consistent \citep{2016MNRAS.456.1960S}. However, the time-delay uncertainties for this sample are not completely homogeneous and a consistent treatment of the continuum and the line-emission light-curve correlation and the time-delay inference is needed to homogenize the sample of the best \mq\ time delays and their uncertainties.

\item[]{$H(z)\ +\ BAO\ data$.} There are 32 $H(z)$ and 12 BAO measurements listed in Tables 1 and 2 of \cite{CaoRatra2022}, spanning the redshift ranges $0.07 \leq z \leq 1.965$ and $0.122 \leq z \leq 2.334$, respectively.

\end{itemize}

\section{Data Analysis Methodology}
\label{sec:analysis}

We utilize the $R-L$ relation parametrization according to \citet{Bentzetal2013}, where we replace the monochromatic luminosity and rest-frame time-delay taking into account the \civ-region emission properties
\begin{equation}
    \label{eq:civ}
    \log{\frac{\tau}{\rm days}}=\beta_{\rm\textsc{c}}+\gamma_{\rm\textsc{c}} \log{\frac{L_{1350}}{10^{44}\,{\rm erg\ s^{-1}}}},
\end{equation}
where $\tau$, $\beta_{\rm\textsc{c}}$, and $\gamma_{\rm\textsc{c}}$ are the \civ\ time-lag, the intercept parameter, and the slope parameter, respectively, and the monochromatic luminosity at 1350\,\AA\
\be
\label{eq:L1350}
    L_{1350}=4\pi D_L^2F_{1350},
\ee
with measured quasar flux $F_{1350}$ at 1350\,\AA\ in units of $\rm erg\ s^{-1}\ cm^{-2}$. The luminosity distance is a function of redshift $z$ and the cosmological parameters,
\begin{equation}
  \label{eq:DL}
\resizebox{0.475\textwidth}{!}{%
    $D_L(z) = 
    \begin{cases}
    \frac{c(1+z)}{H_0\sqrt{\Omega_{\rm k0}}}\sinh\left[\frac{H_0\sqrt{\Omega_{\rm k0}}}{c}D_C(z)\right] & \text{if}\ \Omega_{\rm k0} > 0, \\
    \vspace{1mm}
    (1+z)D_C(z) & \text{if}\ \Omega_{\rm k0} = 0,\\
    \vspace{1mm}
    \frac{c(1+z)}{H_0\sqrt{|\Omega_{\rm k0}|}}\sin\left[\frac{H_0\sqrt{|\Omega_{\rm k0}|}}{c}D_C(z)\right] & \text{if}\ \Omega_{\rm k0} < 0,
    \end{cases}$%
    }
\end{equation}
where the comoving distance is
\begin{equation}
\label{eq:gz}
   D_C(z) = c\int^z_0 \frac{dz'}{H(z')},
\end{equation}
with $c$ being the speed of light.

The \mq\ $R-L$ relation is
\begin{equation}
    \label{eq:mq}
    \log{\frac{\tau^{\prime}}{\rm days}}=\beta_{\rm\textsc{m}}+\gamma_{\rm\textsc{m}} \log{\frac{L_{3000}}{10^{44}\,{\rm erg\ s^{-1}}}},
\end{equation}
where $\tau^{\prime}$, $\beta_{\rm\textsc{m}}$, and $\gamma_{\rm\textsc{m}}$ are the \mii\ time-lag, the intercept parameter, and the slope parameter, respectively, and the monochromatic luminosity at 3000\,\AA
\be
\label{eq:L3000}
    L_{3000}=4\pi D_L^2F_{3000},
\ee
with measured quasar flux $F_{3000}$ at 3000\,\AA\ in units of $\rm erg\ s^{-1}\ cm^{-2}$.

The natural log of the \civ\ likelihood function \citep{D'Agostini_2005} is
\be
\label{eq:LH_civ}
    \ln\mathcal{L}_{\rm C\,\textsc{iv}}= -\frac{1}{2}\Bigg[\chi^2_{\rm C\,\textsc{iv}}+\sum^{N}_{i=1}\ln\left(2\pi\sigma^2_{\mathrm{tot,\textsc{c}},i}\right)\Bigg],
\ee
where
\be
\label{eq:chi2_civ}
    \chi^2_{\rm C\,\textsc{iv}} = \sum^{N}_{i=1}\bigg[\frac{(\log \tau_{\mathrm{obs},i} - \beta_{\rm\textsc{c}}  -\gamma_{\rm\textsc{c}}\log L_{1350,i})^2}{\sigma^2_{\mathrm{tot,\textsc{c}},i}}\bigg]
\ee
with total uncertainty
\be
\label{eq:sigma_civ}
\sigma^2_{\mathrm{tot,\textsc{c}},i}=\sigma_{\rm int,\,\textsc{c}}^2+\sigma_{{\log \tau_{\mathrm{obs},i}}}^2+\gamma_{\rm\textsc{c}}^2\sigma_{{\log F_{1350,i}}}^2,
\ee
where $\sigma_{\rm int,\,\textsc{c}}$ is the \cq\ intrinsic scatter parameter which also contains the unknown systematic uncertainty, and $N$ is the number of data points.

The natural log of the \mii\ likelihood function is
\be
\label{eq:LH_mq}
    \ln\mathcal{L}_{\rm Mg\,\textsc{ii}}= -\frac{1}{2}\Bigg[\chi^2_{\rm Mg\,\textsc{ii}}+\sum^{N}_{i=1}\ln\left(2\pi\sigma^2_{\mathrm{tot,\textsc{m}},i}\right)\Bigg],
\ee
where
\be
\label{eq:chi2_mq}
    \chi^2_{\rm Mg\,\textsc{ii}} = \sum^{N}_{i=1}\bigg[\frac{(\log \tau^{\prime}_{\mathrm{obs},i} - \beta_{\rm\textsc{m}}  -\gamma_{\rm\textsc{m}}\log L_{3000,i})^2}{\sigma^2_{\mathrm{tot,\textsc{m}},i}}\bigg]
\ee
with total uncertainty
\be
\label{eq:sigma_mq}
\sigma^2_{\mathrm{tot,\textsc{m}},i}=\sigma_{\rm int,\,\textsc{m}}^2+\sigma_{{\log \tau^{\prime}_{\mathrm{obs},i}}}^2+\gamma_{\rm\textsc{m}}^2\sigma_{{\log F_{3000,i}}}^2,
\ee
where $\sigma_{\rm int,\,\textsc{m}}$ is the \mq\ intrinsic scatter parameter which also contains the unknown systematic uncertainty.

The $\tau$ error bars are typically asymmetric. As used for \mq{s} in \cite{Khadkaetal_2021a} and \cite{CaoRatra2022}, in what follows, $\tau$ symmetrized errors mean that we are using symmetrized $\sigma_{\tau}$'s defined as $\sigma_{\tau}=[2\sigma_{\tau,+}\sigma_{\tau,-}/(\sigma_{\tau,+}+\sigma_{\tau,-})+\sqrt{\sigma_{\tau,+}\sigma_{\tau,-}}]/2$, where $\sigma_{\tau,+}$ and $\sigma_{\tau,-}$ are the upper and lower errors of $\tau$, respectively. On the other hand, $\tau$ asymmetric errors mean that we directly use asymmetric $\sigma_{\tau,+}$ and $\sigma_{\tau,-}$ as follows: when the theoretical prediction for $\log{\tau}$ is larger (smaller) than the observed value, $\sigma_{\tau}=\sigma_{\tau,+}$ ($\sigma_{\tau}=\sigma_{\tau,-}$).

The detailed descriptions for the likelihood functions of $H(z)$ and BAO data can be found in \cite{CaoRyanRatra2020}. 

\begin{table}
\centering
\resizebox{\columnwidth}{!}{%
\begin{threeparttable}
\caption{Flat priors of the constrained parameters.}
\label{tab:priors}
\begin{tabular}{lcc}
\toprule
Parameter & & Prior\\
\midrule
 & Cosmological-Model Parameters & \\
\midrule
$H_0$\,\tnote{a} &  & [None, None]\\
\obhs\,\tnote{b} &  & [0, 1]\\
\ochs\,\tnote{c} &  & [0, 1]\\
\ok &  & [-2, 2]\\
$\alpha$ &  & [0, 10]\\
\wx &  & [-5, 0.33]\\
\midrule
 & $R-L$ Relation Parameters & \\
\midrule
$\gamma$ &  & [0, 5]\\
$\beta$ &  & [0, 10]\\
$\sigma_{\rm int}$ &  & [0, 5]\\
\bottomrule
\end{tabular}
\begin{tablenotes}[flushleft]
\item [a] \hunit. In the \cq\ and \mq\ alone cases, $H_0$ is set to be 70 \hunit, while in other cases, the prior range is irrelevant (unbounded).
\item [b] In the \cq\ and \mq\ alone cases, \obhs\ is set to be 0.0245, i.e. $\Omega_{b}=0.05$.
\item [c] In the \cq\ and \mq\ alone cases, $\Om\in[0,1]$ is ensured.
\end{tablenotes}
\end{threeparttable}%
}
\end{table}

We list the flat priors of the free cosmological-model and $R-L$ relation parameters in Table \ref{tab:priors}. By maximizing the likelihood functions, we obtain the unmarginalized best-fitting values and posterior distributions of all free cosmological-model and $R-L$ relation parameters. The Markov chain Monte Carlo (MCMC) code \textsc{MontePython} \citep{Audrenetal2013,Brinckmann2019}, the \textsc{class} code, and the \textsc{python} package \textsc{getdist} \citep{Lewis_2019} are used to perform our analyses.

One can find the definitions of the Akaike Information Criterion (AIC), the Bayesian Information Criterion (BIC), and the Deviance Information Criterion (DIC) in our previous paper \citep[see, e.g.][]{CaoDainottiRatra2022}. $\Delta \mathrm{AIC}$, $\Delta \mathrm{BIC}$, and $\Delta \mathrm{DIC}$ are computed as the differences between the AIC, BIC, and DIC values of the other five cosmological dark energy models and those of the flat \lcdm\ reference model. Positive (negative) values of these $\Delta \mathrm{IC}$s show that the model under investigation fits the data worse (better) than does the flat \lcdm\ reference model. In comparison with the model with the minimum IC, $\Delta \mathrm{IC} \in (0, 2]$ indicates weak evidence against the model under investigation, $\Delta \mathrm{IC} \in (2, 6]$ indicates positive evidence against the model under investigation, $\Delta \mathrm{IC} \in (6, 10] $ indicates strong evidence against the model under investigation, and $\Delta \mathrm{IC}>10$ indicates very strong evidence against the model under investigation.

\section{Results}
\label{sec:results}

\begin{figure*}
\centering
 \subfloat[]{%
    \includegraphics[width=0.33\textwidth,height=0.4\textwidth]{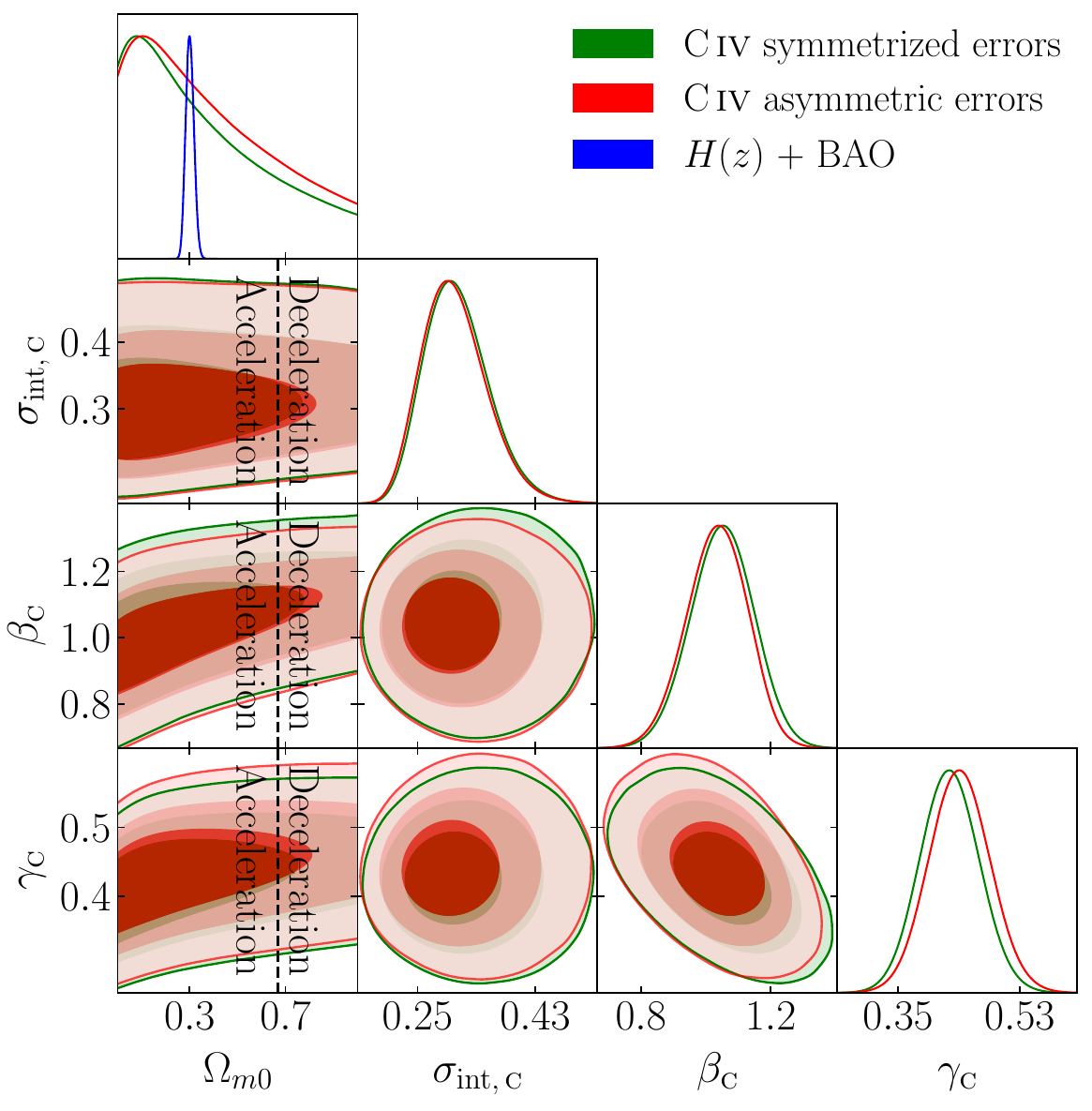}}
 \subfloat[]{%
    \includegraphics[width=0.33\textwidth,height=0.4\textwidth]{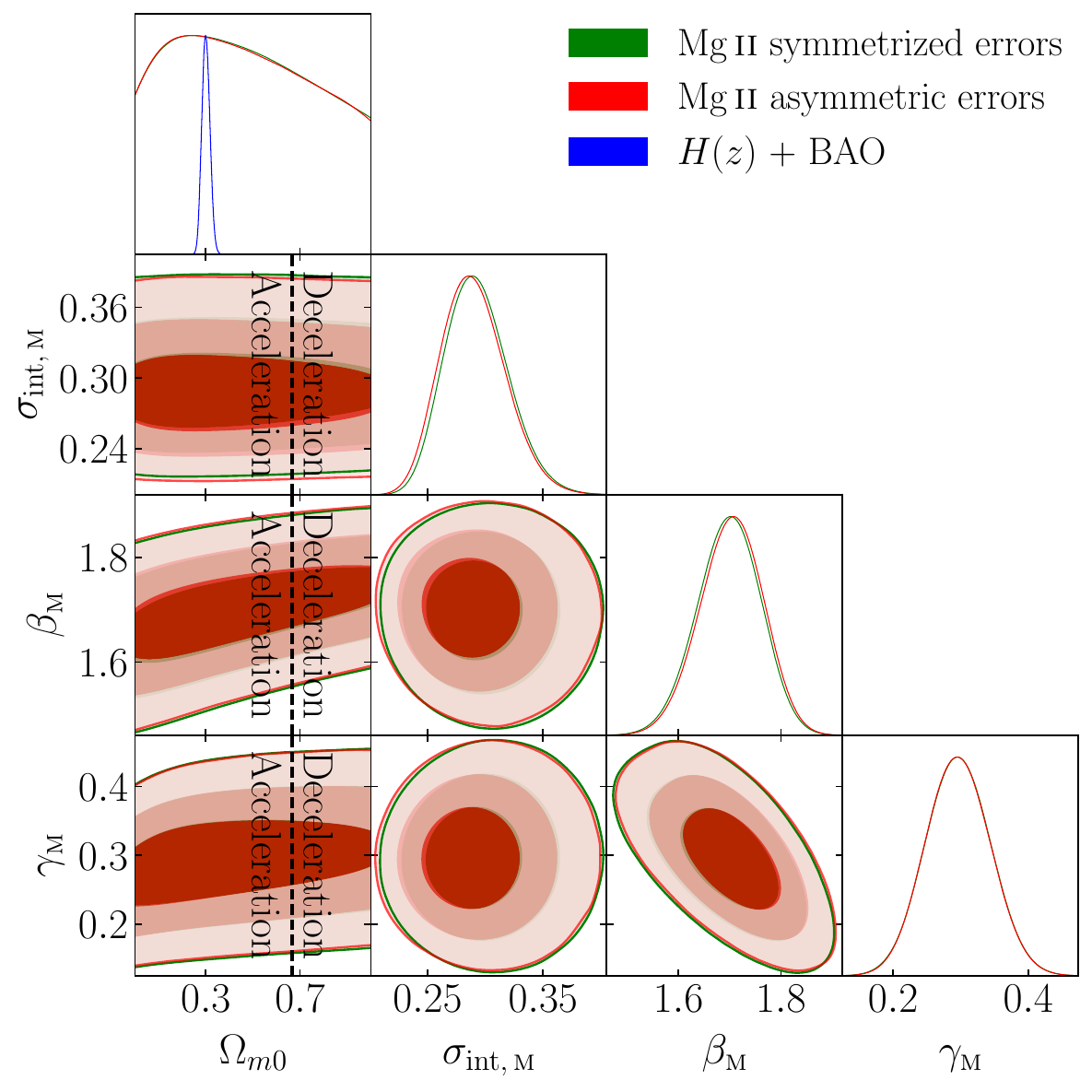}}
 \subfloat[]{%
    \includegraphics[width=0.33\textwidth,height=0.4\textwidth]{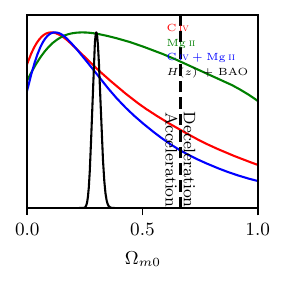}}\\
 \subfloat[]{%
    \includegraphics[width=0.5\textwidth,height=0.55\textwidth]{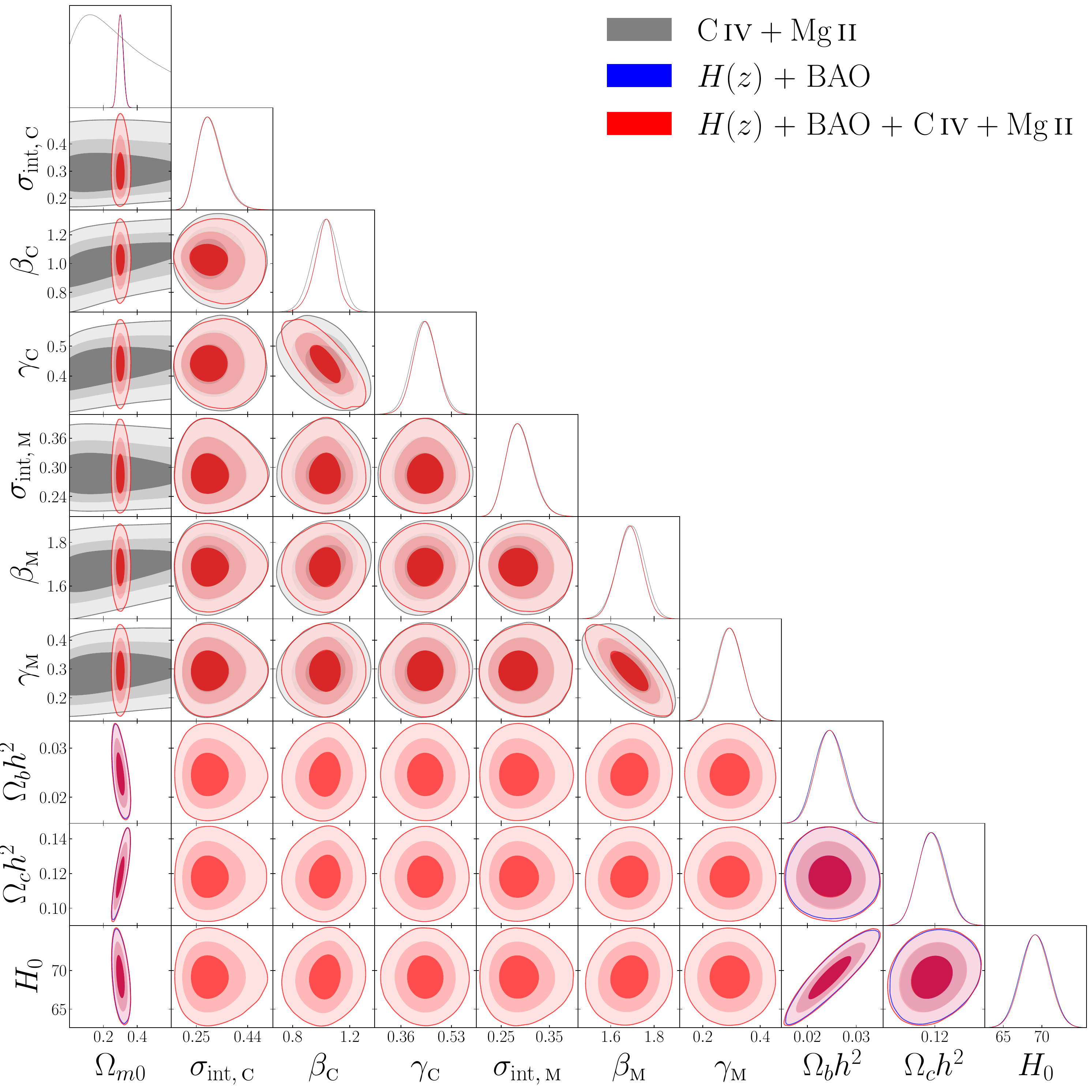}}
 \subfloat[]{%
    \includegraphics[width=0.5\textwidth,height=0.55\textwidth]{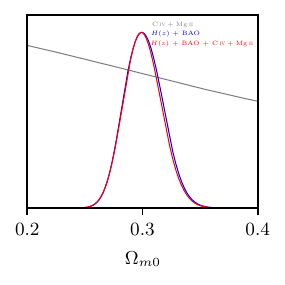}}\\
\caption{One-dimensional likelihood distributions and 1$\sigma$, 2$\sigma$, and 3$\sigma$ two-dimensional likelihood confidence contours for flat \lcdm\ from various combinations of data. The zero-acceleration black dashed lines in panels (a) and (b) divide the parameter space into regions associated with currently-accelerating (left) and currently-decelerating (right) cosmological expansion.}
\label{fig1}
\end{figure*}

\begin{figure*}
\centering
 \subfloat[]{%
    \includegraphics[width=0.33\textwidth,height=0.4\textwidth]{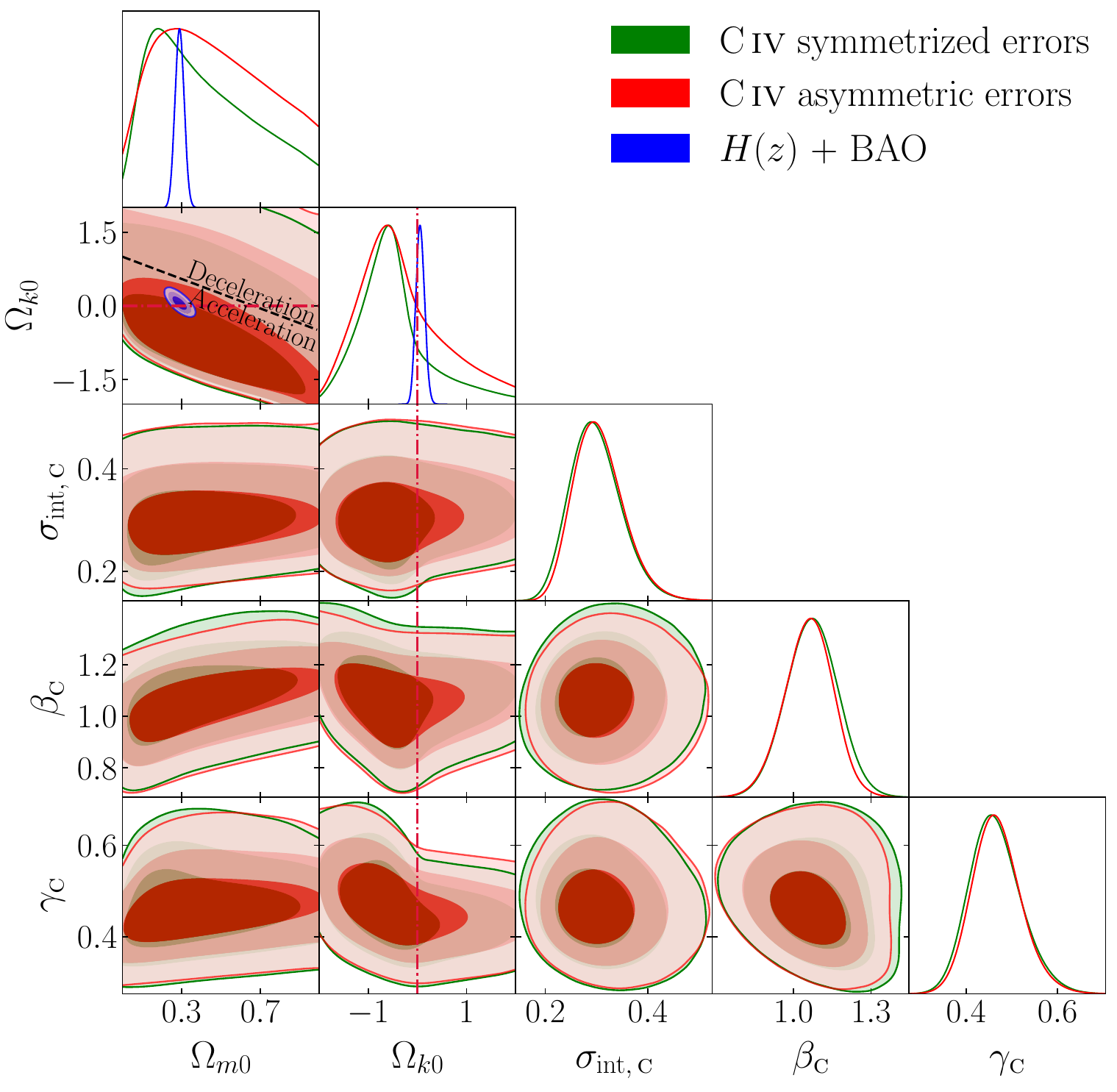}}
 \subfloat[]{%
    \includegraphics[width=0.33\textwidth,height=0.4\textwidth]{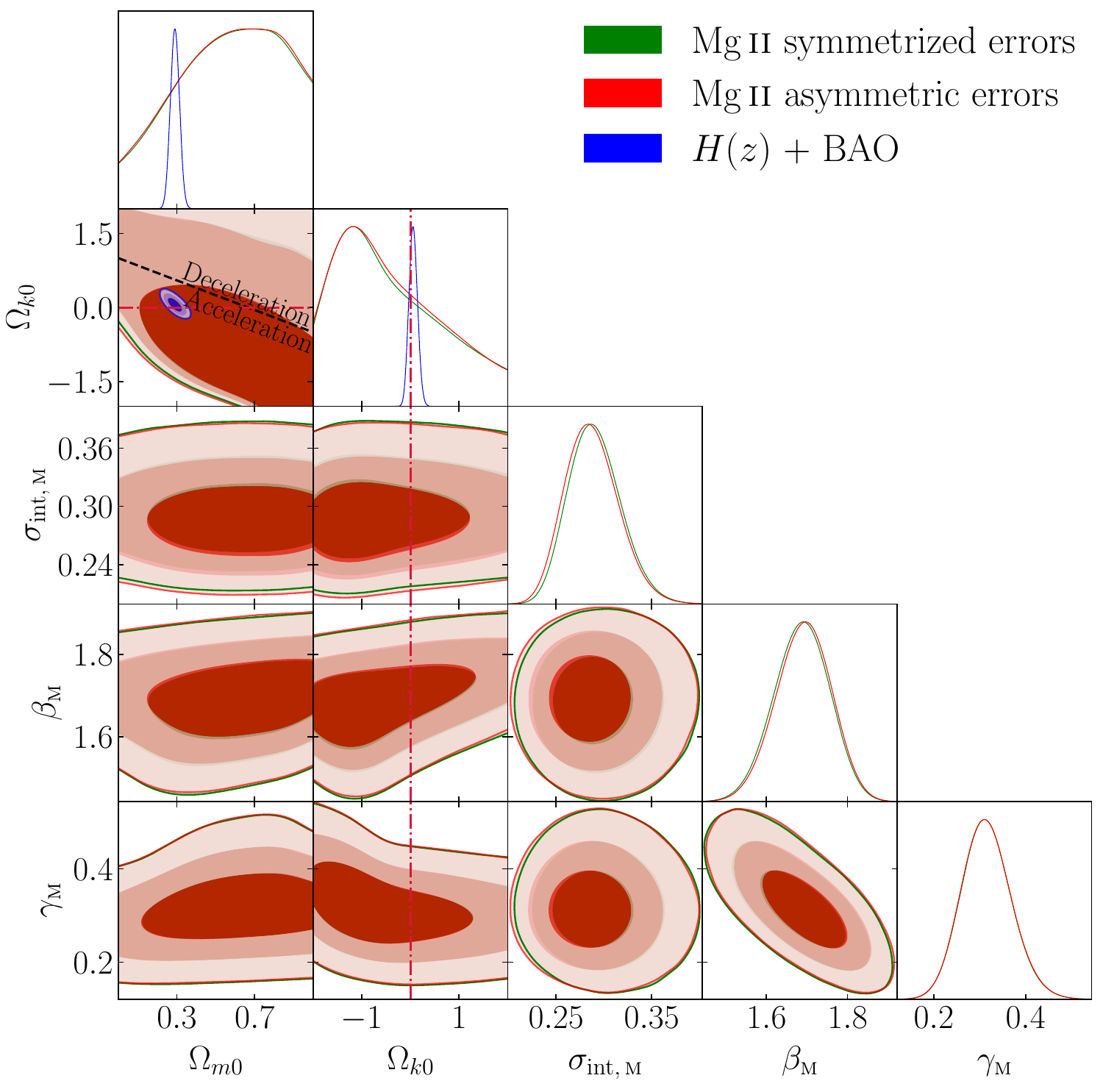}}
 \subfloat[]{%
    \includegraphics[width=0.33\textwidth,height=0.4\textwidth]{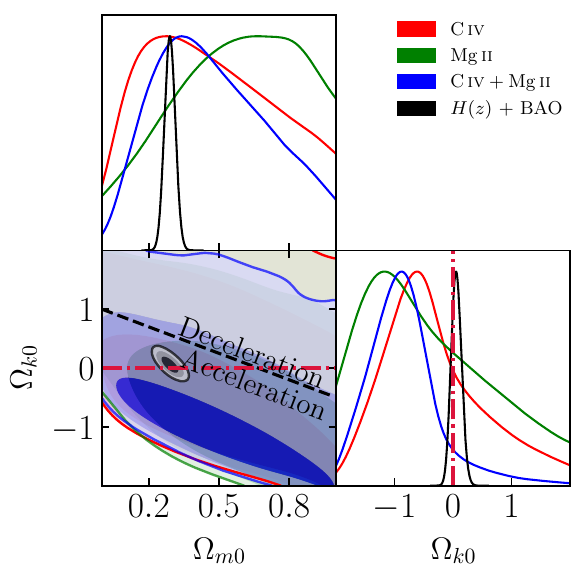}}\\
 \subfloat[]{%
    \includegraphics[width=0.5\textwidth,height=0.55\textwidth]{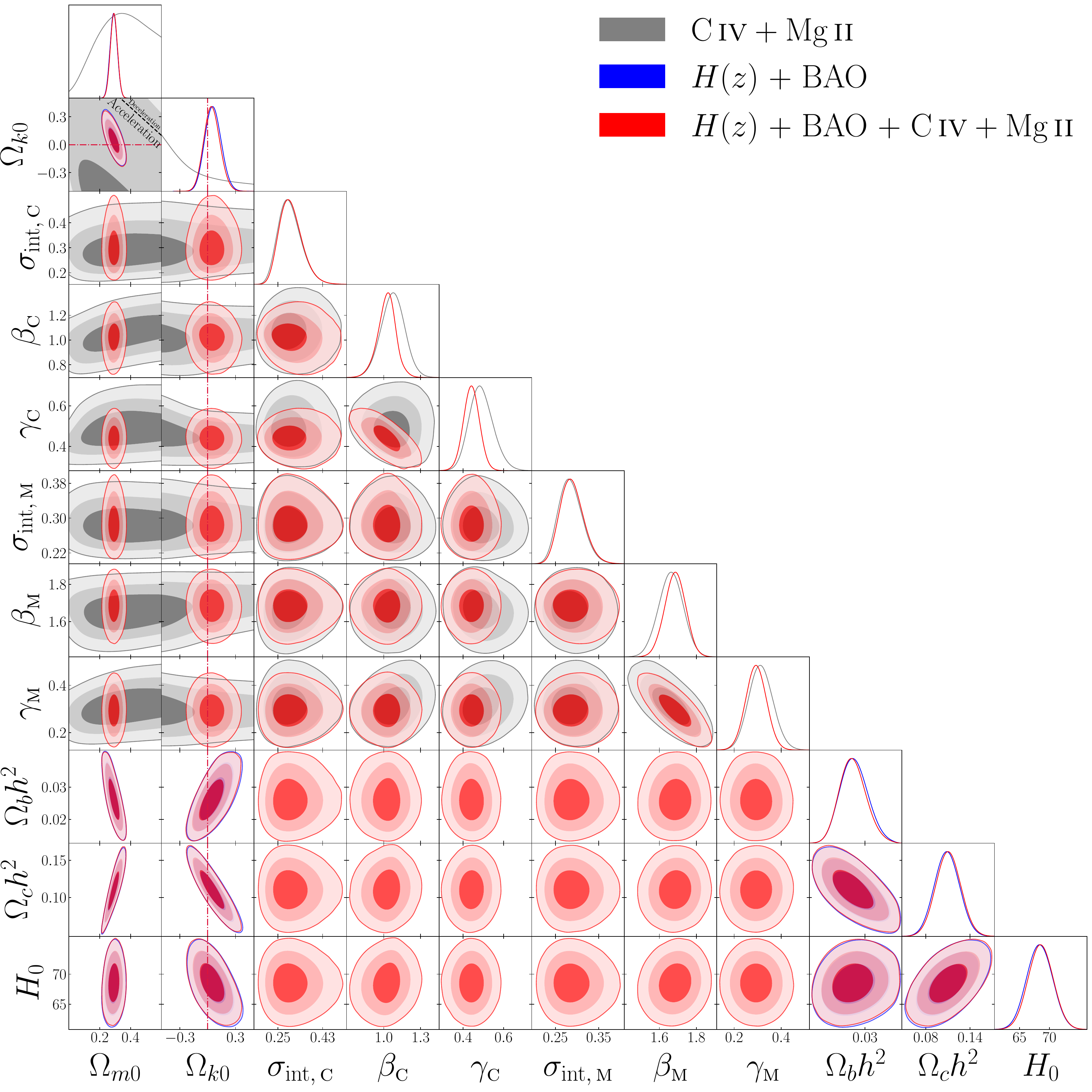}}
 \subfloat[]{%
    \includegraphics[width=0.5\textwidth,height=0.55\textwidth]{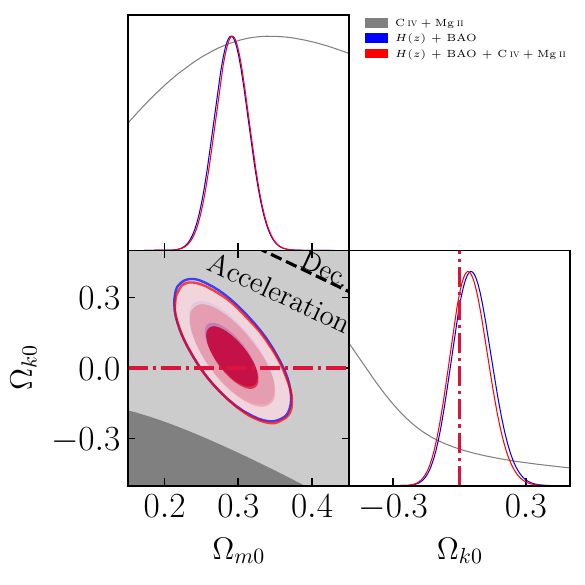}}\\
\caption{Same as Fig.\ \ref{fig1} but for non-flat \lcdm. The zero-acceleration black dashed lines divide the parameter space into regions associated with currently-accelerating (below left) and currently-decelerating (above right) cosmological expansion.}
\label{fig2}
\end{figure*}

\begin{figure*}
\centering
 \subfloat[]{%
    \includegraphics[width=0.33\textwidth,height=0.4\textwidth]{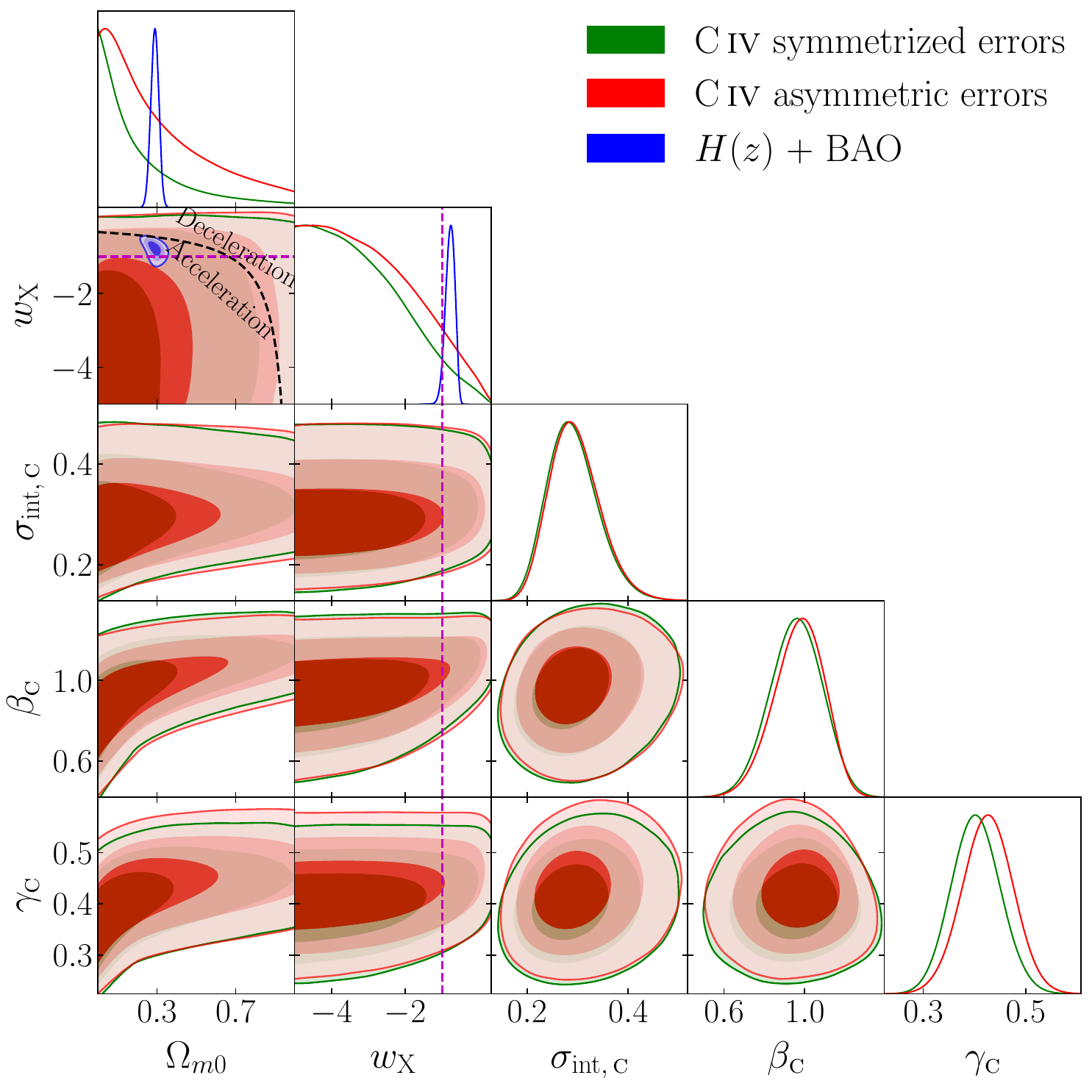}}
 \subfloat[]{%
    \includegraphics[width=0.33\textwidth,height=0.4\textwidth]{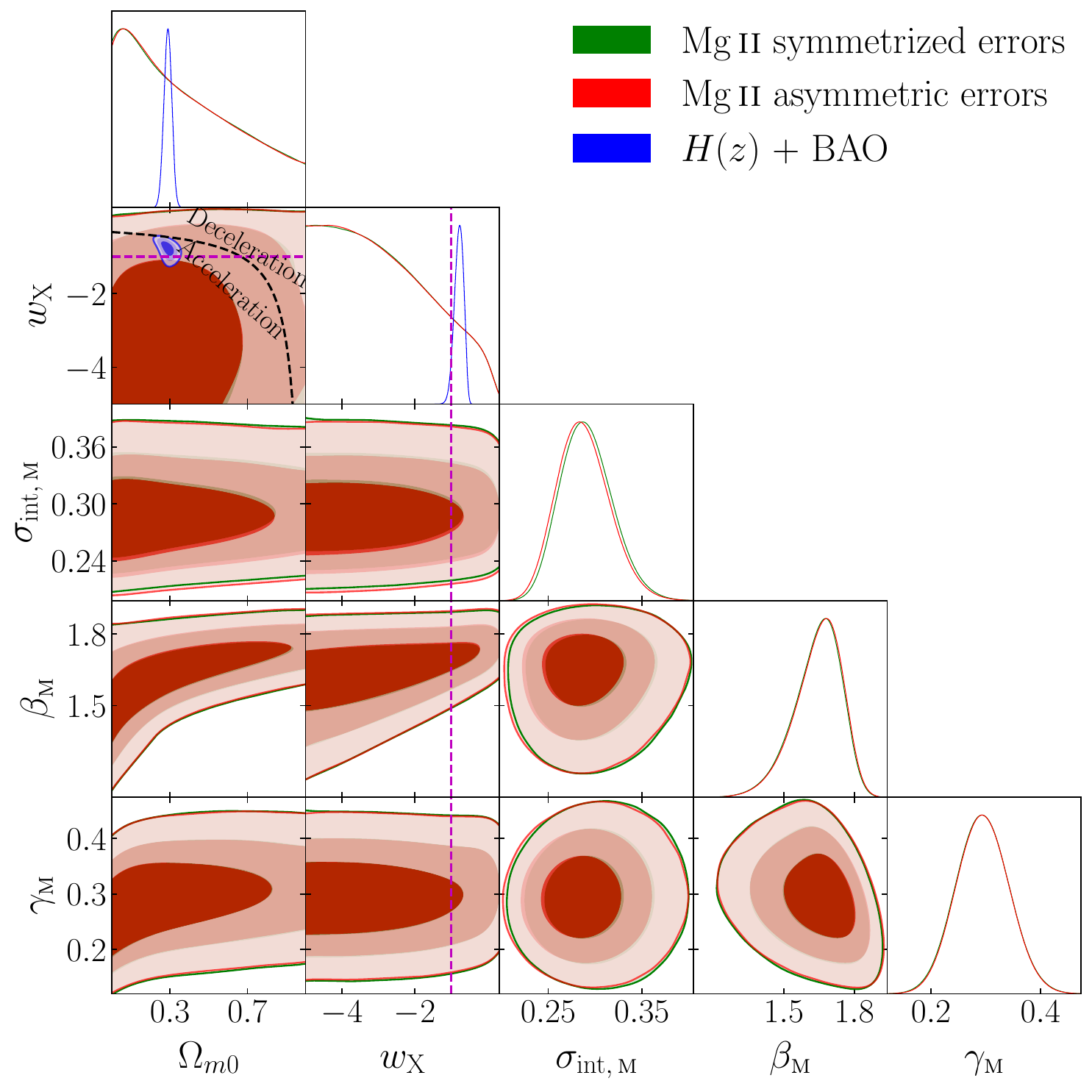}}
 \subfloat[]{%
    \includegraphics[width=0.33\textwidth,height=0.4\textwidth]{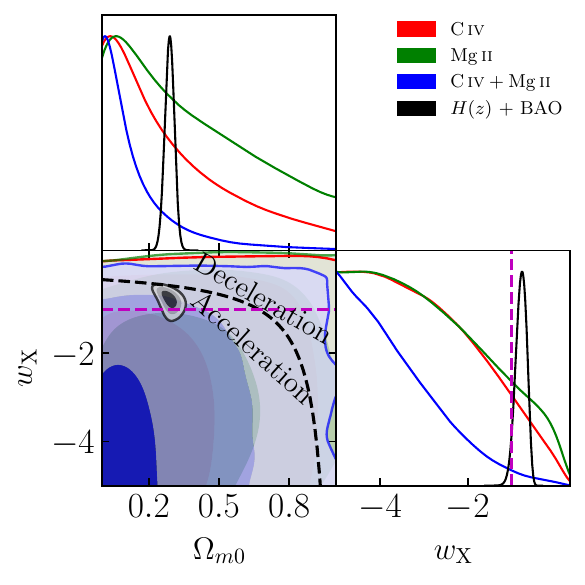}}\\
 \subfloat[]{%
    \includegraphics[width=0.5\textwidth,height=0.55\textwidth]{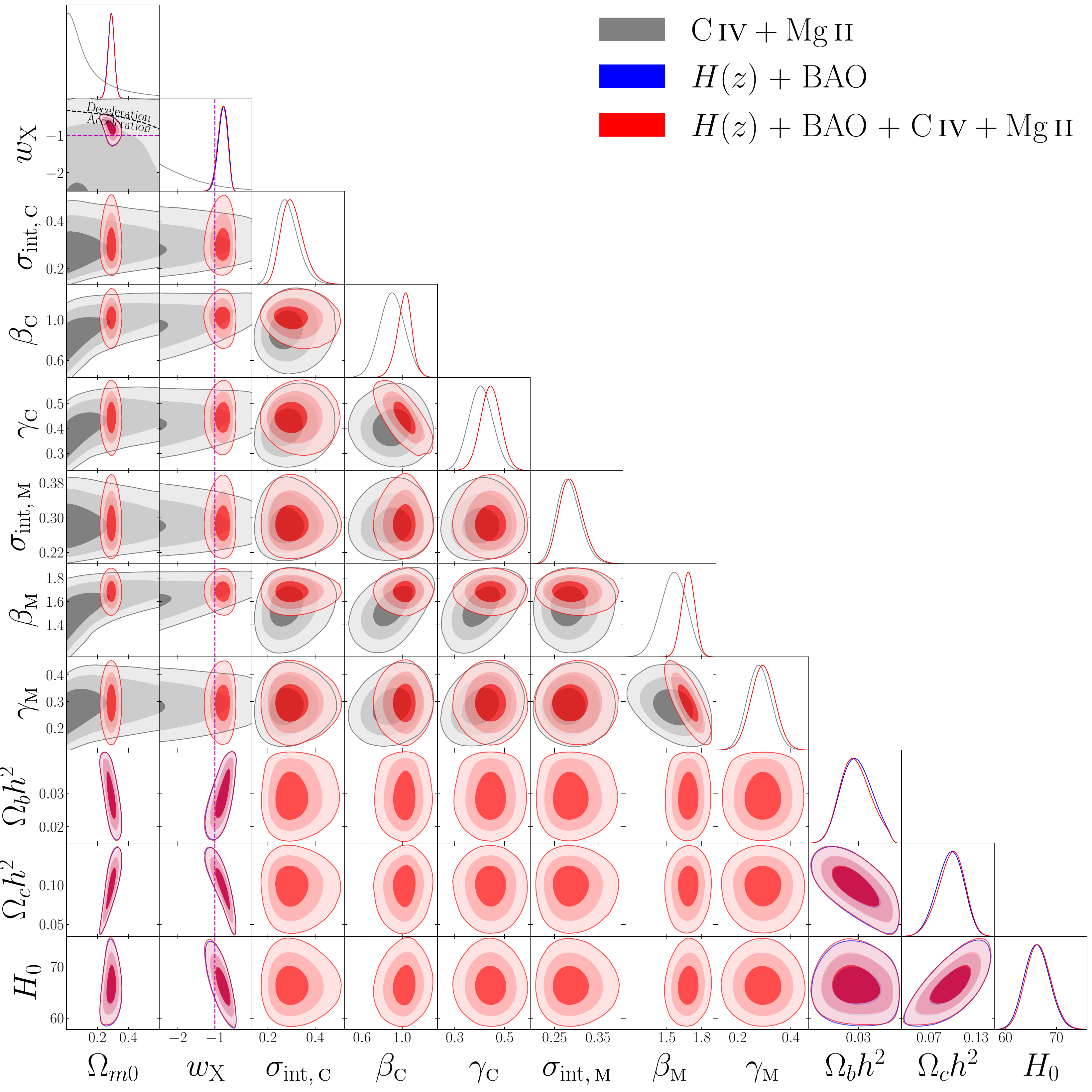}}
 \subfloat[]{%
    \includegraphics[width=0.5\textwidth,height=0.55\textwidth]{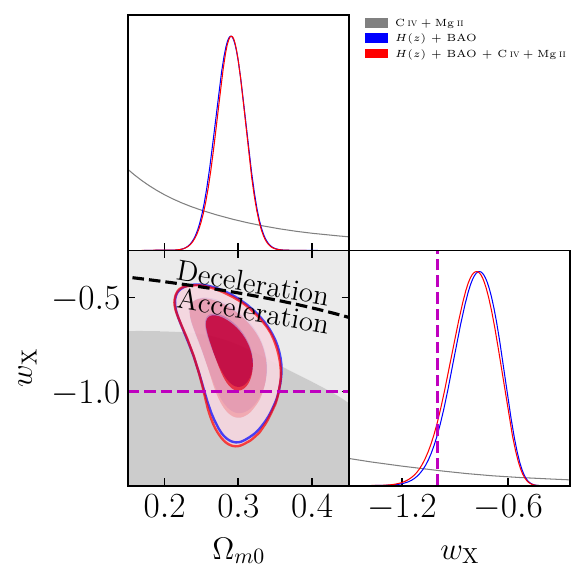}}\\
\caption{One-dimensional likelihood distributions and 1$\sigma$, 2$\sigma$, and 3$\sigma$ two-dimensional likelihood confidence contours for flat XCDM from various combinations of data. The zero-acceleration black dashed lines divide the parameter space into regions associated with currently-accelerating (either below left or below) and currently-decelerating (either above right or above) cosmological expansion. The magenta dashed lines represent $w_{\rm X}=-1$, i.e.\ flat \lcdm.}
\label{fig3}
\end{figure*}

\begin{figure*}
\centering
 \subfloat[]{%
    \includegraphics[width=0.33\textwidth,height=0.4\textwidth]{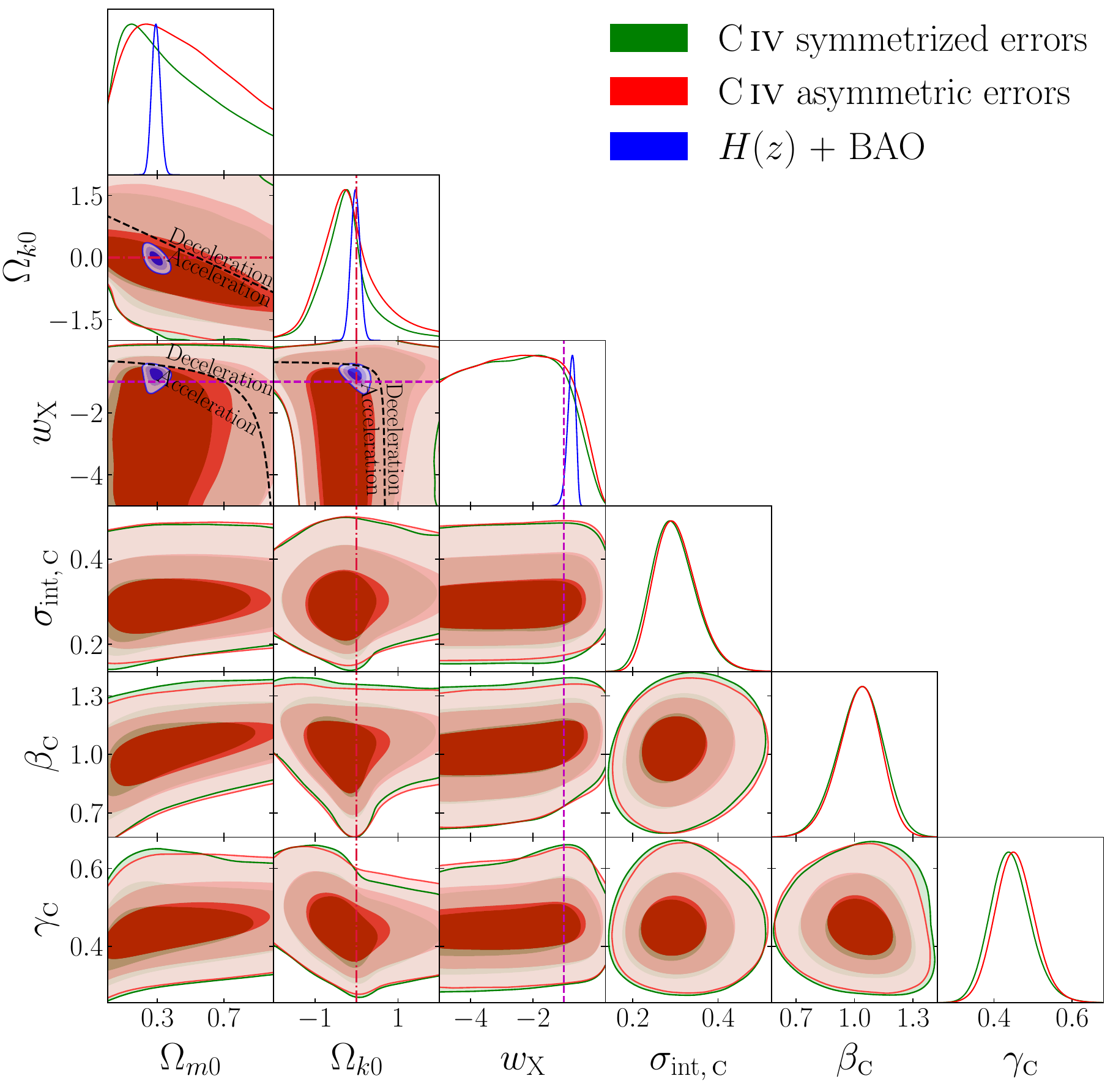}}
 \subfloat[]{%
    \includegraphics[width=0.33\textwidth,height=0.4\textwidth]{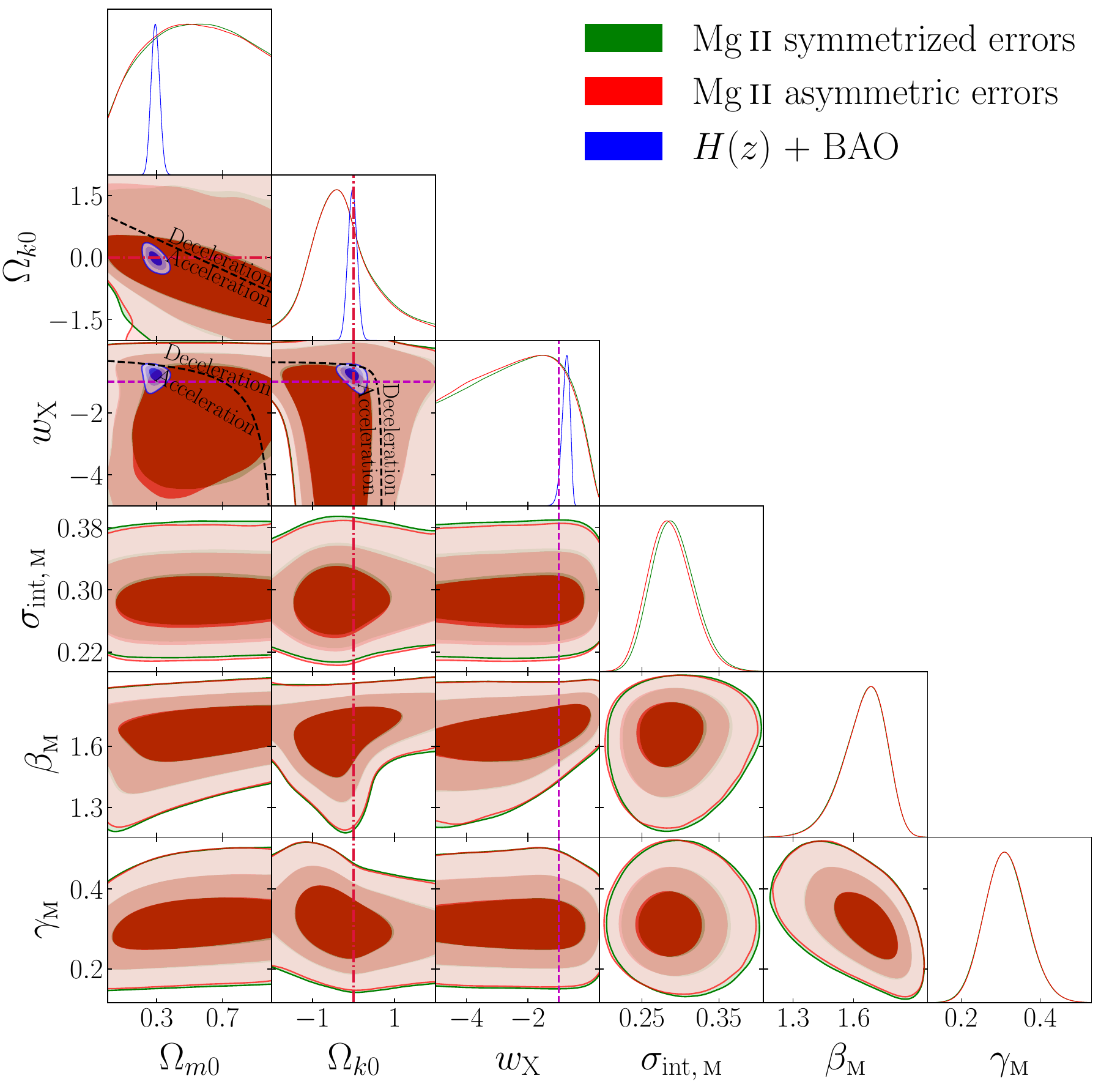}}
 \subfloat[]{%
    \includegraphics[width=0.33\textwidth,height=0.4\textwidth]{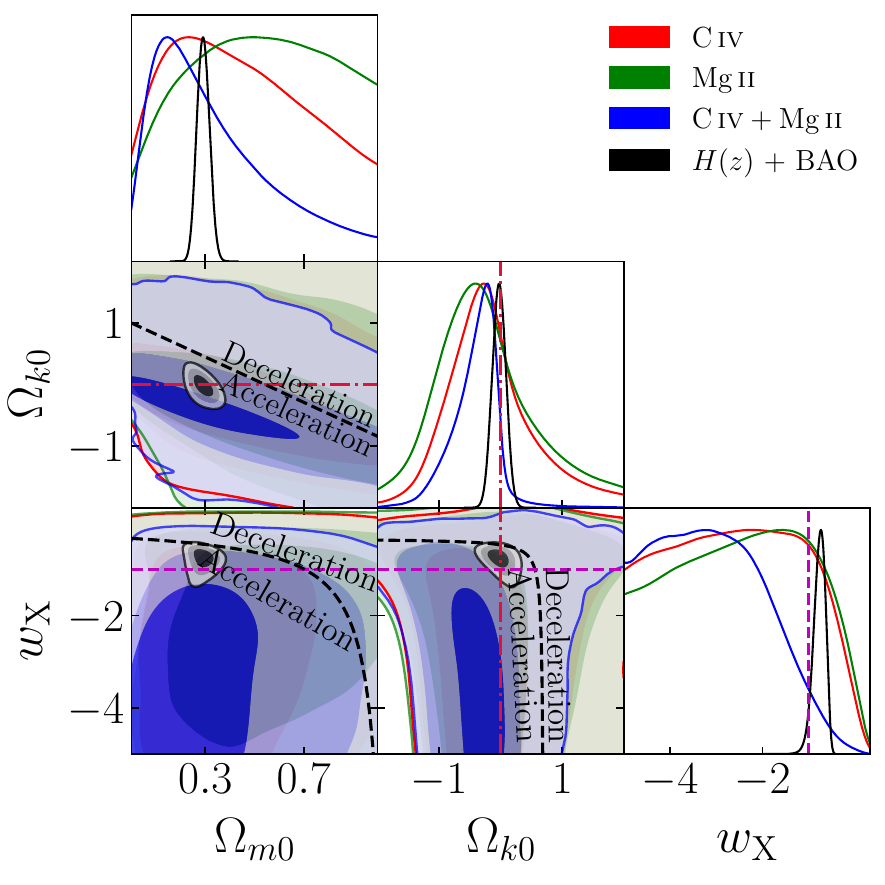}}\\
 \subfloat[]{%
    \includegraphics[width=0.5\textwidth,height=0.55\textwidth]{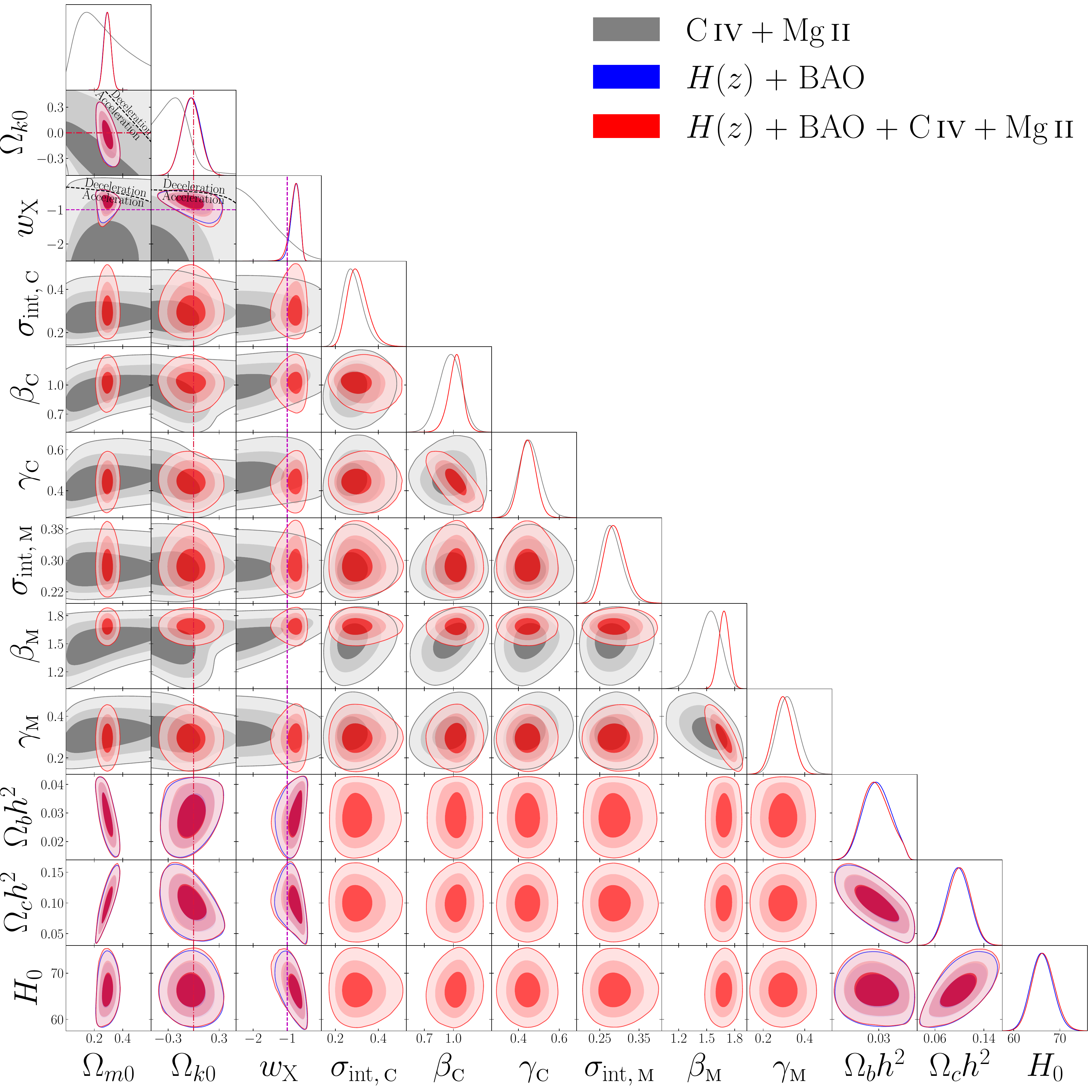}}
 \subfloat[]{%
    \includegraphics[width=0.5\textwidth,height=0.55\textwidth]{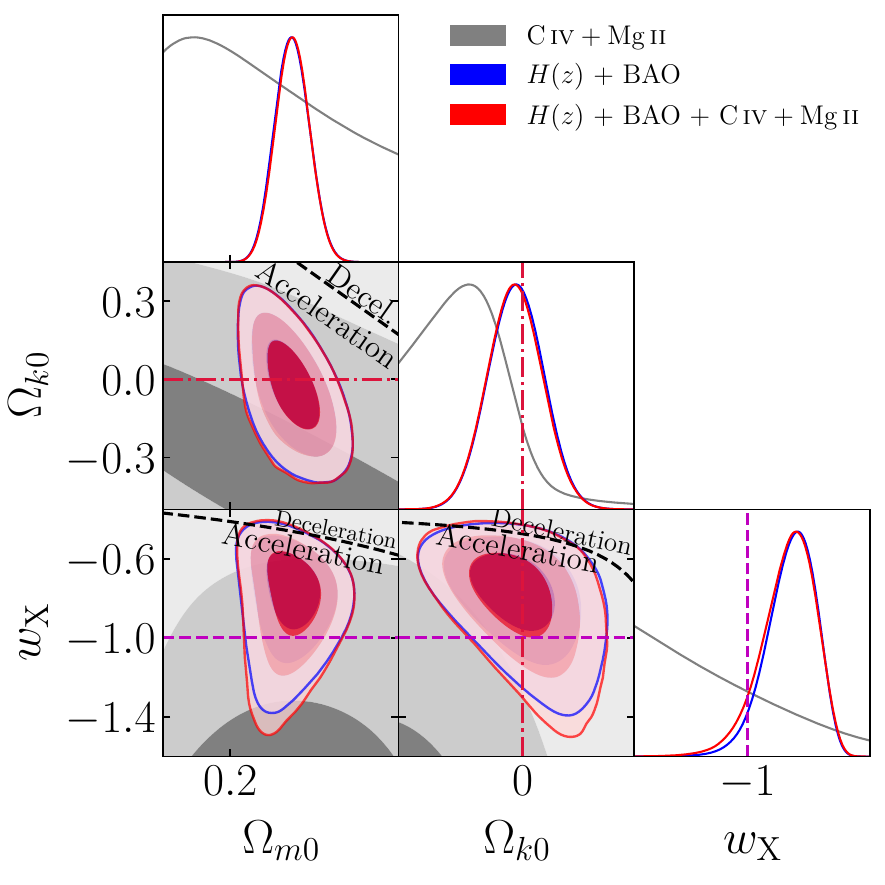}}\\
\caption{Same as Fig.\ \ref{fig3} but for non-flat XCDM. The zero-acceleration black dashed lines are computed for the third cosmological parameter set to the $H(z)$ + BAO data best-fitting values listed in Table \ref{tab:BFP}, and divide the parameter space into regions associated with currently-accelerating (either below left or below) and currently-decelerating (either above right or above) cosmological expansion. The crimson dash-dot lines represent flat hypersurfaces, with closed spatial hypersurfaces either below or to the left. The magenta dashed lines represent $w_{\rm X}=-1$, i.e.\ non-flat \lcdm.}
\label{fig4}
\end{figure*}

\begin{figure*}
\centering
\centering
 \subfloat[]{%
    \includegraphics[width=0.33\textwidth,height=0.4\textwidth]{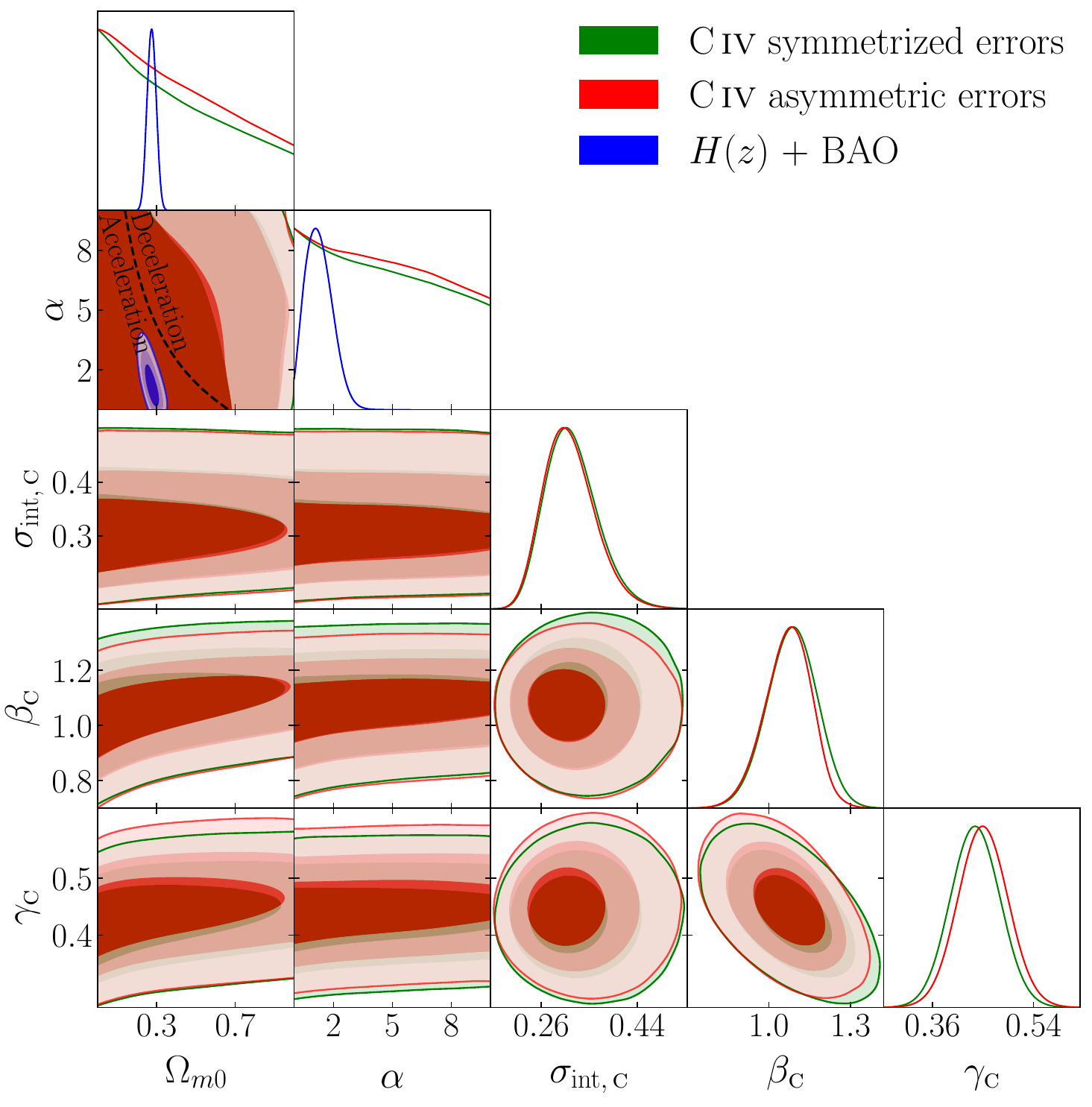}}
 \subfloat[]{%
    \includegraphics[width=0.33\textwidth,height=0.4\textwidth]{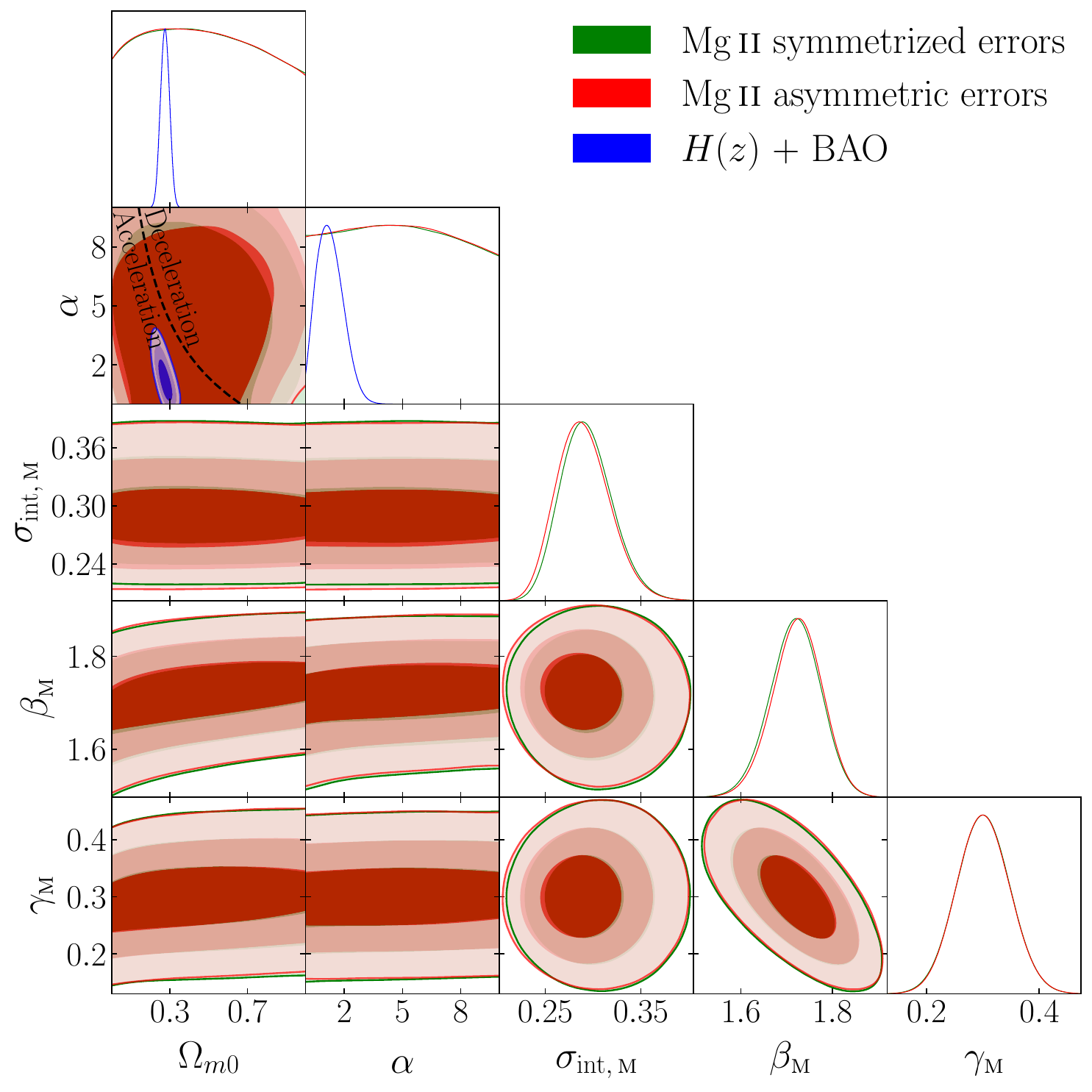}}
 \subfloat[]{%
    \includegraphics[width=0.33\textwidth,height=0.4\textwidth]{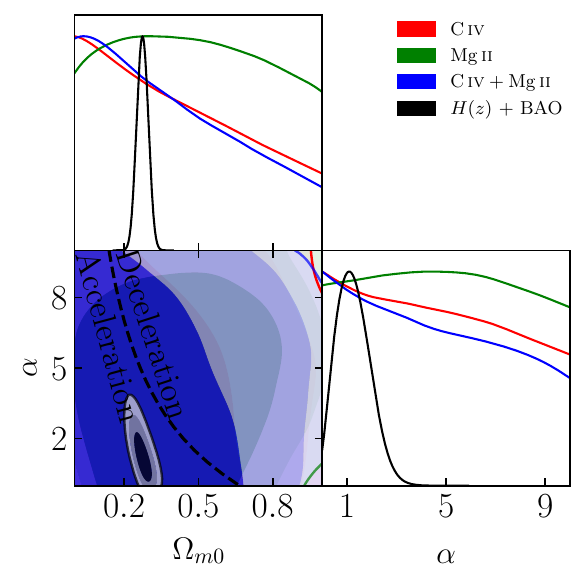}}\\
 \subfloat[]{%
    \includegraphics[width=0.5\textwidth,height=0.55\textwidth]{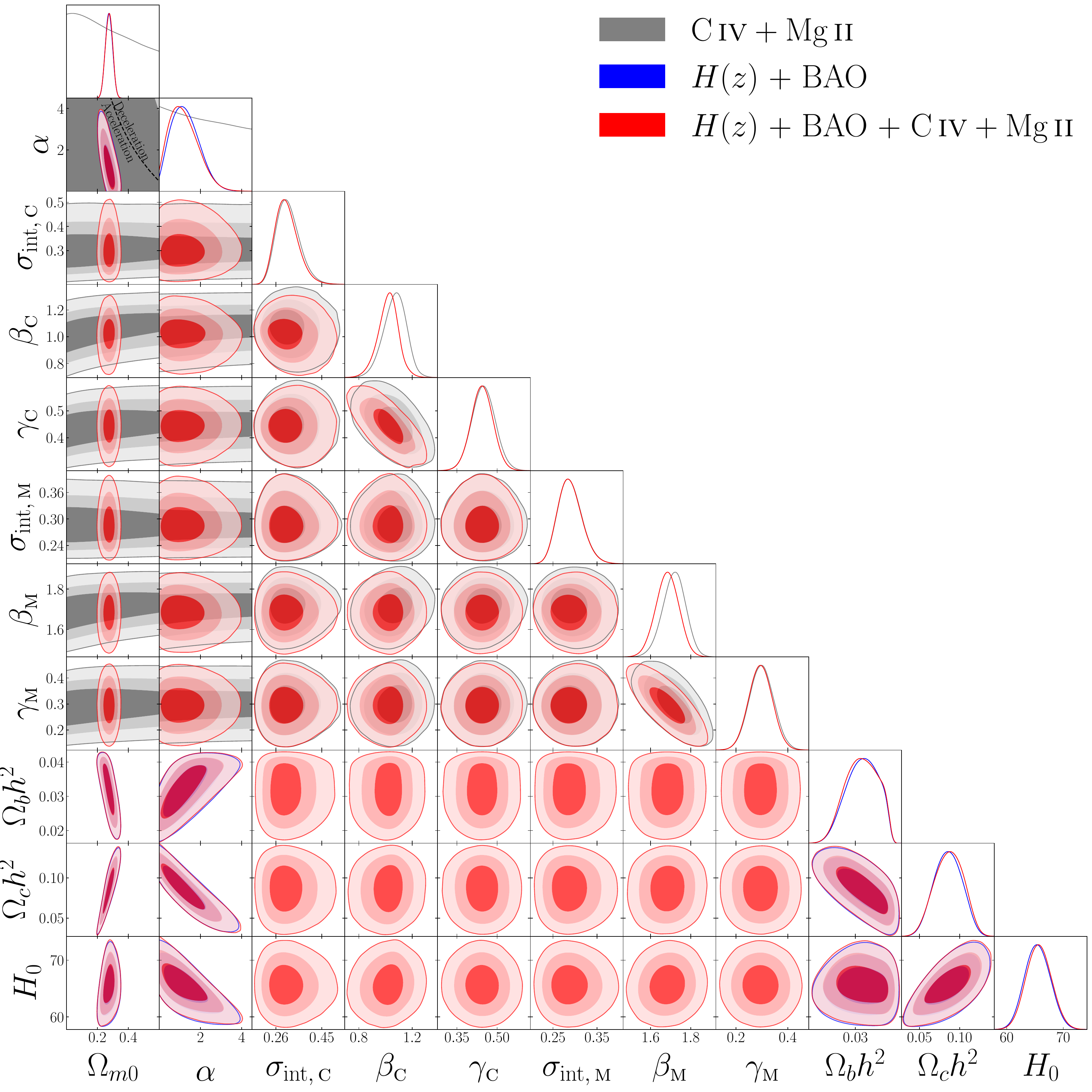}}
 \subfloat[]{%
    \includegraphics[width=0.5\textwidth,height=0.55\textwidth]{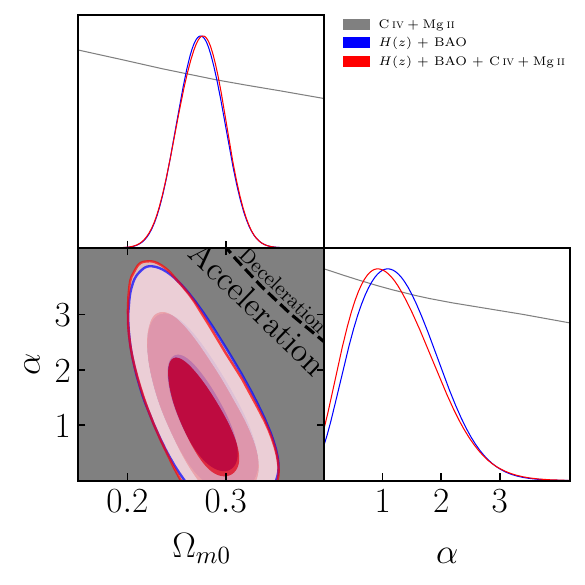}}\\
\caption{One-dimensional likelihood distributions and 1$\sigma$, 2$\sigma$, and 3$\sigma$ two-dimensional likelihood confidence contours for flat \pcdm\ from various combinations of data. The zero-acceleration black dashed lines divide the parameter space into regions associated with currently-accelerating (below left) and currently-decelerating (above right) cosmological expansion. The $\alpha = 0$ axes correspond to flat \lcdm.}
\label{fig5}
\end{figure*}

\begin{figure*}
\centering
 \subfloat[]{%
    \includegraphics[width=0.33\textwidth,height=0.4\textwidth]{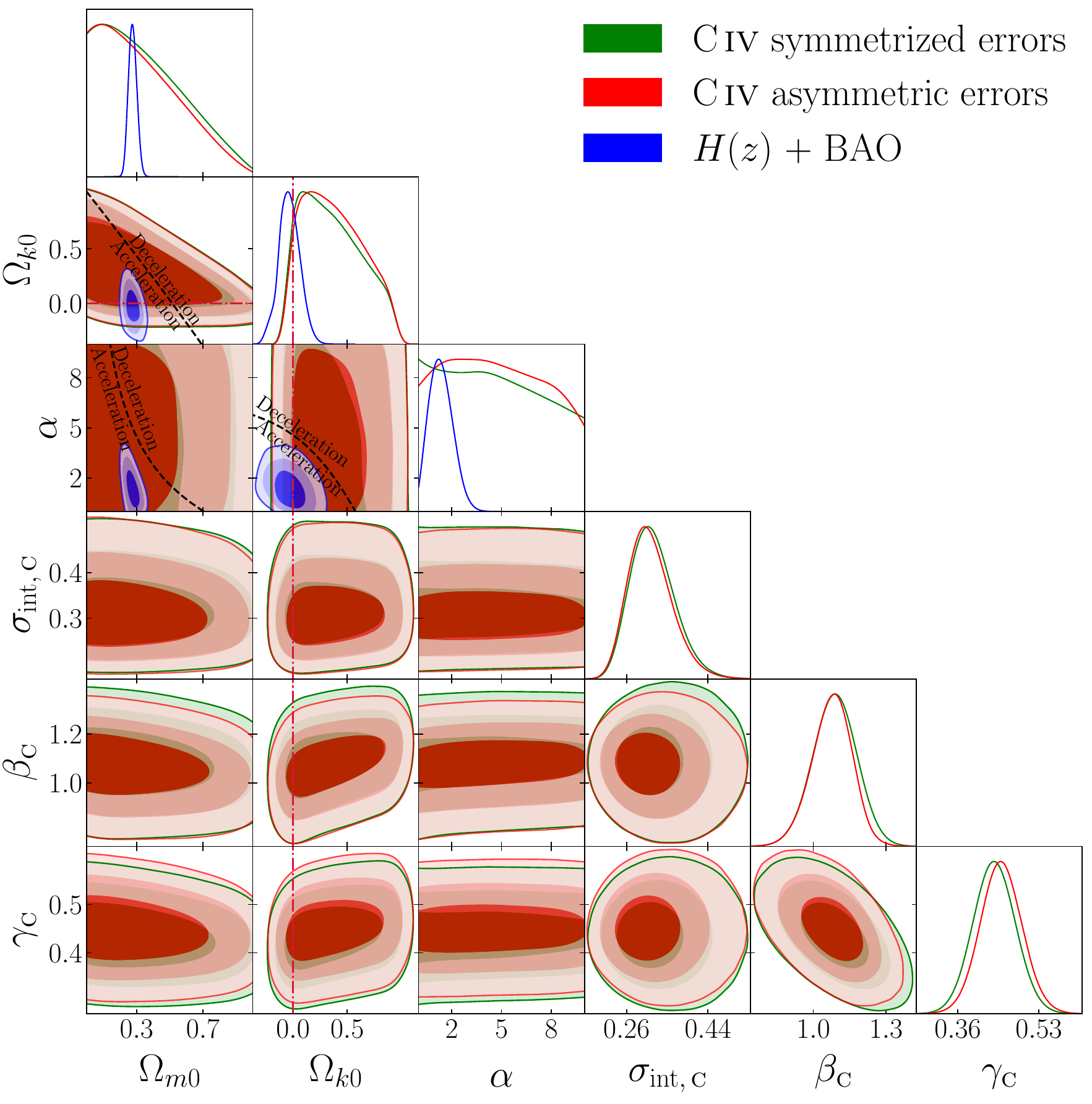}}
 \subfloat[]{%
    \includegraphics[width=0.33\textwidth,height=0.4\textwidth]{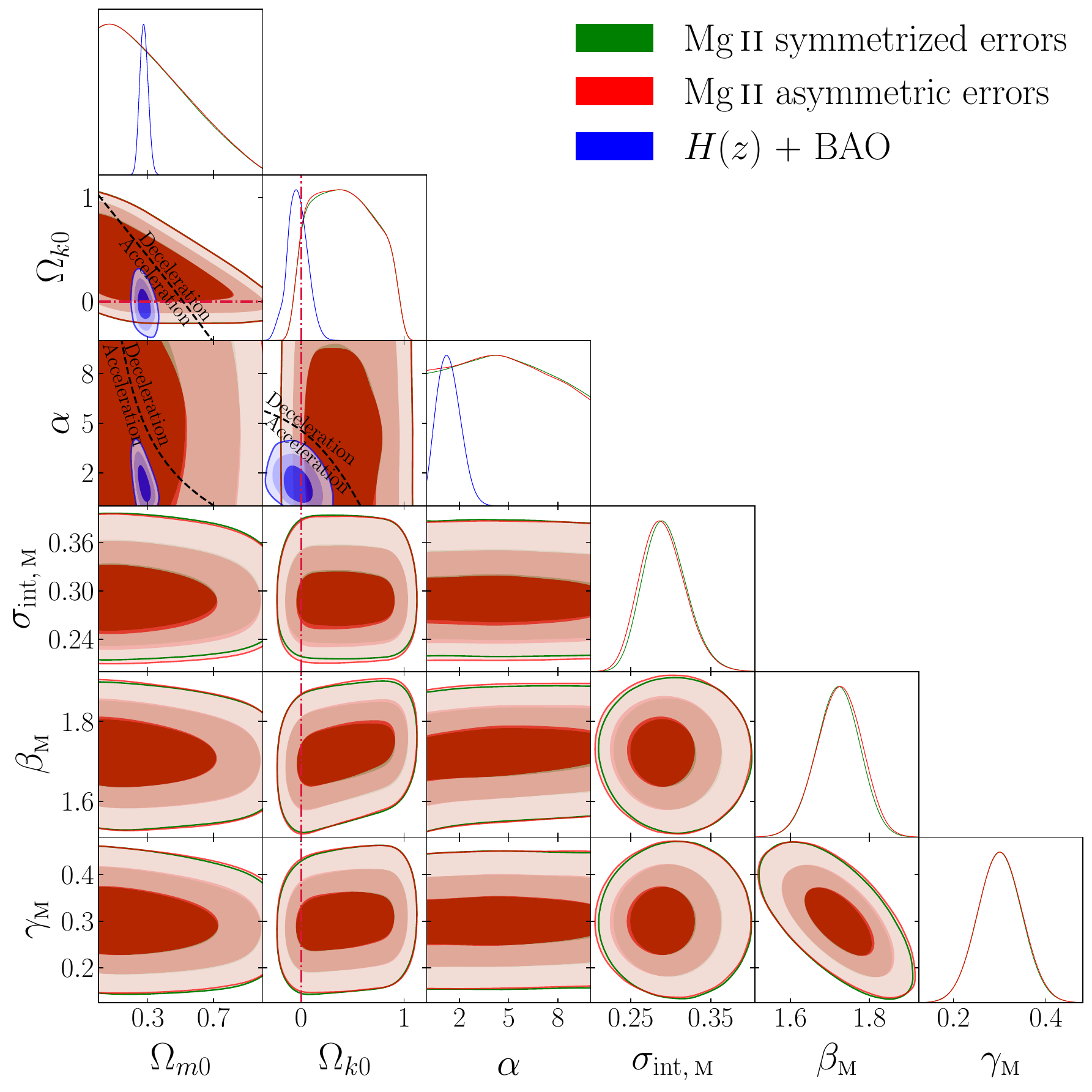}}
 \subfloat[]{%
    \includegraphics[width=0.33\textwidth,height=0.4\textwidth]{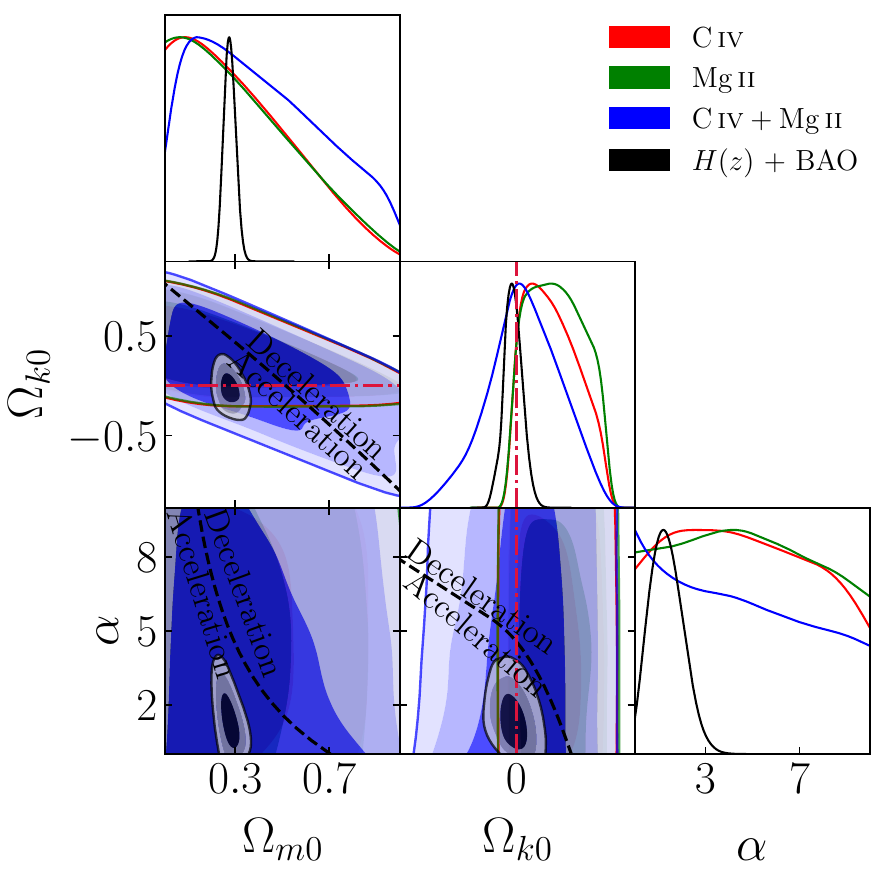}}\\
 \subfloat[]{%
    \includegraphics[width=0.5\textwidth,height=0.55\textwidth]{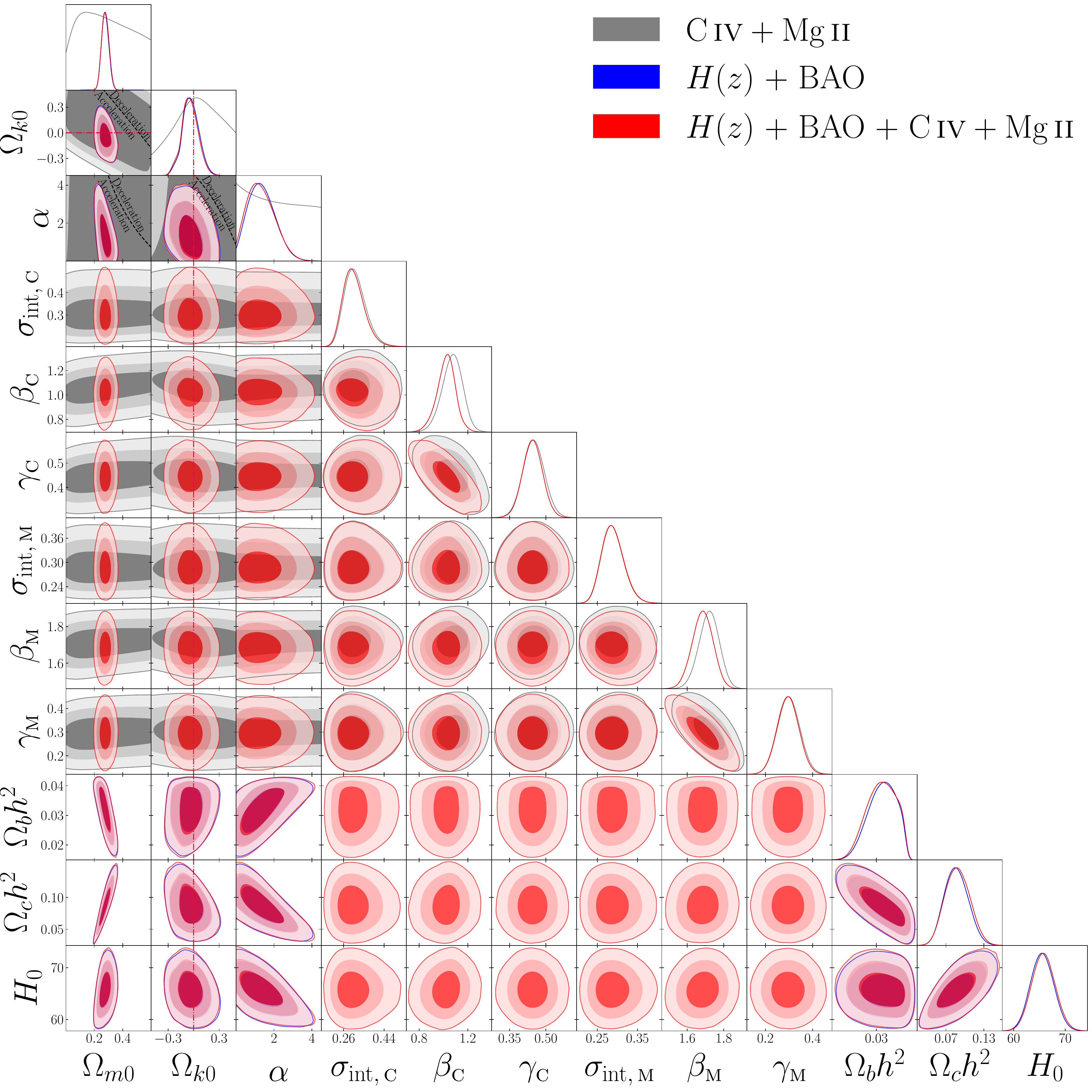}}
 \subfloat[]{%
    \includegraphics[width=0.5\textwidth,height=0.55\textwidth]{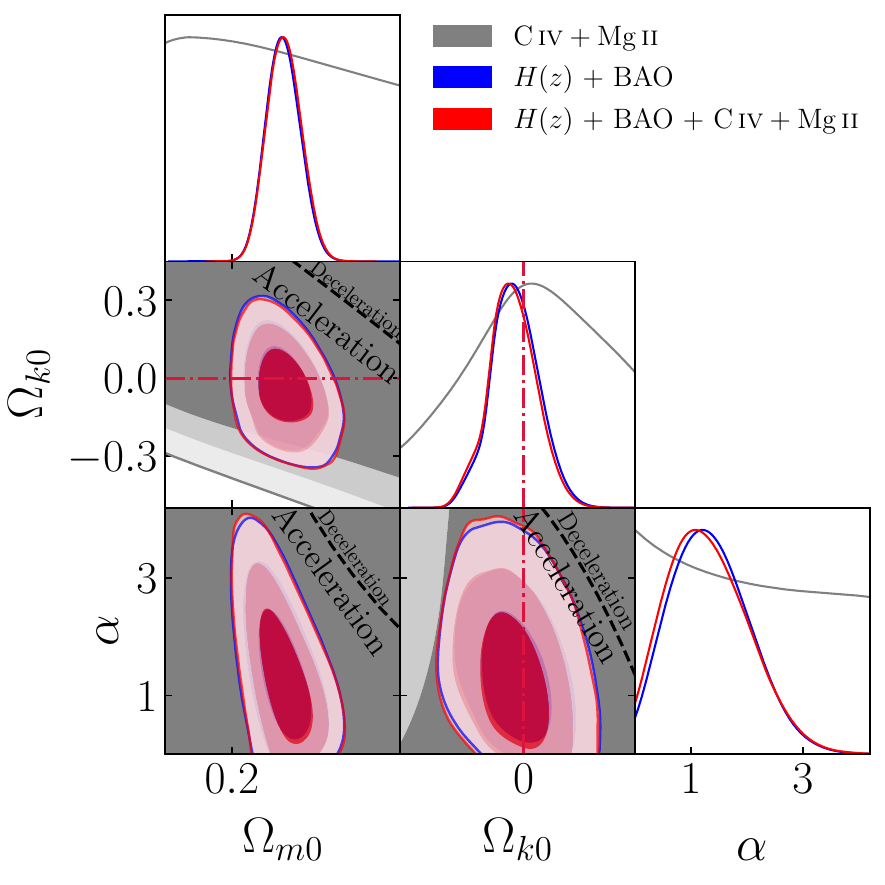}}\\
\caption{Same as Fig.\ \ref{fig5} but for non-flat \pcdm. The zero-acceleration black dashed lines are computed for the third cosmological parameter set to the $H(z)$ + BAO data best-fitting values listed in Table \ref{tab:BFP}, and divide the parameter space into regions associated with currently-accelerating (below left) and currently-decelerating (above right) cosmological expansion. The crimson dash-dot lines represent flat hypersurfaces, with closed spatial hypersurfaces either below or to the left. The $\alpha = 0$ axes correspond to non-flat \lcdm.}
\label{fig6}
\end{figure*}

The posterior one-dimensional probability distributions and two-dimensional confidence regions of cosmological-model and $R-L$ relation parameters for the six cosmological models are shown in Figs.\ \ref{fig1}--\ref{fig6}, where in panel (a) of each figure results of the \civ\ data analyses with symmetrized errors and asymmetric errors are shown in green and red, respectively; in panel (b) of each figure results of the \mii\ data analyses with symmetrized errors and asymmetric errors are shown in green and red, respectively; in panel (c) of each figure the results of the \civ, \mii, and joint \civ\ + \mii\ data analyses with asymmetric errors, and the $H(z)$ + BAO data analysis are shown in red, green, blue, and black, respectively; and in panels (d) and (e) of each figure the results of the joint (asymmetric errors) \civ\ + \mii, $H(z)$ + BAO, and $H(z)$ + BAO + \civ\ + \mii\ data analyses are shown in grey, blue, and red, respectively. The unmarginalized best-fitting parameter values, as well as the values of maximum likelihood $\mathcal{L}_{\rm max}$, AIC, BIC, DIC, $\Delta \mathrm{AIC}$, $\Delta \mathrm{BIC}$, and $\Delta \mathrm{DIC}$, for all models and data sets, are listed in Table \ref{tab:BFP}. The marginalized posterior mean parameter values and uncertainties ($\pm 1\sigma$ error bars and 1 or 2$\sigma$ limits), for all models and data sets, are listed in Table \ref{tab:1d_BFP}. 

In all six cosmological models, all data combinations more favour currently accelerating cosmological expansion. This is also the case with \mii\ QSO data \citep{Khadkaetal_2021a, Khadkaetal2022a}, but differs from what happens with H$\beta$ QSO data, which more favour currently decelerated cosmological expansion \citep{Khadkaetal2021c}.

\subsection{Constraints from \civ, \mii, and \civ\ + \mii\ QSO data}
 \label{subsec:CMQ}

As shown in panels (a) and (b) of Figs.\ \ref{fig1}--\ref{fig6} and Table \ref{tab:1d_BFP}, we find that results from \cq\ and \mq\ symmetrized errors data analyses are only mildly different (less than 1$\sigma$) from those of \cq\ and \mq\ asymmetric errors data analyses. Since the analyses with asymmetric errors are the more correct ones, we summarize the asymmetric errors \civ\ results in what follows. The symmetric errors \mii\ results are discussed in \cite{Khadkaetal_2021a}. 

The \om\ constraints from \civ\ data range from a low of $<0.840$ (2$\sigma$, flat XCDM) to a high of $0.467^{+0.199}_{-0.378}$ (1$\sigma$, non-flat \lcdm). 

The \ok\ constraints from \civ\ data are $-0.330^{+0.534}_{-1.060}$, $-0.168^{+0.451}_{-0.789}$, and $0.096^{+0.359}_{-0.337}$ for non-flat \lcdm, XCDM, and \pcdm, respectively. Although \civ\ data favour closed hypersurfaces in non-flat \lcdm\ and non-flat XCDM, and favour open hypersurfaces in non-flat \pcdm, flat hypersurfaces are well within 1$\sigma$. \civ\ data only provide very weak constraints of \wx\ and $\alpha$. 

From panels (c) of Figs.\ \ref{fig1}--\ref{fig6}, we see that the cosmological-model parameter constraints from \cq\ data and from \mq\ data are mutually consistent, as are the cosmological constraints from \cq\ and $H(z)$ + BAO data and from \mq\ and $H(z)$ + BAO data.\footnote{\citet{Khadkaetal_2021a, Khadkaetal2022a} had earlier shown that the symmetric errors \mii\ cosmological constraints are mutually consistent with those from $H(z)$ + BAO data. This differs from the H$\beta$ QSOs cosmological constraints, which are $\sim 2\sigma$ inconsistent with those from $H(z)$ + BAO data \citep{Khadkaetal2021c}.} It is therefore reasonable to perform joint analyses of \cq\ and \mq\ data. As shown in panels (d) and (e) of Figs.\ \ref{fig1}--\ref{fig6}, the cosmological-model parameter constraints from \civ\ + \mq\ data and from $H(z)$ + BAO data are mutually consistent at $\lesssim 1.5\sigma$,\footnote{I.e., in the two-dimensional contour plots all or most of the 1$\sigma$ $H(z)$ + BAO data contour always lie inside the 2$\sigma$ \civ\ + \mq\ data contour.} so these data sets can be jointly analyzed. 

The constraints on the \civ\ $R-L$ relation parameters in the six different cosmological models are mutually consistent, so the $R-L$ relation \civ\ data set is standardizable. The constraints on the intrinsic scatter parameter of the \civ\ $R-L$ relation, $\sigma_{\rm int,\,\textsc{c}}$, range from a low of $0.296^{+0.041}_{-0.056}$ (flat XCDM) to a high of $0.314^{+0.038}_{-0.055}$ (flat \pcdm), with a difference of $0.26\sigma$, which are $\sim0.1-0.4\sigma$ larger than those of \mii\ ($\sigma_{\rm int,\,\textsc{m}}$). The constraints on the slope parameter of the \civ\ $R-L$ relation, $\gamma_{\rm\textsc{c}}$, range from a low of $0.426\pm0.049$ (flat XCDM) to a high of $0.468^{+0.046}_{-0.056}$ (non-flat \lcdm), with a difference of $0.56\sigma$. The constraints on the intercept parameter of the \civ\ $R-L$ relation, $\beta_{\rm\textsc{c}}$, range from a low of $0.970^{+0.136}_{-0.108}$ (flat XCDM) to a high of $1.076^{+0.087}_{-0.074}$ (non-flat \pcdm), with a difference of $0.68\sigma$.

A summary value of the measured \civ\ $R-L$ relation slope, $\gamma_{\rm\textsc{c}} \simeq 0.45 \pm 0.04$, indicates that it is about $1\sigma$ lower than and consistent with the prediction of simple photoionization theory \citep[$\gamma = 0.5$,][]{Karasetal2021, Panda2022}, on the other hand a summary value of the measured \mii\ $R-L$ relation slope, $\gamma_{\rm\textsc{m}} \simeq 0.3 \pm 0.05$ \citep[also see][]{Khadkaetal_2021a, Khadkaetal2022a}, indicates that it is about $4\sigma$ lower than $\gamma = 0.5$.

The \om\ constraints from (asymmetric errors) \civ\ + \mii\ data range from a low of $<0.563$ (2$\sigma$, flat XCDM) to a high of $<0.537$ (1$\sigma$, flat \pcdm).

The \ok\ constraints from \civ\ + \mii\ data are $-0.818^{+0.391}_{-0.637}$ ($<0.474$, 2$\sigma$), $-0.410^{+0.368}_{-0.222}$ ($-0.410^{+0.698}_{-0.846}$, 2$\sigma$), and $0.088^{+0.384}_{-0.364}$  ($0.088^{+0.732}_{-0.722}$, 2$\sigma$) for non-flat \lcdm, XCDM, and \pcdm, respectively. \civ\ + \mii\ data favour closed hypersurfaces in non-flat \lcdm\ and non-flat XCDM, being $>1\sigma$ (but $<2\sigma$) and $\sim1.1\sigma$ away from flat hypersurfaces, respectively, and favour open hypersurfaces in non-flat \pcdm, with flat hypersurfaces being within 1$\sigma$. \civ\ + \mii\ data still provide weak constraints on \wx\ and $\alpha$, but \wx\ constraints are more than 2$\sigma$ away from $\wX=-1$ (\lcdm).

\civ\ + \mii\ data provide consistent (within $1\sigma$) constraints on the intrinsic scatter, slope, and intercept parameters of both \civ\ and \mii\ $R-L$ relations, which confirms that \civ\ and \mii\ QSOs are standardizable through different $R-L$ relations.

\subsection{Constraints from $H(z)$ + BAO + \civ\ + \mii\ data}
\label{subsec:HzBCM}

The mutually consistent cosmological-model parameter constraints allow us to jointly analyze $H(z)$ + BAO and \civ\ + \mii\ data. In what follows we summarize the cosmological-model parameter constraints from $H(z)$ + BAO + \civ\ + \mii\ data and contrast them with those from $H(z)$ + BAO data.

The $H(z)$ + BAO + \civ\ + \mii\ data provide \om\ constraints ranging from a low of $0.275\pm0.023$ (flat \pcdm) to a high of $0.301^{+0.015}_{-0.017}$ (flat \lcdm), with a difference of $0.91\sigma$, which only slightly differ from those determined using only $H(z)$ + BAO data.

The $H_0$ constraints from $H(z)$ + BAO + \civ\ + \mii\ data range from a low of $65.68^{+2.20}_{-2.19}$ \hunit\ (flat \pcdm) to a high of $69.15\pm1.77$ \hunit\ (flat \lcdm), with a difference of $1.23\sigma$, which are $0.65\sigma$ (flat \pcdm) lower than and $0.35\sigma$ (flat \lcdm) higher than the median statistics estimate of $H_0=68\pm2.8$ \hunit\ \citep{chenratmed}, and $2.94\sigma$ (flat \pcdm) and $1.84\sigma$ (flat \lcdm) lower than the local Hubble constant measurement of $H_0 = 73.2 \pm 1.3$ \hunit\ \citep{Riess_2021}. The $H(z)$ + BAO + \civ\ + \mii\ data provide $H_0$ constraints that are slightly higher ($\sim0.1\sigma$ at most) and mostly more restrictive ($\sim4\%$ at most) than those from $H(z)$ + BAO data.

The \ok\ constraints from $H(z)$ + BAO + \civ\ + \mii\ data are $0.047^{+0.079}_{-0.089}$, $-0.031\pm0.108$, and $-0.044^{+0.090}_{-0.094}$ for non-flat \lcdm, XCDM, and \pcdm, respectively, which are slightly lower ($\sim0.1\sigma$ at most) than those from $H(z)$ + BAO data. Similar to the $H(z)$ + BAO data results, non-flat \lcdm\ mildly favours open hypersurfaces, whereas non-flat XCDM and non-flat \pcdm\ mildly favour closed hypersurfaces. However, flat hypersurfaces are well within 1$\sigma$.

Dark energy dynamics is favoured. For flat (non-flat) XCDM, $w_{\rm X}=-0.799^{+0.143}_{-0.111}$ ($w_{\rm X}=-0.787^{+0.165}_{-0.102}$), with central values being $1.81\sigma$ ($<2\sigma$) higher than $w_{\rm X}=-1$; and for flat (non-flat) \pcdm, $\alpha=1.202^{+0.490}_{-0.862}$ ($\alpha=1.320^{+0.572}_{-0.869}$), with central values being $1.39\sigma$ ($1.52\sigma$) away from $\alpha=0$. The addition of \civ\ + \mii\ data to $H(z)$ + BAO data bring \wx\ and $\alpha$ values lower, and closer to \lcdm\ model values.

As expected, the constraints on the \civ\ and \mii\ $R-L$ relation parameters are consistent with those from the individual data sets and the \civ\ + \mii\ data.

\subsection{Model Comparison}
 \label{subsec:comp}

From the AIC, BIC, and DIC values listed in Table \ref{tab:BFP}, we find the following results (from the more correct \civ\ and \mii\ asymmetric errors analyses):
\begin{itemize}
    \item[1)]{\bf AIC.} $H(z)$ + BAO and $H(z)$ + BAO + \civ\ + \mii\ data favour flat \pcdm\ the most, and the evidence against the rest of the models/parametrizations is either only weak or positive. 
    
    \civ, \mii, and \civ\ + \mii\ data favour non-flat XCDM the most, however, in the \civ\ case, the evidence against non-flat \lcdm\ and flat XCDM is only weak, the evidence against flat \lcdm\ is positive, and the evidence against flat and non-flat \pcdm\ is strong; in the \mii\ case, the evidence against non-flat \lcdm\ and flat XCDM is positive, the evidence against flat \lcdm\ and flat \pcdm\ is strong, and the evidence against non-flat \pcdm\ is very strong; and in the \civ\ + \mii\ case, the evidence against flat XCDM is positive, the evidence against non-flat \lcdm\ is strong, and other models are very strongly disfavoured.
    
    \item[2)] {\bf BIC.} $H(z)$ + BAO and $H(z)$ + BAO + \civ\ + \mii\ data favour flat \lcdm\ the most, and in the former case, the evidence against the rest of the models/parametrizations is either only weak or positive, while in the latter case, the evidence against the rest of the models/parametrizations is either positive or strong (non-flat XCDM and non-flat \pcdm). 
    
    \civ\ data favour non-flat \lcdm\ the most, and the evidence against flat and non-flat XCDM is only weak, the evidence against flat \lcdm\ is positive, and the evidence against flat and non-flat \pcdm\ is strong.
    
    \mii\ and \civ\ + \mii\ data favour non-flat XCDM the most, however, in the \mii\ case, the evidence against non-flat \lcdm, flat XCDM, and non-flat XCDM is only weak, the evidence against flat \pcdm\ is strong, and the evidence against non-flat \pcdm\ is very strong; and in the \civ\ + \mii\ case, the evidence against non-flat \lcdm\ and flat XCDM is positive, the evidence against flat \lcdm\ is strong, and non-flat XCDM and non-flat \pcdm\ are very strongly disfavoured.

    \item[3)] {\bf DIC.} $H(z)$ + BAO and $H(z)$ + BAO + \civ\ + \mii\ data favour flat \pcdm\ the most, and the evidence against the rest of the models/parametrizations is either only weak or positive. 
    
    \civ\ and \mii\ data favour flat \lcdm\ the most, however, in the former case, the evidence against the rest of the models/parametrizations is either only weak or positive, whereas in the latter case, the evidence against flat and non-flat \pcdm\ is only weak, the evidence against non-flat \lcdm\ and flat XCDM is positive, and the evidence against non-flat XCDM is strong.
    
    \civ\ + \mii\ data favour non-flat XCDM the most, and the evidence against the rest of the models/parametrizations is either only weak or positive. 
\end{itemize}

Based on the more reliable DIC \citep{eacc21461d7e4419a9dee07b7fa8f657,90}, except for the \mii\ data set, these data sets do not provide strong evidence against any of the cosmological models/parametrizations.

\section{Discussion}
\label{sec:Discussion}

We have shown that QSOs with measured \civ\ time-delays can be standardized and so can be used as cosmological probes. \mii\ QSOs are also standardizable and so can be jointly analyzed with \civ\ QSOs to constrain cosmological model parameters. This is not true for current H$\beta$ QSOs \citep{Khadkaetal2021c} and more work is needed to clarify the H$\beta$ QSO situation.

However, we find a 2.3$\sigma$ difference in the measured slopes of the \civ\ and \mii\ $R-L$ relations, in the flat $\Lambda$CDM model asymmetric error bars results of Table \ref{tab:1d_BFP}, or from 
the summary values of the measured \civ\ and \mii\ $R-L$ relation slopes, $\gamma_{\rm\textsc{c}} \simeq 0.45 \pm 0.04$ and $\gamma_{\rm\textsc{m}} \simeq 0.3 \pm 0.05$. And while the \civ\ slope is only about $1\sigma$ lower than the $\gamma = 0.5$ slope predicted by simple photoionization theory \citep{Bentzetal2013, Karasetal2021, Panda2022} or a dust-based model of the BLR \citep{czerny2011,2021ApJ...920...30N,2022ApJ...931...39M,2022A&A...663A..77N}, the \mii\ slope is about $4\sigma$ lower than $\gamma = 0.5$, which is more statistically significant. 

In this section we examine potential (selection effect produced) differences between the \civ\ and \mii\ compilations and conclude that the ones we study here are not very significant. We begin by computing the Eddington ratio $\lambda_{\rm Edd}$ for the 78 \mii\ sources and the 38 \civ\ sources. We use the definition $\lambda_{\rm Edd}=L_{\rm bol}/L_{\rm Edd}$ where the bolometric luminosity is estimated as a multiple of the corresponding monochromatic luminosity ($L_{1350}$ for the \civ\ sample and $L_{3000}$ for the \mii\ sample; for the calculations in this section we adopt the fixed flat $\Lambda$CDM model with $\Om=0.3$, $\Omega_{\Lambda}=0.7$, and $H_0=70\,{\rm km\,s^{-1}\,Mpc^{-1}}$), taking into account the monochromatic luminosity-dependent bolometric correction factors according to \citet{2019MNRAS.488.5185N}. Subsequently, we evaluate the correlations between $\lambda_{\rm Edd}$ and the monochromatic luminosity and the redshift. 

\begin{figure}
    \centering
    \includegraphics[width=\columnwidth]{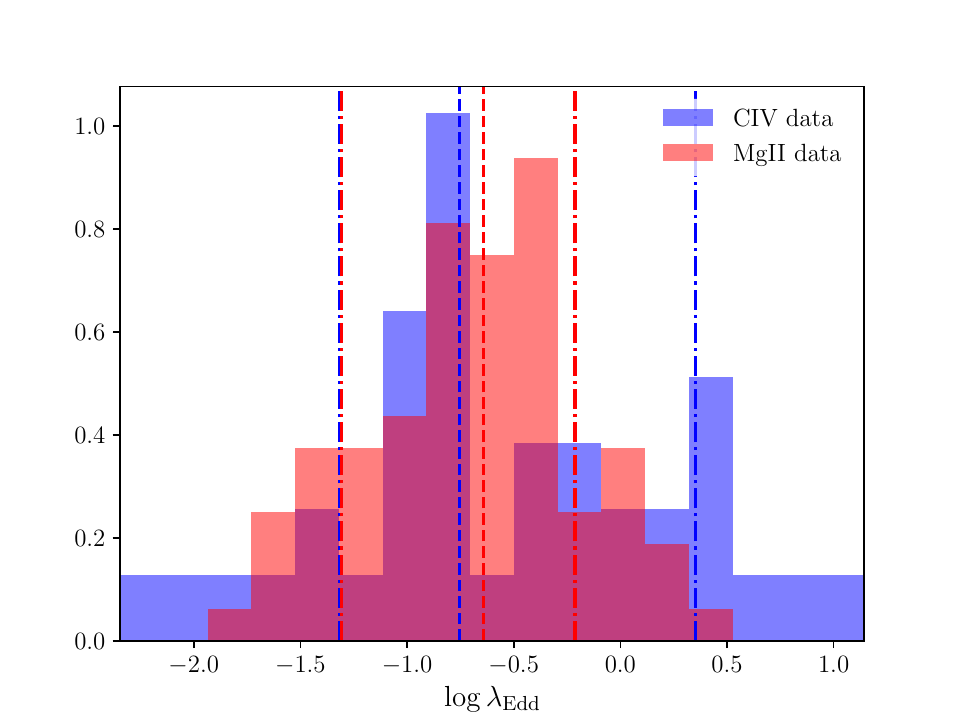}
    \caption{Normalized distributions of the Eddington ratio ($\log{\lambda_{\rm Edd}}$) for the \civ\ sample (blue histogram) and for the \mii\ sample (pink histogram). Vertical dashed lines stand for median values, while the dot-dashed lines represent 16- and 84-\% percentiles. The bin size is $\Delta(\log{\lambda_{\rm Edd}})\sim 0.205$.}
    \label{fig_lambda_dist}
\end{figure}

The Eddington-ratio normalized distributions for the \civ\ and \mii\ samples are shown in Fig.~\ref{fig_lambda_dist}, represented by blue and pink histograms, respectively. The vertical dashed lines stand for the distribution medians, while vertical dot-dashed lines mark 16- and 84-\% percentiles. For the \mii\ sample, the $\lambda_{\rm Edd}$ distribution median is $-0.64$, while 16- and 84-\% percentiles are $-1.31$ and $-0.21$, respectively. For the \civ\ sample, the median of the $\lambda_{\rm Edd}$ distribution is $-0.75$, while 16- and 84-\% percentiles are $-1.32$ and $0.35$, respectively.  The median $\lambda_{\rm Edd}$ of the \mii\ sources is larger than the median $\lambda_{\rm Edd}$ of the \civ\ sources. However, the \civ\ distribution is skewed significantly towards higher $\lambda_{\rm Edd}$ values, with a total range of $(-2.35, 1.14)$, in comparison with the range of $(-1.93,0.50)$ for the \mii\ sample. Also note that the Eddington ratios of the super-Eddington sources generally have large error bars, being consistent with the Eddington limit as well.

\begin{figure*}
    \centering
    \includegraphics[width=0.48\textwidth]{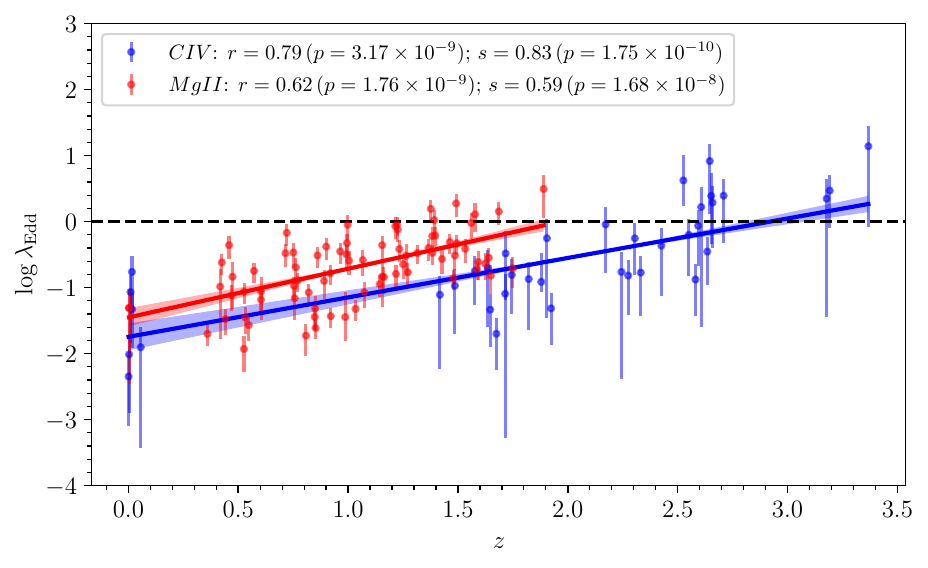}
    \includegraphics[width=0.48\textwidth]{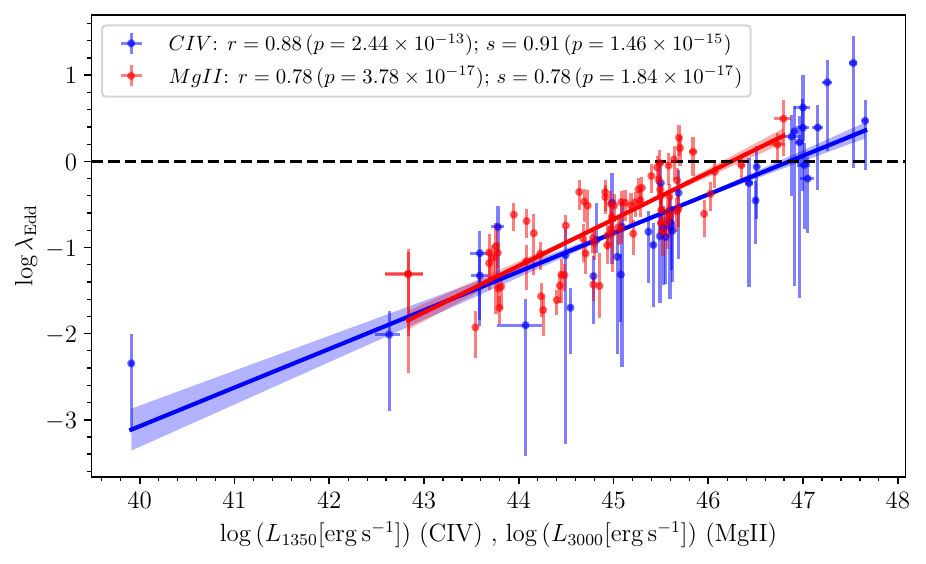}
    \caption{The Eddington ration $\lambda_{\rm Edd}=L_{\rm bol}/L_{\rm Edd}$ as a function of the source redshift (left panel) and as a function of the monochromatic luminosities $L_{1350}$ or $L_{3000}$ (right panel) for \civ\ (blue points) and \mii\ (red points) sources. In both panels, we include the Pearson ($r$) and the Spearman rank-order correlation coefficients ($s$), which indicate significant positive correlations between $\lambda_{\rm Edd}$ and $z$ as well as $\lambda_{\rm Edd}$ and $L_{1350}$ (or $L_{3000}$). The horizontal dashed line stands for the Eddington limit ($\log{\lambda_{\rm Edd}}=0$). The solid blue and red lines stand for the best-fitting linear relations for the \civ\ and \mii\ datasets, respectively, see Eqs.~\eqref{eq_lambda_z} and Eqs.~\eqref{eq_lambda_lum} for the best-fitting slopes and intercepts including 1$\sigma$ uncertainties.}
    \label{fig_lambda_correlations}
\end{figure*}

The $\lambda_{\rm Edd}$-$z$ and $\lambda_{\rm Edd}$-$L_{1350}$ or $\lambda_{\rm Edd}$-$L_{3000}$  correlations are positive, see Fig.~\ref{fig_lambda_correlations}, left and right panels, respectively. The correlation significance is evaluated using Pearson and Spearman rank-order correlation coefficients, see the legend in Fig.~\ref{fig_lambda_correlations}. For both \civ\ and \mii\ samples, the $\lambda_{\rm Edd}$-$z$ correlation is positive and significant, being slightly stronger and more significant for \civ\ sources. The correlation between $\lambda_{\rm Edd}$ and the corresponding monochromatic luminosity is also positive and significant for both samples. Here we stress that the correlation between the Eddington ratio $\lambda_{\rm Edd}$ and the corresponding monochromatic luminosities is enhanced intrinsically due to the definition of $\lambda_{\rm Edd}=L_{\rm bol}/L_{\rm Edd}$, where the bolometric luminosity is proportional to the monochromatic luminosity. However, the relative comparison of the correlation slopes provides hints about the similarities/differences of the two samples. 

The slopes and the intercepts of the best-fitting linear relations between $\lambda_{\rm Edd}$ and $z$ are
\begin{align}
    \log{\lambda_{\rm Edd}}(\mathrm{\civ})&=(0.60 \pm 0.10)z-(1.75\pm 0.20)\,,\notag\\
    \log{\lambda_{\rm Edd}}(\mathrm{\mii})&=(0.74 \pm 0.12)z-(1.45\pm 0.14)\,,\label{eq_lambda_z}
\end{align}
while for the best-fitting linear relations between $\lambda_{\rm Edd}$ and the monochromatic luminosity we obtain
\begin{align}
    \log{\lambda_{\rm Edd}}(\mathrm{\civ})&=(0.45 \pm 0.04)\log{L_{1350}}-(21.03\pm 1.97)\,,\notag\\
    \log{\lambda_{\rm Edd}}(\mathrm{\mii})&=(0.54 \pm 0.05)\log{L_{3000}}-(25.01\pm 2.25)\,.\label{eq_lambda_lum}
\end{align}
The best-fitting relations are depicted in Fig.~\ref{fig_lambda_correlations} including 1$\sigma$ uncertainties of the best-fitting parameters. Compared to the \civ\ case, the \mii\ Eddington ratios for the sample of 78 sources appear to increase more steeply with both redshift and monochromatic luminosity, though the \mii\ and \civ\ slopes are consistent within 1.0$\sigma$ (for the $\lambda_{\rm Edd}$ vs. $z$ correlations) and 1.4$\sigma$ (for the $\lambda_{\rm Edd}$ vs. monochromatic luminosity correlations), respectively. The differences in slopes may merely be due to the limited number of sources in each sample, and hence a selection effect.

\begin{figure}
    \centering
    \includegraphics[width=\columnwidth]{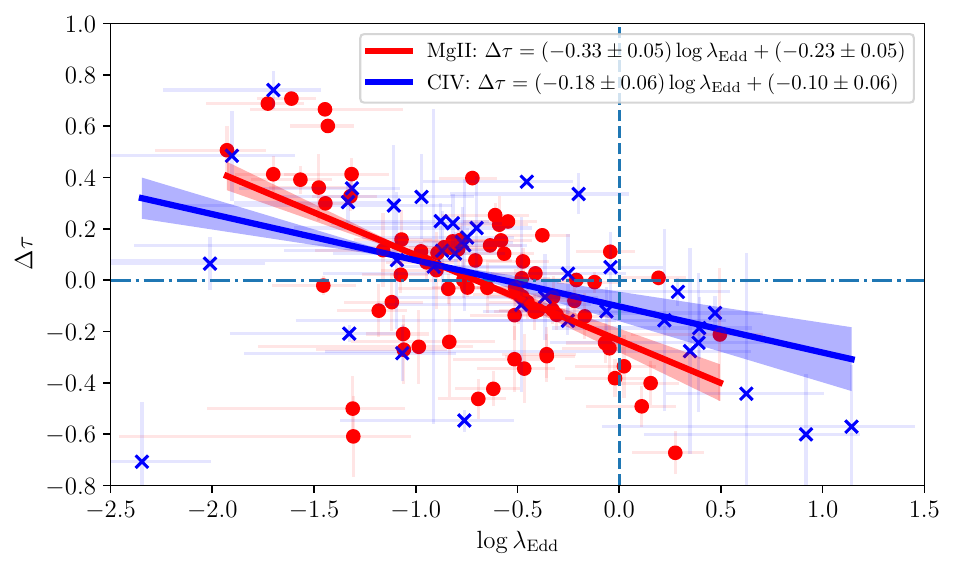}
    \caption{(Anti)correlation between $\Delta \tau \equiv \log{(\tau/\tau_{\rm RL})}$ and the Eddington ratio $\lambda_{\rm Edd}$ for both \civ\ sources (blue points) and \mii\ sources (red points). The correlation is more signifcant for the \mii\ sample with a Spearman rank-order correlation coefficient of $-0.53$  $(p=4.76\times 10^{-7})$ in comparison with $-0.50$ ($p=0.0015$) for the \civ\ sample. The slope of the anticorrelation is steeper for \mii\ sources, see Eq.~\eqref{eq_delta_tau} for the comparison.}
    \label{fig_delta_tau}
\end{figure}

However, when we study how the time-delay difference between the measured value and the value predicted from the best-fit $R-L$ relation (in the fixed flat $\Lambda$CDM model) --- $\Delta \tau \equiv \log{(\tau/\tau_{\rm RL})}$ \citep{2019ApJ...883..170M,2020ApJ...896..146Z} --- correlates with the Eddington ratio, the (anti)correlation is more significant and steeper for \mii\ sources. This is shown in Fig.~\ref{fig_delta_tau} for both \civ\ (blue points) and \mii\ sources (red points). As a caveat, we note that unless the SMBH mass is inferred independently of the RM, $\Delta \tau$ and $\log{\lambda_{\rm Edd}}$ are intrinsically correlated as $\Delta \tau \propto -\log{\tau}$, which implies anticorrelation.\footnote{It is generally thought advisable to use independent accretion-rate proxies, such as the relative \Feii\ strength \rfe\, the variability factor $F_{\rm var}$ \citep{2020ApJ...903...86M}, or the shape of the ionizing continuum \citep{Panda_etal_2019,2020ApJ...899...73F, Ferland_etal_2020}, rather than the Eddington ratio, due to the interdependency on the rest-frame time delay and the monochromatic luminosity that appear in the $R-L$ correlation. However, here we focus on a relative comparison between the \mii\ and \civ\ samples instead of the absolute values of the correlation slopes and normalizations.} Therefore, we can only assess the different behaviour between $\Delta \tau$ and $\lambda_{\rm Edd}$ based on the relative comparison of the correlation coefficients and the correlation slopes for the two samples. The anticorrelation between $\Delta \tau$ and $\log{\lambda_{\rm Edd}}$ is more significant for the \mii\ sample with a Spearman rank-order correlation coefficient of $s=-0.53$ $(p=4.76\times 10^{-7})$, while for the \civ\ sample we obtain $s=-0.50$ ($p=0.0015$). When we fit a linear function to the anticorrelations we get
\begin{align}
    \Delta \tau (\mathrm{\civ})&=(-0.18 \pm 0.06)\log{\lambda_{\rm Edd}}+(-0.10\pm 0.06)\,,\notag\\
    \Delta \tau (\mathrm{\mii})&=(-0.33 \pm 0.05)\log{\lambda_{\rm Edd}}+(-0.23\pm 0.05)\,.\label{eq_delta_tau}
\end{align}
Hence, the anticorrelation is significantly steeper for the \mii\ sample (by nearly a factor of two) in comparison with the current \civ\ sample. A combination of steeper $\Delta \tau$ -- $\log{\lambda_{\rm Edd}}$ anticorrelation with a slightly steeper correlation between $\log{\lambda_{\rm Edd}}$ and the monochromatic luminosity for the \mii\ sources can qualitatively account for the flatter \mii\ $R-L$ relation with respect to the \civ\ one. However, with a future increase in the number of \mii\ and \civ\ RM sources, both towards lower and higher redshifts, the difference in the correlation slopes is expected to become progressively weaker, since the general trends appear to be consistent between the \mii\ and the \civ\ samples and the differences can be traced to a few outlying sources. 

There may also be a potential problem in the way $\lambda_{\rm Edd}$ was measured in the \civ\ QSOs. Such measurements are uncertain, particularly for \civ. The {\civ}-emitting material is likely not completely virialized as revealed by its blueshift and blueward asymmetry with respect to low-ionization lines, which hints at an outflow approaching the observer. Therefore, the SMBH mass determination may be biased \citep{2005MNRAS.356.1029B}. The anticorrelation between the virial factor and the \civ\ FWHM is also the most uncertain among broad lines due to the smallest correlation coefficient as investigated by \citet{2018NatAs...2...63M}, which results in larger errorbars of \civ\ Eddington ratios in comparison with \mii\ ones in Fig.~\ref{fig_lambda_correlations}. Looking at how the \civ\ sample was selected might suggest that the histogram in Fig.~\ref{fig_lambda_dist} may not well represent the true \civ\ $\lambda_{\rm Edd}$ values. One of the criteria used when selecting the final \civ\ QSOs from the SDSS-RM sample is the variability of the light curve. Sources with no significant variability were excluded from the final sample to ensure a reliable time-lag estimate \citep{2019ApJ...887...38G}. Several papers claim that there is a negative correlation between the variability and the Eddington ratio \citep[e.g.][and references therein]{2022arXiv220512275D}. This suggests that the criterion of the variability applied to the SDSS-RM sample excludes highly-radiating QSOs. Such QSOs flatten the $R-L$ relation and increase the scatter in the case of the H$\beta$ $R-L$ relation, which is not observed in the \civ\ $R-L$ relation. In addition, \citet{2021ApJ...915..129K} claim that their three sources do not show strong outflows, which are usually found in QSOs with high Eddington ratios \citep{2017MNRAS.465.2120C, 2018A&A...618A.179M}. We performed a visual inspection of the rest of the sources and most of them show spectral features associated with low Eddington ratios (symmetric profiles in \civ\ $\lambda$1549, moderate He\,\textsc{ii} $\lambda1640$ contribution, strong contribution of C\,\textsc{iii}] $\lambda1909$, and low contributions of \Feii\ and Fe\,\textsc{iii}). All these facts, in contradiction to Fig.~\ref{fig_lambda_dist}, suggest that the \civ\ sample mainly includes low Eddington sources. However, a detailed analysis of variability and spectral properties is needed to confirm the accretion state of sources in the \civ\ sample.

The \civ\ sample spans eight orders of luminosity ($L_{1350}\sim 10^{40-48}$ erg s$^{-1}$) and covers a large redshift range ($0.001<z<3.37$), and is standardizable and suitable for constraining cosmological models. The \mii\ sample has smaller luminosity and redshift ranges ($L_{3000}\sim 10^{43-47}$ erg s$^{-1}$, $0<z<1.8$), but is standardizable and also suitable for constraining cosmological parameters \citep{Khadkaetal_2021a, Khadkaetal2022a}, and can be jointly analyzed with the \civ\ sample. The same is not true for the H$\beta$ sample. Although the H$\beta$ sample spans almost five orders of luminosity ($L_{5100}=10^{41.5-46}$ erg s$^{-1}$), the redshift range is narrow ($0.002<z<0.9$), but most importantly the current H$\beta$ sample appears to be not standardizable \citep{Khadkaetal2021c}. Therefore, we cannot jointly use the \mii, \civ\ and H$\beta$ samples, which is unfortunate since they together contain more sources over large luminosity and redshift ranges, all of which are beneficial if such QSOs are to be useful for cosmological purposes.

With future missions and surveys, such as the Black Hole Mapper by the SDSS-V  \citep{SDSS_BH_Mapper_2017arXiv171103234K}, long-term spectroscopic and photometric measurements of reverberation lags for the continuum and the ``major" emission lines (\civ, \mii, and H$\beta$/H$\alpha$) will be obtained for a large sample of quasars, adding wide-area, multi-epoch optical spectroscopy to the era of time-domain imaging. With the photometric Legacy Survey of Space and Time performed by the Vera C.\ Rubin observatory \citep{LSST_2019ApJ...873..111I}, many more RM objects will be obtained, e.g.\ the Rubin Deep Drilling Field should result in a few thousand measurements at redshifts $\sim 1 - 2$ \citep{Ksubm22, 2018arXiv181106542B}. This will lead to the decrease in the statistical error. The source monitoring will be performed photometrically in six broad optical bands, which will be used to probe the continuum accretion-disc response as well as the BLR response using the photometric RM \citep[see e.g., ][for the assessment of the photometric RM for BLR RM and the future application in cosmology]{2019FrASS...6...75P,2020mbhe.confE..10M}. Currently, we obtain a difference with respect to BAO+$H(z)$ data constraints by at most $0.1\sigma$ when 116 \civ\ + \mii\ sources are included, hence a QSO dataset larger by an order of magnitude can naturally influence the overall cosmological constraints more. It is more challenging to lower the dispersion in the individual time-delay measurements. Using a consistent methodology for all the RM QSOs to infer time-delays will be necessary to avoid systematic errors introduced by combining different subsets of RM sources into one large dataset. Specifically, surveys and RM projects typically select sources in a narrow luminosity range, which can result in a problem. For instance, if the time-delay method used for higher-luminosity sources is systematically susceptible to yield smaller lags in comparison with the method used for lower-luminosity sources, it can lead to a systematically smaller slope of the R-L relation compared to the case when the same method is applied to all the sources across several orders of magnitude in luminosity.

\section{Conclusions}
\label{sec:conclusion}

In this paper, for the first time, we use 38 highest-quality \cq\ data to simultaneously constrain cosmological model parameters, in six cosmological models, and \civ\ $R-L$ relation parameters. We use a new technique we have developed to more correctly take into account the asymmetric time-lag $\tau$ error bars and applied it to both \civ\ and \mii\ QSO data \citep{Khadkaetal_2021a} for the first time. We find that similar to \mq\ data, \cq\ data are also standardizable through the $R-L$ relation since the \civ\ $R-L$ relation parameters are cosmological model-independent. Unlike the H$\beta$ QSOs cosmological constraints \citep{Khadkaetal2021c}, those from \civ\ and \mii\ QSO data are consistent with cosmological constraints from better established $H(z)$ + BAO data.

The mutually consistent cosmological parameter constraints from \civ, \mii, and $H(z)$ + BAO data allow us to perform joint analyses on \civ\ + \mii\ data as well as on $H(z)$ + BAO + \civ\ + \mii\ data. Although the joint \civ\ + \mii\ cosmological constraints are still weak, they do slightly ($\sim0.1\sigma$ at most) alter the $H(z)$ + BAO constraints when jointly analyzed.

The Rubin Observatory Legacy Survey of Space and Time \citep{LSST_2019ApJ...873..111I}, as well as the SDSS-V Black Hole Mapper \citep{SDSS_BH_Mapper_2017arXiv171103234K}, will find many more \civ\ and \mii\  QSOs. These new QSOs will result in significantly more restrictive \civ\ (and \mii) cosmological constraints than the first ones we have derived here.

\begin{sidewaystable*}
\centering
\resizebox*{\columnwidth}{0.75\columnwidth}{%
\begin{threeparttable}
\caption{Unmarginalized best-fitting parameter values for all models from various combinations of data.}\label{tab:BFP}
\begin{tabular}{lcccccccccccccccccccc}
\toprule
Model & Data set & $\Omega_{b}h^2$ & $\Omega_{c}h^2$ & $\Omega_{m0}$ & $\Omega_{k0}$ & $w_{\mathrm{X}}$/$\alpha$\tnote{a} & $H_0$\tnote{b} & $\sigma_{\mathrm{int,\,\textsc{c}}}$ & $\gamma_{\rm\textsc{c}}$ & $\beta_{\rm\textsc{c}}$ & $\sigma_{\mathrm{int,\,\textsc{m}}}$ & $\gamma_{\rm\textsc{m}}$ & $\beta_{\rm\textsc{m}}$ & $-2\ln\mathcal{L}_{\mathrm{max}}$ & AIC & BIC & DIC & $\Delta \mathrm{AIC}$ & $\Delta \mathrm{BIC}$ & $\Delta \mathrm{DIC}$ \\
\midrule
 & $H(z)$ + BAO & 0.0244 & 0.1181 & 0.301 & -- & -- & 68.98  & -- & -- & -- & -- & -- & -- & 25.64 & 31.64 & 36.99 & 32.32 & 0.00 & 0.00 & 0.00\\
 & \mq\ symm & -- & 0.0518 & 0.157 & -- & -- & -- & -- & -- & -- & 0.282 & 0.285 & 1.670 & 30.16 & 38.16 & 47.59 & 37.68 & 0.00 & 0.00 & 0.00\\
Flat & \mq\ asymm & -- & 0.0472 & 0.148 & -- & -- & -- & -- & -- & -- & 0.278 & 0.282 & 1.676 & 30.18 & 38.18 & 47.61 & 37.86 & 0.00 & 0.00 & 0.00\\%
\lcdm & \cq\ symm & -- & $-0.0007$ & 0.050 & -- & -- & -- & 0.275 & 0.403 & 0.989 & -- & -- & -- & 20.61 & 28.61 & 35.16 & 31.10 & 0.00 & 0.00 & 0.00\\
 & \cq\ asymm & -- & $-0.0026$ & 0.046 & -- & -- & -- & 0.265 & 0.411 & 0.980 & -- & -- & -- & 20.51 & 28.51 & 35.06 & 32.74 & 0.00 & 0.00 & 0.00\\
 & \cq\ asymm + \mq\ asymm & -- & 0.0082 & 0.068 & -- & -- & -- & 0.274 & 0.412 & 0.995 & 0.280 & 0.286 & 1.647 & 50.94 & 64.94 & 84.21 & 69.10 & 0.00 & 0.00 & 0.00\\
 & $H(z)$ + BAO + \cq\ asymm + \mq\ asymm & 0.0245 & 0.1148 & 0.295 & -- & -- & 68.86 & 0.284 & 0.421 & 1.079 & 0.279 & 0.288 & 1.688 & 78.31 & 96.31 & 123.98 & 98.13 & 0.00 & 0.00 & 0.00\\
\midrule
 & $H(z)$ + BAO & 0.0260 & 0.1098 & 0.292 & 0.048 & -- & 68.35 & -- & -- & -- & -- & -- & -- & 25.30 & 33.30 & 40.43 & 33.87 & 1.66 & 3.44 & 1.54\\
 & \mq\ symm & -- & 0.1703 & 0.399 & $-1.135$ & -- & -- & -- & -- & -- & 0.275 & 0.352 & 1.625 & 25.39 & 35.39 & 47.17 & 40.12 & $-2.77$ & $-0.42$ & 2.44\\
Non-flat & \mq\ asymm & -- & 0.1708 & 0.400 & $-1.134$ & -- & -- & -- & -- & -- & 0.270 & 0.349 & 1.635 & 25.62 & 35.62 & 47.40 & 40.25 & $-2.56$ & $-0.20$ & 2.38\\
\lcdm & \cq\ symm & -- & 0.0147 & 0.081 & $-0.358$ & -- & -- & 0.237 & 0.486 & 0.968 & -- & -- & -- & 9.38 & 19.38 & 27.57 & 36.94 & $-9.23$ & $-7.59$ & 5.84\\
 & \cq\ asymm & -- & 0.0378 & 0.128 & $-0.471$ & -- & -- & 0.242 & 0.498 & 0.939 & -- & -- & -- & 14.44 & 24.44 & 32.63 & 36.77 & $-4.07$ & $-2.43$ & 4.03\\
 & \cq\ asymm + \mq\ asymm & -- & 0.0791 & 0.213 & $-0.678$ & -- & -- & 0.269 & 0.512 & 1.018 & 0.279 & 0.293 & 1.642 & 42.92 & 58.92 & 80.95 & 68.83 & $-6.01$ & $-3.26$ & $-0.27$\\
 & $H(z)$ + BAO + \cq\ asymm + \mq\ asymm & 0.0260 & 0.1112 & 0.293 & 0.036 & -- & 68.64 & 0.278 & 0.430 & 1.045 & 0.278 & 0.293 & 1.684 & 78.09 & 98.09 & 128.84 & 99.73 & 1.79 & 4.86 & 1.60\\
\midrule
 & $H(z)$ + BAO & 0.0296 & 0.0951 & 0.290 & -- & $-0.754$ & 65.79 & -- & -- & -- & -- & -- & -- & 22.39 & 30.39 & 37.52 & 30.63 & $-1.25$ & 0.53 & $-1.69$\\
 & \mq\ symm & -- & $-0.0234$ & 0.003 & -- & $-4.949$ & -- & -- & -- & -- & 0.270 & 0.241 & 1.354 & 24.39 & 34.39 & 46.18 & 41.19 & $-3.77$ & $-1.41$ & 3.51\\
Flat & \mq\ asymm & -- & $-0.0226$ & 0.005 & -- & $-4.983$ & -- & -- & -- & -- & 0.267 & 0.249 & 1.361 & 24.35 & 34.35 & 46.13 & 41.14 & $-3.83$ & $-1.48$ & 3.28\\
XCDM & \cq\ symm & -- & $-0.0237$ & 0.003 & -- & $-4.990$ & -- & 0.234 & 0.321 & 0.727 & -- & -- & -- & 12.69 & 22.69 & 30.88 & 31.64 & $-5.92$ & $-4.29$ & 0.53\\
 & \cq\ asymm & -- & $-0.0203$ & 0.010 & -- & $-4.988$ & -- & 0.232 & 0.355 & 0.736 & -- & -- & -- & 15.58 & 25.58 & 33.77 & 33.67 & $-2.93$ & $-1.29$ & 0.93\\
 & \cq\ asymm + \mq\ asymm & -- & $-0.0212$ & 0.008 & -- & $-4.875$ & -- & 0.225 & 0.337 & 0.757 & 0.262 & 0.248 & 1.399 & 40.24 & 56.24 & 78.27 & 66.94 & $-8.70$ & $-5.95$ & $-2.16$\\
 & $H(z)$ + BAO + \cq\ asymm + \mq\ asymm & 0.0283 & 0.1007 & 0.297 & -- & $-0.792$ & 66.05 & 0.283 & 0.425 & 1.074 & 0.279 & 0.286 & 1.691 & 75.85 & 95.85 & 126.60 & 97.19 & $-0.46$ & 2.62 & $-0.94$\\
\midrule
 & $H(z)$ + BAO & 0.0289 & 0.0985 & 0.296 & $-0.053$ & $-0.730$ & 65.76 & -- & -- & -- & -- & -- & -- & 22.13 & 32.13 & 41.05 & 32.51 & 0.49 & 4.06 & 0.19\\
 & \mq\ symm & -- & $-0.0103$ & 0.030 & $-0.060$ & $-3.207$ & -- & -- & -- & -- & 0.257 & 0.284 & 1.382 & 18.29 & 30.29 & 44.43 & 46.23 & $-7.87$ & $-3.16$ & 8.54\\
Non-flat & \mq\ asymm & -- & $-0.0016$ & 0.048 & $-0.095$ & $-2.947$ & -- & -- & -- & -- & 0.260 & 0.303 & 1.395 & 19.66 & 31.66 & 45.80 & 45.06 & $-6.52$ & $-1.81$ & 7.19\\
XCDM & \cq\ symm & -- & $-0.0161$ & 0.019 & $-0.035$ & $-3.815$ & -- & 0.220 & 0.428 & 0.827 & -- & -- & -- & 6.26 & 18.26 & 28.09 & 38.74 & $-10.35$ & $-7.08$ & 7.64\\
 & \cq\ asymm & -- & $-0.0130$ & 0.025 & $-0.042$ & $-4.329$ & -- & 0.228 & 0.386 & 0.828 & -- & -- & -- & 12.08 & 24.08 & 33.90 & 37.86 & $-4.43$ & $-1.16$ & 5.12\\
 & \cq\ asymm + \mq\ asymm & -- & $-0.0149$ & 0.021 & $-0.034$ & $-5.000$ & -- & 0.226 & 0.374 & 0.808 & 0.270 & 0.285 & 1.262 & 32.43 & 50.43 & 75.21 & 68.46 & $-14.51$ & $-9.00$ & $-0.71$\\
 & $H(z)$ + BAO + \cq\ asymm + \mq\ asymm & 0.0281 & 0.1046 & 0.296 & $-0.065$ & $-0.779$ & 67.10 & 0.276 & 0.434 & 1.064 & 0.274 & 0.298 & 1.676 & 75.81 & 97.81 & 131.64 & 98.92 & 1.51 & 7.66 & 0.79\\
\midrule
 & $H(z)$ + BAO & 0.0310 & 0.0900 & 0.280 & -- & 1.010 & 65.89 & -- & -- & -- & -- & -- & -- & 22.31 & 30.31 & 37.45 & 29.90 & $-1.33$ & 0.46 & $-2.42$\\
 & \mq\ symm & -- & 0.0414 & 0.136 & -- & 0.018 & -- & -- & -- & -- & 0.282 & 0.282 & 1.668 & 30.17 & 40.17 & 51.96 & 38.18 & 2.01 & 4.37 & 0.50\\
Flat & \mq\ asymm & -- & 0.0532 & 0.160 & -- & 0.047 & -- & -- & -- & -- & 0.279 & 0.284 & 1.679 & 30.20 & 40.20 & 51.99 & 38.40 & 2.02 & 4.38 & 0.52\\
\pcdm & \cq\ symm & -- & 0.0003 & 0.052 & -- & 0.030 & -- & 0.274 & 0.399 & 0.996 & -- & -- & -- & 20.67 & 30.67 & 38.85 & 33.47 & 2.05 & 3.69 & 2.37\\
 & \cq\ asymm & -- & $-0.0059$ & 0.039 & -- & 0.061 & -- & 0.270 & 0.410 & 0.977 & -- & -- & -- & 20.59 & 30.59 & 38.78 & 34.56 & 2.08 & 3.72 & 1.82\\
 & \cq\ asymm + \mq\ asymm & -- & $-0.0078$ & 0.035 & -- & 0.014 & -- & 0.263 & 0.399 & 0.984 & 0.283 & 0.265 & 1.662 & 51.19 & 67.19 & 89.21 & 72.04 & 2.25 & 5.00 & 2.94\\
 & $H(z)$ + BAO + \cq\ asymm + \mq\ asymm & 0.0289 & 0.0985 & 0.288 & -- & 0.663 & 66.68 & 0.275 & 0.430 & 1.050 & 0.278 & 0.304 & 1.668 & 75.78 & 95.78 & 126.54 & 96.48 & $-0.52$ & 2.55 & $-1.65$\\
\midrule
 & $H(z)$ + BAO & 0.0306 & 0.0920 & 0.284 & $-0.058$ & 1.200 & 65.91 & -- & -- & -- & -- & -- & -- & 22.05 & 32.05 & 40.97 & 31.30 & 0.41 & 3.98 & $-1.02$\\
 & \mq\ symm & -- & 0.1167 & 0.290 & $-0.277$ & 0.000 & -- & -- & -- & -- & 0.283 & 0.290 & 1.678 & 29.81 & 41.81 & 55.95 & 38.70 & 3.65 & 8.36 & 1.02\\
Non-flat & \mq\ asymm & -- & 0.1354 & 0.328 & $-0.310$ & 0.042 & -- & -- & -- & -- & 0.275 & 0.292 & 1.675 & 29.91 & 41.91 & 56.05 & 38.65 & 3.73 & 8.45 & 0.78\\
\pcdm & \cq\ symm & -- & 0.0489 & 0.151 & $-0.146$ & 0.065 & -- & 0.272 & 0.417 & 1.010 & -- & -- & -- & 20.42 & 32.42 & 42.24 & 34.07 & 3.80 & 7.08 & 2.96\\
 & \cq\ asymm & -- & 0.0367 & 0.126 & $-0.125$ & 0.121 & -- & 0.268 & 0.423 & 1.006 & -- & -- & -- & 20.41 & 32.41 & 42.24 & 34.92 & 3.91 & 7.18 & 2.18\\
 & \cq\ asymm + \mq\ asymm & -- & 0.0611 & 0.176 & $-0.173$ & 0.115 & -- & 0.264 & 0.421 & 1.034 & 0.271 & 0.289 & 1.659 & 50.74 & 68.74 & 93.52 & 72.86 & 3.80 & 9.31 & 3.76\\
 & $H(z)$ + BAO + \cq\ asymm + \mq\ asymm & 0.0313 & 0.0882 & 0.278 & $-0.044$ & 1.256 & 65.76 & 0.281 & 0.455 & 1.035 & 0.276 & 0.292 & 1.687 & 75.73 & 97.73 & 131.55 & 97.61 & 1.42 & 7.57 & $-0.52$\\
\bottomrule
\end{tabular}
\begin{tablenotes}[flushleft]
\item [a] \wx\ corresponds to flat/non-flat XCDM and $\alpha$ corresponds to flat/non-flat \pcdm.
\item [b] \hunit. $\Omega_b$ and $H_0$ are set to be 0.05 and 70 \hunit, respectively.
\end{tablenotes}
\end{threeparttable}%
}
\end{sidewaystable*}

\begin{sidewaystable*}
\centering
\resizebox*{\columnwidth}{0.75\columnwidth}{%
\begin{threeparttable}
\caption{One-dimensional marginalized posterior mean values and uncertainties ($\pm 1\sigma$ error bars or $2\sigma$ limits) of the parameters for all models from various combinations of data.}\label{tab:1d_BFP}
\begin{tabular}{lccccccccccccc}
\toprule
Model & Data set & $\Omega_{b}h^2$ & $\Omega_{c}h^2$ & $\Omega_{m0}$ & $\Omega_{k0}$ & $w_{\mathrm{X}}$/$\alpha$\tnote{a} & $H_0$\tnote{b} & $\sigma_{\mathrm{int,\,\textsc{c}}}$ & $\gamma_{\rm\textsc{c}}$ & $\beta_{\rm\textsc{c}}$ & $\sigma_{\mathrm{int,\,\textsc{m}}}$ & $\gamma_{\rm\textsc{m}}$ & $\beta_{\rm\textsc{m}}$ \\
\midrule
 & $H(z)$ + BAO & $0.0247\pm0.0030$ & $0.1186^{+0.0076}_{-0.0083}$ & $0.301^{+0.016}_{-0.018}$ & -- & -- & $69.14\pm1.85$ & -- & -- & -- & -- & -- & -- \\
 & \mq\ symm & -- & -- & $0.470^{+0.199}_{-0.426}$ & -- & -- & -- & -- & -- & -- & $0.293^{+0.023}_{-0.030}$ & $0.297\pm0.047$ & $1.698^{+0.063}_{-0.058}$ \\
Flat & \mq\ asymm & -- & -- & $0.470^{+0.196}_{-0.430}$ & -- & -- & -- & -- & -- & -- & $0.290^{+0.024}_{-0.030}$ & $0.296\pm0.047$ & $1.703^{+0.064}_{-0.057}$ \\
\lcdm & \cq\ symm & -- & -- & $<0.471$\tnote{c} & -- & -- & -- & $0.311^{+0.039}_{-0.056}$ & $0.427\pm0.043$ & $1.051\pm0.096$ & -- & -- & -- \\
 & \cq\ asymm & -- & -- & $<0.503$\tnote{c} & -- & -- & -- & $0.307^{+0.039}_{-0.056}$ & $0.441\pm0.044$ & $1.034^{+0.097}_{-0.087}$ & -- & -- & -- \\
 & \cq\ asymm + \mq\ asymm & -- & -- & $<0.444$\tnote{c} & -- & -- & -- & $0.305^{+0.037}_{-0.054}$ & $0.440\pm0.042$ & $1.030\pm0.089$ & $0.289^{+0.023}_{-0.030}$ & $0.292\pm0.045$ & $1.691\pm0.061$ \\
 & $H(z)$ + BAO + \cq\ asymm + \mq\ asymm & $0.0247\pm0.0028$ & $0.1183^{+0.0073}_{-0.0080}$ & $0.301^{+0.015}_{-0.017}$ & -- & -- & $69.15\pm1.77$ & $0.303^{+0.036}_{-0.053}$ & $0.442\pm0.039$ & $1.026^{+0.075}_{-0.065}$ & $0.288^{+0.023}_{-0.029}$ & $0.294\pm0.044$ & $1.686\pm0.056$ \\
\midrule
 & $H(z)$ + BAO & $0.0266^{+0.0039}_{-0.0045}$ & $0.1088\pm0.0166$ & $0.291\pm0.023$ & $0.059^{+0.081}_{-0.091}$ & -- & $68.37\pm2.10$ & -- & -- & -- & -- & -- & -- \\
 & \mq\ symm & -- & -- & $0.568^{+0.359}_{-0.196}$ & $-0.427^{+0.546}_{-1.453}$ & -- & -- & -- & -- & -- & $0.291^{+0.024}_{-0.030}$ & $0.315^{+0.049}_{-0.056}$ & $1.687\pm0.067$ \\
Non-flat & \mq\ asymm & -- & -- & $0.567^{+0.365}_{-0.192}$ & $-0.419^{+0.560}_{-1.449}$ & -- & -- & -- & -- & -- & $0.288^{+0.024}_{-0.031}$ & $0.316^{+0.048}_{-0.057}$ & $1.692^{+0.070}_{-0.064}$ \\
\lcdm & \cq\ symm & -- & -- & $0.417^{+0.153}_{-0.353}$ & $-0.478^{+0.346}_{-0.851}$ & -- & -- & $0.300^{+0.041}_{-0.055}$ & $0.465^{+0.047}_{-0.062}$ & $1.072\pm0.098$ & -- & -- & -- \\
 & \cq\ asymm & -- & -- & $0.467^{+0.199}_{-0.378}$ & $-0.330^{+0.534}_{-1.060}$ & -- & -- & $0.305^{+0.039}_{-0.055}$ & $0.468^{+0.046}_{-0.056}$ & $1.059^{+0.097}_{-0.086}$ & -- & -- & -- \\
 & \cq\ asymm + \mq\ asymm & -- & -- & $0.473^{+0.187}_{-0.311}$ & $-0.818^{+0.391}_{-0.637}$ & -- & -- & $0.299^{+0.036}_{-0.053}$ & $0.491^{+0.050}_{-0.064}$ & $1.073^{+0.093}_{-0.094}$ & $0.285^{+0.023}_{-0.030}$ & $0.314^{+0.048}_{-0.052}$ & $1.662\pm0.065$ \\
 & $H(z)$ + BAO + \cq\ asymm + \mq\ asymm & $0.0262^{+0.0037}_{-0.0044}$ & $0.1105\pm0.0165$ & $0.292\pm0.022$ & $0.047^{+0.079}_{-0.089}$ & -- & $68.55\pm2.05$ & $0.303^{+0.035}_{-0.052}$ & $0.441\pm0.039$ & $1.023^{+0.075}_{-0.064}$ & $0.288^{+0.023}_{-0.030}$ & $0.294\pm0.043$ & $1.685\pm0.055$ \\
\midrule
 & $H(z)$ + BAO & $0.0295^{+0.0042}_{-0.0050}$ & $0.0969^{+0.0178}_{-0.0152}$ & $0.289\pm0.020$ & -- & $-0.784^{+0.140}_{-0.107}$ & $66.22^{+2.31}_{-2.54}$ & -- & -- & -- & -- & -- & -- \\
 & \mq\ symm & -- & -- & $<0.515$\tnote{c} & -- & $<-0.371$ & -- & -- & -- & -- & $0.291^{+0.024}_{-0.030}$ & $0.294\pm0.048$ & $1.640^{+0.118}_{-0.073}$ \\
Flat & \mq\ asymm & -- & -- & $<0.512$\tnote{c} & -- & $<-0.370$ & -- & -- & -- & -- & $0.288^{+0.024}_{-0.030}$ & $0.295\pm0.047$ & $1.644^{+0.118}_{-0.072}$ \\
XCDM & \cq\ symm & -- & -- & $<0.737$ & -- & $<-0.926$ & -- & $0.292^{+0.041}_{-0.056}$ & $0.403\pm0.047$ & $0.957\pm0.128$ & -- & -- & -- \\
 & \cq\ asymm & -- & -- & $<0.840$ & -- & $<-0.636$ & -- & $0.296^{+0.041}_{-0.056}$ & $0.426\pm0.049$ & $0.970^{+0.136}_{-0.108}$ & -- & -- & -- \\
 & \cq\ asymm + \mq\ asymm & -- & -- & $<0.563$ & -- & $<-1.509$ & -- & $0.282^{+0.038}_{-0.054}$ & $0.405\pm0.047$ & $0.900^{+0.122}_{-0.121}$ & $0.283^{+0.023}_{-0.029}$ & $0.282^{+0.042}_{-0.046}$ & $1.557^{+0.117}_{-0.101}$ \\
 & $H(z)$ + BAO + \cq\ asymm + \mq\ asymm & $0.0292^{+0.0040}_{-0.0050}$ & $0.0983^{+0.0181}_{-0.0144}$ & $0.290^{+0.020}_{-0.018}$ & -- & $-0.799^{+0.143}_{-0.111}$ & $66.45^{+2.28}_{-2.53}$ & $0.305^{+0.036}_{-0.053}$ & $0.443\pm0.039$ & $1.021^{+0.077}_{-0.065}$ & $0.289^{+0.023}_{-0.030}$ & $0.295\pm0.044$ & $1.684\pm0.056$ \\
\midrule
 & $H(z)$ + BAO & $0.0294^{+0.0047}_{-0.0050}$ & $0.0980^{+0.0186}_{-0.0187}$ & $0.292\pm0.025$ & $-0.027\pm0.109$ & $-0.770^{+0.149}_{-0.098}$ & $66.13^{+2.35}_{-2.36}$ & -- & -- & -- & -- & -- & -- \\
 & \mq\ symm & -- & -- & $0.526^{+0.351}_{-0.268}$ & $-0.235^{+0.555}_{-0.936}$ & $-2.416^{+1.894}_{-1.260}$ & -- & -- & -- & -- & $0.291^{+0.024}_{-0.031}$ & $0.313^{+0.049}_{-0.055}$ & $1.649^{+0.124}_{-0.079}$ \\
 & \mq\ asymm & -- & -- & $0.523^{+0.341}_{-0.280}$ & $-0.245^{+0.544}_{-0.917}$ & $-2.448^{+1.834}_{-1.321}$ & -- & -- & -- & -- & $0.287^{+0.024}_{-0.030}$ & $0.314^{+0.049}_{-0.055}$ & $1.651^{+0.122}_{-0.079}$ \\
Non-flat & \cq\ symm & -- & -- & $0.399^{+0.140}_{-0.375}$ & $-0.247^{+0.360}_{-0.645}$ & $-2.636^{+1.523}_{-1.570}$ & -- & $0.297^{+0.042}_{-0.057}$ & $0.443^{+0.045}_{-0.055}$ & $1.028^{+0.121}_{-0.106}$ & -- & -- & -- \\
XCDM & \cq\ asymm & -- & -- & $0.452^{+0.186}_{-0.394}$ & $-0.168^{+0.451}_{-0.789}$ & $-2.573^{+1.568}_{-1.605}$ & -- & $0.302^{+0.040}_{-0.056}$ & $0.453^{+0.046}_{-0.051}$ & $1.025^{+0.115}_{-0.096}$ & -- & -- & -- \\
 & \cq\ asymm + \mq\ asymm & -- & -- & $0.338^{+0.101}_{-0.299}$ & $-0.410^{+0.368}_{-0.222}$ & $<-1.124$ & -- & $0.282^{+0.037}_{-0.053}$ & $0.456^{+0.047}_{-0.056}$ & $0.966^{+0.119}_{-0.110}$ & $0.280^{+0.023}_{-0.030}$ & $0.319^{+0.048}_{-0.054}$ & $1.526^{+0.132}_{-0.108}$ \\
 & $H(z)$ + BAO + \cq\ asymm + \mq\ asymm & $0.0290^{+0.0044}_{-0.0053}$ & $0.0997^{+0.0186}_{-0.0188}$ & $0.293\pm0.025$ & $-0.031\pm0.108$ & $-0.787^{+0.165}_{-0.102}$ & $66.41^{+2.26}_{-2.49}$ & $0.305^{+0.036}_{-0.053}$ & $0.444\pm0.040$ & $1.023^{+0.077}_{-0.065}$ & $0.289^{+0.023}_{-0.030}$ & $0.295\pm0.045$ & $1.685\pm0.056$ \\
\midrule
 & $H(z)$ + BAO & $0.0320^{+0.0054}_{-0.0041}$ & $0.0855^{+0.0175}_{-0.0174}$ & $0.275\pm0.023$ & -- & $1.267^{+0.536}_{-0.807}$ & $65.47^{+2.22}_{-2.21}$ & -- & -- & -- & -- & -- & -- \\
 & \mq\ symm & -- & -- & -- & -- & -- & -- & -- & -- & -- & $0.294^{+0.023}_{-0.029}$ & $0.301\pm0.046$ & $1.719\pm0.054$ \\
Flat & \mq\ asymm & -- & -- & -- & -- & -- & -- & -- & -- & -- & $0.290^{+0.024}_{-0.030}$ & $0.301\pm0.046$ & $1.724^{+0.055}_{-0.051}$ \\
\pcdm & \cq\ symm & -- & -- & $<0.553$\tnote{c} & -- & $<6.291$\tnote{c} & -- & $0.318^{+0.039}_{-0.055}$ & $0.436\pm0.043$ & $1.086\pm0.090$ & -- & -- & -- \\
 & \cq\ asymm & -- & -- & $<0.565$\tnote{c} & -- & -- & -- & $0.314^{+0.038}_{-0.055}$ & $0.450\pm0.044$ & $1.070^{+0.092}_{-0.075}$ & -- & -- & -- \\
 & \cq\ asymm + \mq\ asymm & -- & -- & $<0.537$\tnote{c} & -- & $<6.202$\tnote{c} & -- & $0.312^{+0.037}_{-0.054}$ & $0.449\pm0.043$ & $1.069^{+0.091}_{-0.074}$ & $0.289^{+0.023}_{-0.030}$ & $0.299\pm0.046$ & $1.717^{+0.059}_{-0.053}$ \\
 & $H(z)$ + BAO + \cq\ asymm + \mq\ asymm & $0.0318^{+0.0053}_{-0.0045}$ & $0.0866^{+0.0190}_{-0.0171}$ & $0.275\pm0.023$ & -- & $1.202^{+0.490}_{-0.862}$ & $65.68^{+2.20}_{-2.19}$ & $0.306^{+0.036}_{-0.053}$ & $0.444\pm0.040$ & $1.019^{+0.078}_{-0.066}$ & $0.289^{+0.023}_{-0.030}$ & $0.295\pm0.044$ & $1.683\pm0.056$ \\
\midrule
 & $H(z)$ + BAO & $0.0320^{+0.0057}_{-0.0038}$ & $0.0865^{+0.0172}_{-0.0198}$ & $0.277^{+0.023}_{-0.026}$ & $-0.034^{+0.087}_{-0.098}$ & $1.360^{+0.584}_{-0.819}$ & $65.53\pm2.19$ & -- & -- & -- & -- & -- & -- \\
 & \mq\ symm & -- & -- & $0.476^{+0.230}_{-0.396}$ & $0.042^{+0.390}_{-0.378}$ & -- & -- & -- & -- & -- & $0.294^{+0.023}_{-0.029}$ & $0.300\pm0.046$ & $1.721\pm0.054$ \\
Non-flat & \mq\ asymm & -- & -- & $0.473^{+0.216}_{-0.409}$ & $0.040^{+0.395}_{-0.383}$ & -- & -- & -- & -- & -- & $0.291^{+0.024}_{-0.030}$ & $0.301\pm0.047$ & $1.724\pm0.056$ \\
\pcdm & \cq\ symm & -- & -- & $<0.542$\tnote{c} & $0.104^{+0.381}_{-0.362}$ & -- & -- & $0.319^{+0.039}_{-0.055}$ & $0.437\pm0.043$ & $1.089\pm0.089$ & -- & -- & -- \\
 & \cq\ asymm & -- & -- & $0.427^{+0.153}_{-0.408}$ & $0.096^{+0.359}_{-0.337}$ & $4.747^{+2.065}_{-4.304}$ & -- & $0.313^{+0.037}_{-0.055}$ & $0.451\pm0.043$ & $1.076^{+0.087}_{-0.074}$ & -- & -- & -- \\
 & \cq\ asymm + \mq\ asymm & -- & -- & $<0.536$\tnote{c} & $0.088^{+0.384}_{-0.364}$ & $<6.162$\tnote{c} & -- & $0.312^{+0.037}_{-0.054}$ & $0.450\pm0.043$ & $1.072^{+0.088}_{-0.076}$ & $0.290^{+0.023}_{-0.030}$ & $0.299\pm0.046$ & $1.719\pm0.055$\\
 & $H(z)$ + BAO + \cq\ asymm + \mq\ asymm & $0.0317^{+0.0058}_{-0.0043}$ & $0.0884^{+0.0183}_{-0.0203}$ & $0.278^{+0.024}_{-0.026}$ & $-0.044^{+0.090}_{-0.094}$ & $1.320^{+0.572}_{-0.869}$ & $65.77\pm2.21$ & $0.306^{+0.036}_{-0.053}$ & $0.445\pm0.040$ & $1.021^{+0.079}_{-0.066}$ & $0.289^{+0.023}_{-0.030}$ & $0.296\pm0.045$ & $1.684\pm0.056$\\
\bottomrule
\end{tabular}
\begin{tablenotes}[flushleft]
\item [a] \wx\ corresponds to flat/non-flat XCDM and $\alpha$ corresponds to flat/non-flat \pcdm.
\item [b] \hunit. $\Omega_b$ and $H_0$ are set to be 0.05 and 70 \hunit, respectively.
\item [c] This is the 1$\sigma$ limit. The 2$\sigma$ limit is set by the prior and not shown here.
\end{tablenotes}
\end{threeparttable}%
}
\end{sidewaystable*}

\section*{Acknowledgements}

This research was supported in part by US DOE grant DE-SC0011840, by the Polish Funding Agency National Science Centre, project 2017/26/A/ST9/00756 (Maestro 9), by GAČR EXPRO grant 21-13491X, by Millenium Nucleus NCN$19\_058$ (TITANs), and by the Conselho Nacional de Desenvolvimento Científico e Tecnológico (CNPq) Fellowship (164753/2020-6). The authors acknowledge the Czech-Polish mobility program (M\v{S}MT
8J20PL037 and PPN/BCZ/2019/1/00069). Part of the computation for this project was performed on the Beocat Research Cluster at Kansas State University.

\section*{Data Availability}

The data analysed in this article are listed in Table \ref{tab:civdata} of this paper and in table A1 of \cite{Khadkaetal_2021a}.
 


\bibliographystyle{mnras}
\bibliography{references} 




\onecolumn
\begin{appendix}
\section{Golden sample of \civ\ QSO data}
\label{sec:appendix_golden}
\addtolength{\tabcolsep}{0pt}
\LTcapwidth=\linewidth
\begin{longtable}{lccccc}
\caption{Sample of QSOs with a high-quality detection of the \civ\ time-delay. The sample is based on 38 sources compiled by \citet{2021ApJ...915..129K}. In the table, we list from the left to the right column: object name, redshift, flux density at 1350\,\AA, monochromatic luminosity at 1350\,\AA\, for the flat $\Lambda$CDM model ($H_0=70\,{\rm km\,s^{-1}\,Mpc^{-1}}$, $\Om=0.3$, $\Omega_{\Lambda}=0.7$), the rest-frame \civ\ time-lag (in days) determined for all the sources using either the ICCF or the zDCF method (or their combination), and the original reference. }
\label{tab:civdata}\\
\toprule
Object &  $z$ &  $\log \left(F_{1350}/{\rm erg}\,{\rm s^{-1}}{\rm cm^{-2}}\right)$  &  $\log \left(L_{1350}/{\rm erg}\,{\rm s^{-1}}\right)$  &  $\tau$ (days) & Reference\\
\midrule
\endfirsthead
\endhead
\bottomrule
\endfoot
NGC 4395 &  0.001064 &  $-11.4848 \pm     0.0272$ &    $39.9112 \pm     0.0272$ &    $0.040^{+0.024}_{-0.018}$  & \citet{2005ApJ...632..799P,2006ApJ...641..638P} \\
NGC 3783 & 0.00973  &  $-9.7341 \pm 0.0918$ &   $43.5899 \pm 0.0918 $&     $3.80^{+1.0}_{-0.9}$  & \citet{2005ApJ...632..799P,2006ApJ...641..638P}\\
NGC 7469 &  0.01632 &    $-9.9973 \pm 0.0712$  &  $43.7803 \pm 0.0712$ &    $2.5^{+0.3}_{-0.2}$    & \citet{2005ApJ...632..799P,2006ApJ...641..638P}\\
3C 390.3 &  0.0561 &   $-10.8036 \pm 0.2386$ &   $44.0719 \pm 0.2386$&    $35.7^{+11.4}_{-14.6}$ & \citet{2005ApJ...632..799P,2006ApJ...641..638P}\\
NGC 4151 &  0.00332 &    $-9.7544 \pm 0.1329$  &  $42.6314 \pm 0.1329$ &     $3.34^{+0.82}_{-0.77}$   & \citet{2006ApJ...647..901M}\\
NGC 5548  & 0.01676 &  $-10.2111 \pm 0.0894$  &  $43.5899 \pm  0.0894$ &     $4.53^{+0.35}_{-0.34}$ &  \citet{2015ApJ...806..128D}\\
CTS 286 &  2.551 &  $-11.6705 \pm 0.0719$  &  $47.0477 \pm     0.0719$&   $459^{+71}_{-92}$  & \citet{2018ApJ...865...56L} \\
CTS 406 &  3.178 &   $-12.0382 \pm 0.0402$ & $46.9101 \pm 0.0402$ &   $98^{+55}_{-74}$   & \citet{2018ApJ...865...56L}  \\
CTS 564 &  2.653 &   $-11.7615 \pm 0.0664$ & $46.9978 \pm 0.0664$ & $115^{+184}_{-29}$  & \citet{2018ApJ...865...56L} \\
CTS 650 &  2.659 &  $-11.8815 \pm 0.1068$  &  $46.8802 \pm 0.1068$ &  $162^{+33}_{-10}$   & \citet{2018ApJ...865...56L}  \\
CTS 953 &  2.526 &   $-11.7082 \pm 0.0868$ &   $46.9996 \pm 0.0868$ &   $73^{+115}_{-58}$  & \citet{2018ApJ...865...56L}  \\
CTS 1061 &  3.368 &   $-11.4788 \pm 0.0405$ &   $47.5299 \pm 0.0405$ &    $91^{+111}_{-24}$     & \citet{2018ApJ...865...56L}  \\
J 214355 &  2.607 &   $-11.7786 \pm    0.0485$ &   $46.9624 \pm    0.0485$ &  $136^{+100}_{-90}$    &  \citet{2018ApJ...865...56L} \\
J 221516 &  2.709 &   $-11.6263 \pm    0.0569$ &   $47.1550 \pm    0.0569$&   $153^{+91}_{-12}$   &  \citet{2018ApJ...865...56L} \\
DES J0228-04 & 1.905 &   $-11.9791 \pm    0.0405$ &   $46.4298 \pm    0.0405$ &   $123^{+43}_{-42}$   &  \citet{2019MNRAS.487.3650H} \\
DES J0033-42 &  2.593 &   $-12.2248 \pm  0.0201$ &   $46.5105  \pm   0.0201$ &    $95^{+16}_{-23}$    &  \citet{2019MNRAS.487.3650H}  \\
RMID 032 &  1.715 &   $-13.8040 \pm    0.0210$ &   $44.4928 \pm    0.0210$  &  $21.1^{+22.7}_{-8.3}$   & \citet{2019ApJ...887...38G}  \\
RMID 052 &  2.305 &   $-13.1121 \pm    0.0021$  &  $45.4990  \pm   0.0021$ &   $32.6^{+6.9}_{-2.1}$   & \citet{2019ApJ...887...38G}  \\
RMID 181 &  1.675 &   $-13.7265 \pm    0.0149$ &    $44.5451 \pm    0.0149$ &  $102.1^{+26.8}_{-10.0}$  & \citet{2019ApJ...887...38G}  \\
RMID 249 &  1.717 &   $-13.3140  \pm     0.0099$ &    $44.9841 \pm    0.0099$ &    $22.8^{+31.3}_{-11.5}$    & \citet{2019ApJ...887...38G}  \\
RMID 256 &  2.244 &   $-13.4939  \pm     0.0030$ &    $45.0888 \pm     0.0030$ &    $43.1^{+49.0}_{-15.1}$   & \citet{2019ApJ...887...38G}  \\
RMID 275 &  1.577 & $-12.5961  \pm     0.0010 $ &  $45.6110 \pm     0.0010$ &   $76.7^{+10.0}_{-3.9}$   & \citet{2019ApJ...887...38G}  \\
RMID 298 &  1.635  &   $-12.6497  \pm     0.0010$ &   $45.5960 \pm     0.0010$ &    $82.3^{+64.5}_{-24.5}$  & \citet{2019ApJ...887...38G}  \\
RMID 312 &  1.924  &   $-13.3424  \pm     0.0040$ &   $45.0770 \pm     0.0040$ &    $70.9^{+9.6}_{-3.3}$   & \citet{2019ApJ...887...38G}  \\
RMID 332 &  2.581 &   $-13.1795  \pm     0.0020$ &    $45.5510 \pm     0.0020$ &    $83.8^{+23.3}_{-6.5}$ & \citet{2019ApJ...887...38G}  \\
RMID 387 &  2.426 &   $-12.9782   \pm    0.0010$ &    $45.6870  \pm    0.0010$ &    $48.4^{+34.7}_{-10.1}$    & \citet{2019ApJ...887...38G}   \\
RMID 401 &  1.822 &   $-12.8714  \pm     0.0030$ &    $45.4900 \pm     0.0030$ &    $60.6^{+36.7}_{-13.0}$    & \citet{2019ApJ...887...38G}   \\
RMID 418 &  1.418 &   $-13.0533  \pm     0.0030$ &    $45.0398 \pm     0.0030$ &    $58.6^{+51.6}_{-21.3}$   & \citet{2019ApJ...887...38G}  \\
RMID 470 &  1.879 &   $-13.5732   \pm    0.0060$ &    $44.8210 \pm     0.0060$ &    $27.4^{+63.5}_{-22.0}$   & \citet{2019ApJ...887...38G}    \\
RMID 527 &  1.647 &  $-13.4655  \pm     0.0030$ &   $44.7880 \pm     0.0030$  &  $47.3^{+13.3}_{-5.0}$    & \citet{2019ApJ...887...38G}  \\
RMID 549 & 2.275 &   $-13.2283  \pm     0.0020$  &  $45.3690 \pm     0.0020$ &   $68.9^{+31.6}_{-9.6}$     & \citet{2019ApJ...887...38G}  \\
RMID 734 &  2.332 &   $-13.0935  \pm     0.0010$ & $45.5299 \pm     0.0010$ &   $68.0^{+38.2}_{-11.5}$  & \citet{2019ApJ...887...38G}  \\
RMID 363 &  2.635 &   $-12.2525   \pm    0.0206$ & $46.4997 \pm     0.0206$ &  $300.4^{+17.1}_{-4.7}$   & \citet{2019ApJ...883L..14S}  \\
RMID 372 &  1.745 &   $-12.6952  \pm     0.0198$ &    $45.6201 \pm     0.0198$ &    $67.0^{+20.4}_{-7.4}$   & \citet{2019ApJ...883L..14S}  \\
RMID 651 &  1.486 &  $-12.7234   \pm    0.0198$ &   $45.4200  \pm    0.0198$ &   $91.7^{+56.3}_{-22.7}$    & \citet{2019ApJ...883L..14S}  \\
S5 0836+71 &  2.172 &  $-11.5354   \pm    0.0680$ &    $47.0128 \pm     0.0680$ &   $230^{+91}_{-59}$   & \citet{2021ApJ...915..129K}  \\
SBS 1116+603  &  2.646 &   $-11.5013   \pm    0.0485$ &   $47.2553 \pm     0.0485$ &    $65^{+17}_{-37}$  &  \citet{2021ApJ...915..129K}   \\
SBS 1425+606  & 3.192 &   $-11.2978  \pm 0.0356$  &    $47.6551  \pm    0.0356$ &   $285^{+30}_{-53}$    & \citet{2021ApJ...915..129K}    \\
\end{longtable}

\end{appendix}


\bsp	
\label{lastpage}
\end{document}